\documentclass[rmp,aps,twocolumn,nofootinbib]{revtex4}
\usepackage{graphics}
\usepackage{epsfig}
\usepackage[usenames]{color}

\def\inbar{\,\vrule height1.5ex width.4pt depth0pt}
\def\IR{\relax{\rm I\kern-.18em R}}
\def\IC{\relax\hbox{$\inbar\kern-.3em{\rm C}$}}

\newcommand{\pone}{2 $^{3}$P$_{1}\ $}
\newcommand{\met}{2 $^{3}$S$_{1}\ $}

\newcommand{\ptwo}{2 $^{3}$P$_{2}\ $}
\newcommand{\p}{2 $^{3}$P$\ $}

\begin{document}
\title{Cold and trapped metastable noble gases}

\author{Wim Vassen}
\email{w.vassen@vu.nl} \affiliation{LaserLaB Vrije Universiteit,
De Boelelaan 1081, 1081 HV Amsterdam, The Netherlands}%

\author{Claude Cohen-Tannoudji and Michele Leduc}
\email{leduc@lkb.ens.fr} \affiliation{Ecole Normale Superieure and College de
France, Laboratoire Kastler Brossel, 24 rue Lhomond, 75231 Paris Cedex 05, France.}%

\author{Denis Boiron and Christoph I. Westbrook}
\email{christoph.westbrook@institutoptique.fr} \affiliation{Laboratoire Charles
Fabry de l'Institut d'Optique, CNRS, Univ Paris-Sud, Campus Polytechnique RD128 91127 Palaiseau France}%
%

\author{Andrew Truscott and Ken Baldwin}
\email{kenneth.baldwin@anu.edu.au} \affiliation{ARC Centre of Excellence for
Quantum-Atom Optics \\ Research School of Physics and Engineering,\\
Australian National University, \\
Canberra, ACT 0200, Australia.}%

\author{Gerhard Birkl}
\email{gerhard.birkl@physik.tu-darmstadt.de} \affiliation{Institut f\"ur Angewandte Physik,
Technische Universit\"at Darmstadt, Schlossgartenstra\ss e 7, 64289 Darmstadt, Germany}%

\author{Pablo Cancio}
\email{pablo.canciopastor@ino.it} \affiliation{Istituto Nazionale di Otiica (INO-CNR) and
European Laboratory for Non-linear
Spectroscopy (LENS), Via N. Carrara 1, 50019 Sesto Fiorentino FI, Italy}%

\author{Marek Trippenbach} \email{matri@fuw.edu.pl} \affiliation{Wydzial Fizyki, Uniwersytet Warszawski, ul. Hoza 69, 00-681 Warszawa, Polska}%

\begin{abstract}
We review experimental work on cold, trapped metastable noble gases. 
We emphasize the aspects which distinguish work with these atoms from the large body
of work on cold, trapped atoms in general.
These aspects include detection techniques and collision processes unique to 
metastable atoms. 
We describe several experiments exploiting these unique features
in fields including atom optics and statistical physics. 
We also discuss precision measurements on these atoms including
fine structure splittings, isotope shifts, and atomic lifetimes. 
\end{abstract}

\bibliographystyle{apsrmp}
\date{\today}
\maketitle
\tableofcontents

%
%
%

\section{Introduction}
\label{sec:intro}

Cold atom physics came to life three decades ago and has been
expanding ever since.   
It started as a subfield of atomic physics and now extends to other domains such as molecular
physics, statistical physics, condensed matter and quantum information (see Nature Insight on Ultracold Matter, Nature $\bf{416}$, 205-245 (2002), for an
introduction to these topics).
The field began with the demonstration of cooling and trapping methods based on the light-matter interaction. 
This work was recognized in 1997 by a Nobel physics prize \cite{Chu:98,Cohen:98,Phillips:98}. 
The achievement of Bose-Einstein condensation (BEC) in cold dilute gases \cite{Anderson:95,Bradley:95,Davis:95,Cornell:02, Ketterle:02}, followed by the realization of
degenerate Fermi gases \cite{DeMarco:99,Schreck:01b,Truscott:01}
opened up new perspectives to study situations in which particle statistics and interactions play a central role~\cite{Pethick:02,Stringari:03}. 
In addition to providing new connections between atomic and condensed matter physics, it has long been recognized that the production of cold atoms is useful for improving atomic spectroscopy and atomic clocks ~\cite{Chu:02, Lemonde:09, Derevianko:11}. 
Another broadening of the field came a few years ago with the extension of cooling methods to molecules of two or three atoms, including cooling molecules in their ground state \cite{Carr:09}.

Two major advances of the past decade have provided particularly interesting possibilities. 
First, it became possible to modify the interaction strength between ultra-cold atoms by simply tuning a magnetic field, taking advantage of Feshbach resonances \cite{Koehler:06, Chin:10}. 
Second, one can now manipulate cold atoms in optical potentials, change the dimensionality of the
system and load the gas into tailored periodic potentials (optical lattices), created by pairs
of counter-propagating laser beams, offering the opportunity to simulate condensed matter
systems \cite{Bloch:05,Bloch:08,Jessen:96}.

\begin{figure}[h]
\centering
\includegraphics[width=0.9\linewidth]{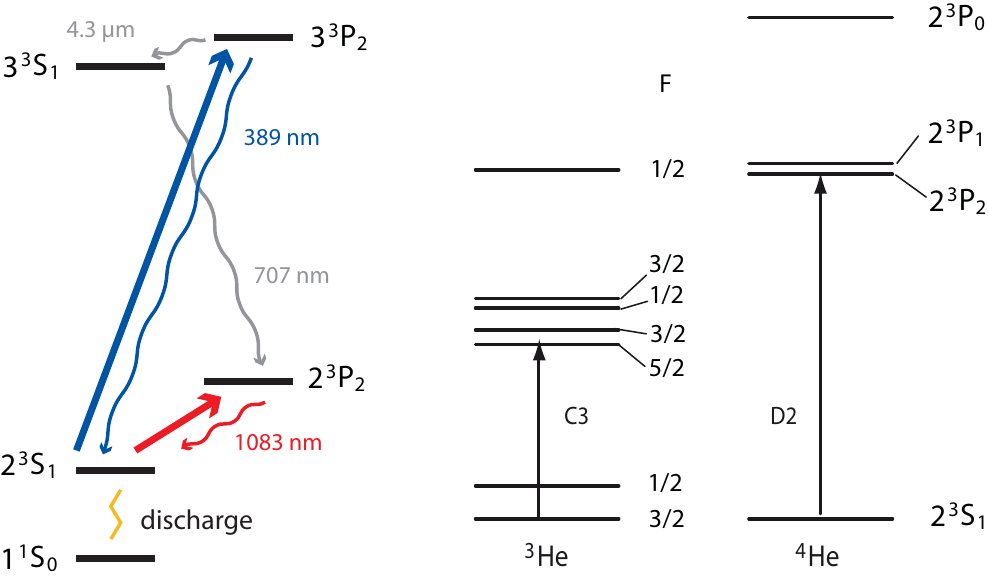}
\caption{Energy levels for helium: (left) principal helium transitions for laser cooling and trapping. The long-lived metastable 2 $^3$S$_1$ state (20 eV above the ground state) is used as an effective ground state and the 2 $^3$S$_1$ - 2 $^3$P$_2$ transition at 1083~nm transition is used for laser cooling and trapping. The 2 $^3$S$_1$ - 3 $^3$P$_2$ transition at 389 nm is used in some studies; (right) excited state manifold for the 1083~nm cooling transition in both $^3$He and $^4$He (not to scale).  For $^4$He, the 2 $^3$S$_1$ - 2 $^3$P$_2$ D2 transition is used and for $^3$He, which shows hyperfine structure as a result of the $I$=1/2 nuclear spin, the C3 transition ($F$=3/2 - $F$=5/2) is used.}
\label{fig:heliumenergylevels}
\end{figure}

\begin{figure}[h]
\centering
\includegraphics[width=8cm]{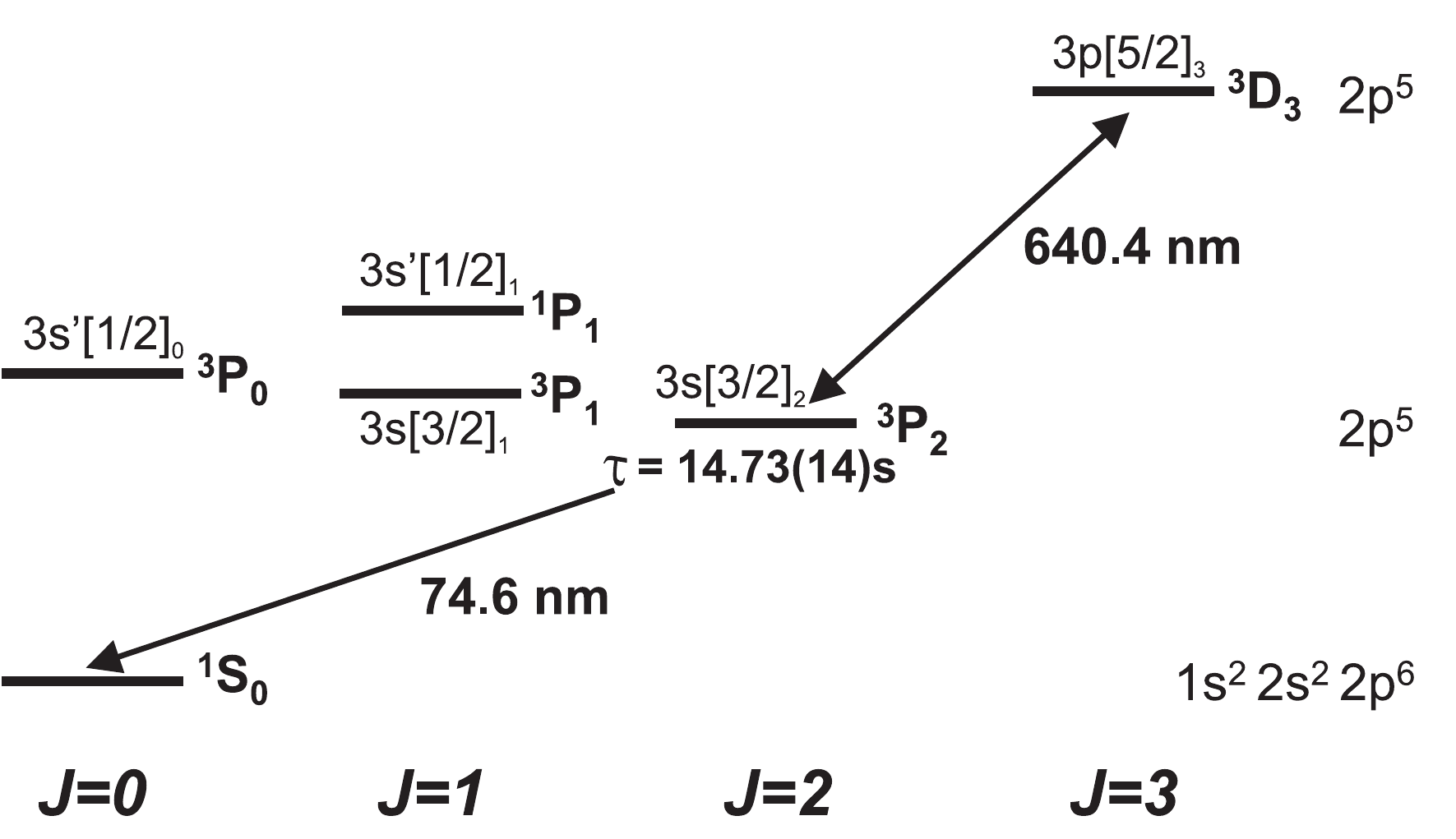}
\caption{Neon energy levels (not to scale). The metastable $^3$P$_2$ state can be manipulated by light on a transition to the upper $^3$D states.}
\label{fig:neonenergylevels}
\end{figure}

For noble gases in their ground state,  
the powerful laser cooling and manipulation methods that have
been developed cannot, unfortunately, be easily applied
because the wavelengths of the resonance lines are far in the ultraviolet.
These atoms however, have metastable excited states which are connected 
to higher-lying levels by allowed transitions and accessible with current lasers. 
With lifetimes ranging from 15 seconds to 7870 seconds 
(see table~\ref{tab:data} for a summary of atomic data of the most common metastable noble gases), these metastable states serve as effective
ground states for optical manipulation and detection.
Figures~\ref{fig:heliumenergylevels} and~\ref{fig:neonenergylevels} show the relevant
level structure of helium and neon.  

Metastable noble gas cooling began in the 1980s, with
pioneering experiments on helium \cite{Aspect:88} and neon~\cite{Shimizu:87}.
The initial work on helium concentrated on velocity selective 
coherent population trapping \cite{Aspect:88, Hack:00},
a technique which permits cooling below the single photon recoil velocity.
The experiments raise
interesting issues in statistical physics because the behavior of the 
velocity distribution is dominated by rare events and
can be interpreted with the theory of Levy flights.
This field was reviewed by \citet{Cohen:98} and \citet{Bardou:02} and we will not
discuss this work in great detail in this review. 

In 2001 
Bose-Einstein condensation of metastable helium atoms was demonstrated
\cite{Pereira:01a, Robert:01} and 2006 saw the realization of the first degenerate Fermi gas of metastable helium \cite{Mcnamara:06}.
It is remarkable that these gases can be cooled to or close to quantum degeneracy
 (kinetic energy $\approx$10$^{-10}$ eV) in spite of their large internal energy ($\approx$ 20~eV). 
In that sense, they are rather exotic and offer unusual features
that have produced many results complementary to those with other atoms,
as outlined in the present article.

At first sight the internal energy of a metastable atom is primarily a drawback, 
because it makes them difficult to produce and intrinsically fragile. 
They tend to deexcite in collisions with surfaces as well as with other atoms. 
Two colliding identical metastable noble gas atoms Ng* 
(we will use an asterisk to denote an atom in a metastable state)
have a large probability of undergoing Penning ionization,
resulting in the destruction of the metastable atoms and the production of one ion, 
one atom in the ground state, and one electron (normal Penning ionization) or a molecular ion and an electron (associative ionization):
\begin{equation}\label{eq:PI}
 \text{Ng*}+\text{Ng*} \rightarrow \bigg\{
\begin{array}{l}
     \text{Ng}+\text{Ng}^++\text{e}^-  \\
     \text{Ng}_2^++\text{e}^-.
 \end{array}
\end{equation}
In most experiments we do not discriminate between these two processes, 
so we will often refer to both of them as Penning ionization in the following.
A review of the general field of Penning ionization can be found in \citet{Siska:93}.
The existence of strong Penning ionization was initially thought to hinder the 
production of dense samples of metastable atoms, 
but, as we discuss in Sec.~\ref{sec:cold_collisions}, an
effective way to overcome this disadvantage was demonstrated, 
at least for helium and to some extent for neon: 
ionization is strongly suppressed in a spin polarized sample \cite{Vassen:96,Shlyapnikov:94,Spoden:05}. 
Furthermore, the study of Penning collisions with
ultracold atoms has turned out to be a fruitful domain in itself.

On the other hand, dealing with metastable atoms offers unique advantages. It
provides new observation methods which are not possible with ground state
atoms. Metastable atoms hitting a surface loose their energy, expelling
an electron which can be detected with good efficiency using electron multipliers.
This detection scheme provides both spatial and temporal resolution 
as well as single atom sensitivity.
Thus the study of two-particle correlations is natural (see
Sec~\ref{sec:correlations}). 
In addition, Penning collisions in the gas result in emission of
ions and electrons which can also be detected with electron multipliers.
This feature gives information about the dynamics of the atoms in real time, 
while keeping them trapped.
Real time, {\it in situ} detection is sometimes advantageous compared to the usual time of flight methods used for detecting cold atoms,
and we will give some examples in later sections.

In this review, metastable helium plays a special role. 
This is in part because it is the only metastable atom to date in which quantum 
degeneracy has been achieved.
This success is partly due to its simplicity: the absence of orbital angular momentum
in the ground state and its low mass
means that relativistic effects are not very important and that 
electron spin is very nearly conserved in collisions.
The simplicity of the helium atom has other ramifications which we will illustrate
in this review. 
Helium is the simplest atom after hydrogen, composed of only three particles 
(one nucleus and two
electrons), and its atomic structure can be calculated {\it ab-initio} with great accuracy. 
Atomic spectroscopy using cold helium is of great interest because of our 
ability to make precision measurements of lifetimes, Lamb shifts, and the fine structure.
The latter is especially important because it
leads to a spectroscopic method to determine the fine structure constant. 

Molecular potentials between interacting helium
atoms can also be calculated {\it ab-initio} with great precision, meaning
that collisions between two helium atoms in the 2 $^3$S$_1$ 
metastable state can be accurately described theoretically. 
Metastable helium dimers can be created by photoassociation and
their spectra provide accurate methods
for measuring the $s$-wave scattering length between two colliding atoms, 
a quantity of critical importance for understanding the properties of
interacting ultracold atoms.

In  the present article we will emphasize the specific features of
cold metastable noble gas atoms and describe the great variety of
experiments that they have enabled.
Our emphasis is on experiments and we therefore do not discuss
exhaustively the theoretical literature, especially that on collisions of cold metastable atoms. 
Even with respect to experiments, we do not cover all possible topics; 
ultracold neutral plasmas and atom lithography are not discussed and atom interferometry only briefly.
Discussions of these three topics can be found elsewhere~\cite{Killian:10,Baldwin:05,Cronin:09}. 

The first section (II) is devoted to the production and detection of cold clouds of metastable 
noble gas atoms, 
describing metastable production, beam slowing and trapping. 
We also discuss in some detail detection techniques which are often what
distinguishes cold, metastable atom experiments from others in the field. 
The following section (III) gives a
detailed description of the different types of collisions taking place between
cold metastable noble gases from helium to xenon, distinguishing between
inelastic collisions resulting in atom losses, and elastic ones that help
thermalizing the cloud at each step of the evaporation process. 
The next sections deal with photoassociation experiments leading to the formation of
exotic giant helium dimers as well as to a very accurate determination of the
$s$-wave scattering length of He* (IV and V). 
We then turn to precision measurements of atomic properties
in particular of atomic lifetimes, fine structure and isotope shifts (Sec.~VI). 
In the final sections (VII, VIII and IX) we discuss several experiments in atom optics and statistical
physics which have been enabled by our ability to produce cold samples of
metastable atoms and to detect them often on a single atom basis. 
In the last section (X) we outline what appear to us to be the most promising
avenues for future research.


%
%
%
%
%
%

\section{Production and detection}
\label{sec:prod}

The experimental techniques for producing cold (mK) and ultracold ($\mu$K) samples of metastable noble gas atoms are 
similar to the ones developed since the 1980's for alkali atoms.
Important steps are (1) slowing, cooling, and trapping of gas-phase atoms with near-resonant laser light
\cite{Cohen:98,Chu:98,Phillips:98,Metcalf:99}, (2) transfer of the atoms to conservative
trapping potentials, such as magnetic traps or optical dipole traps, and (3) evaporative or sympathetic 
cooling for a further 
reduction of the sample temperature and an increase in its phase-space density.
This sequence is followed by a determination of the final sample parameters, 
such as temperature, density, 
and phase space density, and
in the case of aiming at quantum degeneracy, by obtaining evidence for 
the transition to Bose-Einstein condensation (BEC) or to Fermi degeneracy.

We mentioned in the introduction the necessity of studying
noble gas atoms in metastable states. 
This fact has some essential consequences for experimental
implementations: 
(1) atoms in a metastable state have to be created in a discharge or by electron bombardement 
rather than being evaporated in an oven as for other atomic species, and the efficiency of
exciting the atoms into the metastable state in the discharge is
low with a large number of atoms remaining in the ground state;
(2) inelastic collisions between metastable noble gas atoms present a strong
additional loss mechanism since the metastable state energy exceeds
half of the ionization energy; but also (3), the high internal energy allows
the implementation of highly sensitive electronic detection techniques not available for alkali atoms.

\begin{figure}[hbtp]
\centerline{
\includegraphics[width=1.0\linewidth]{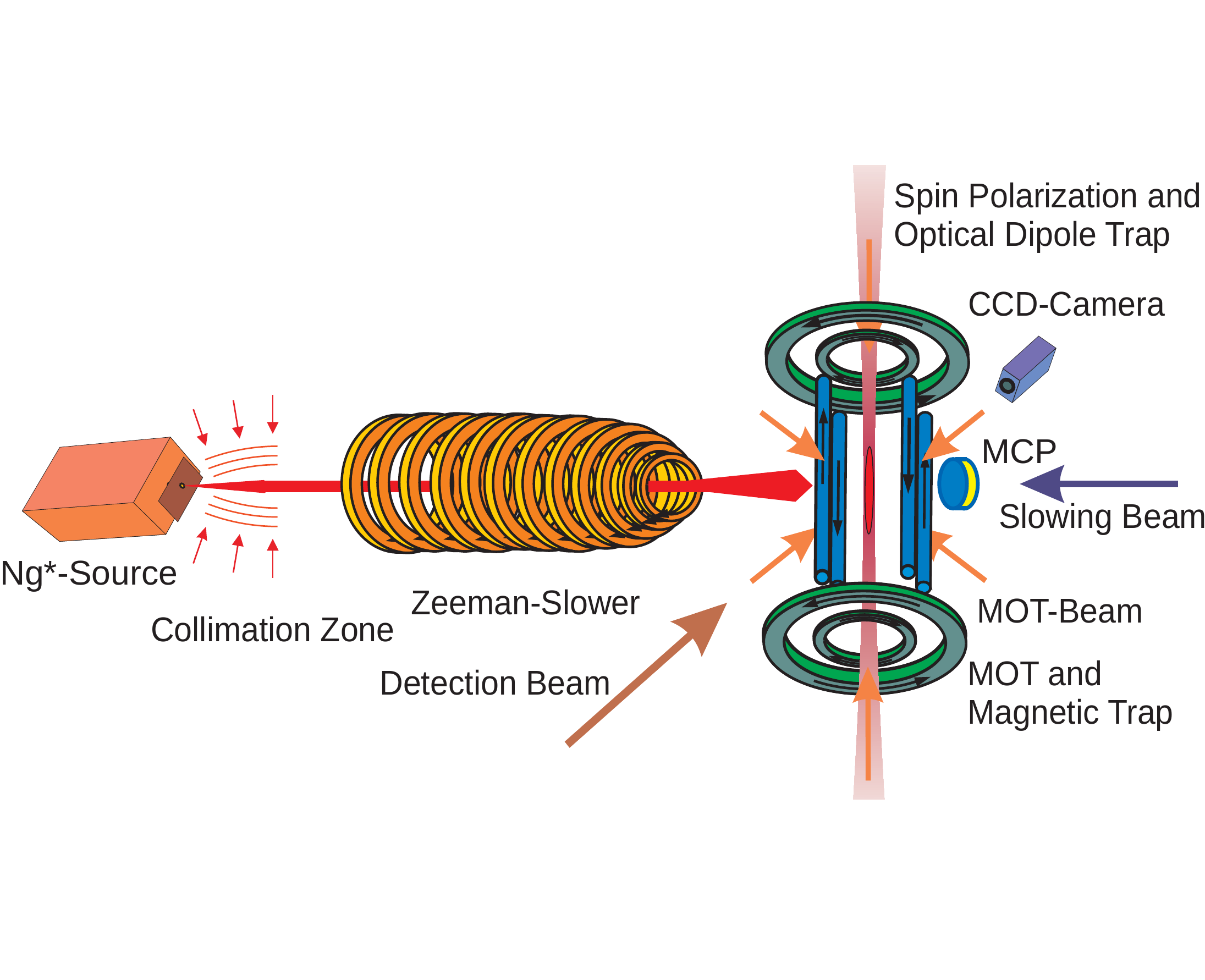}}
\caption{Example of an experimental setup
for producing ultracold metastable noble gas (Ng*) atoms. 
A more detailed example of a source is shown in Fig.~\ref{fig:NeedleSkimmer}.
After the source comes a region of transverse collimation using 
radiation pressure from resonant laser beams, and a 
Zeeman slower in which the atomic beam is slowed down. 
Specific for research on metastable noble gases is the microchannel plate (MCP) detector, 
which allows detection of ions (produced by Penning ionization) 
or metastable atoms, released from a trap.
This particular setup is used with neon atoms.
\label{fig:prod_setup}}
\end{figure}

In Fig.~\ref{fig:prod_setup},
a schematic overview of a typical experimental setup
for the production of ultracold noble gas atoms is presented. Some details may vary between different 
realizations and atomic species, 
but the general configuration remains the same.
Noble gas atoms (Ng*), excited to a metastable state by means of an electric discharge in a liquid nitrogen 
or liquid helium cooled 
gas source, are collimated by radiative pressure forces in a transverse laser field. The collimated 
beam of metastable atoms is then slowed in a Zeeman slower and used to load a magneto-optical trap 
(MOT). 
After the laser cooling phase, the atoms are spin-polarized by optical pumping and loaded into a 
static magnetic trap or into an optical dipole trap. 
Radio-frequency (rf) forced 
evaporative cooling (in magnetic traps) 
or evaporation by lowering the trap depth (in dipole traps) enables a further cooling 
of the atomic cloud. After switching off the trap, the atoms are detected either 
optically by fluorescence and absorption imaging or electronically 
on an electron multiplier detector. 

In the following sections, the different stages in this production process are described in more detail. 
The parameters given are typical values and may vary for the different noble gas species and for different 
experimental realisations.

\subsection{Discharge sources}
\label{sec:prodsource}

\begin{figure}[hbtp]
\centerline{
\includegraphics[width=0.7\linewidth]{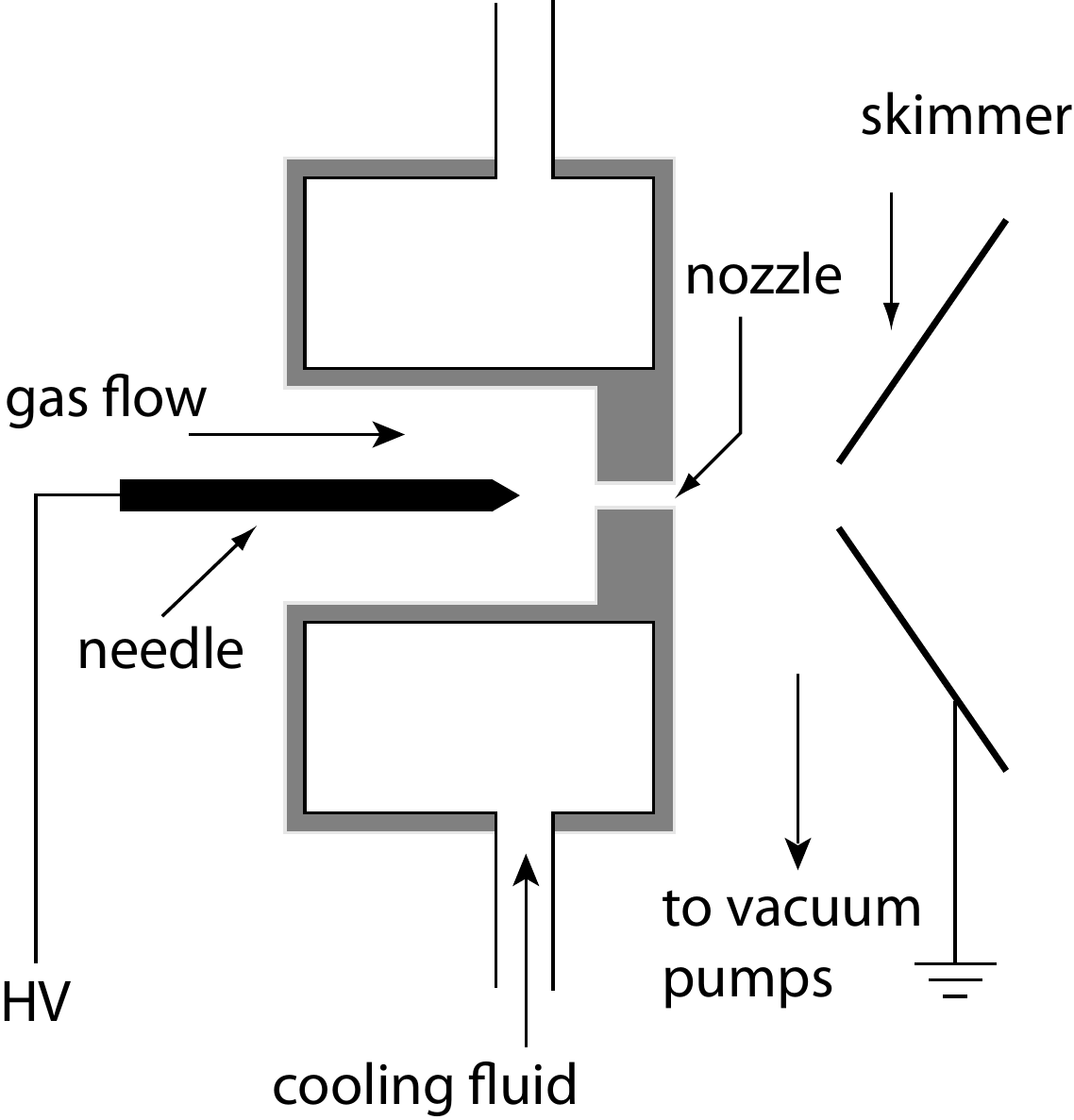}}
\caption{Example of a simple discharge source to produce a beam of metastable atoms, showing a high voltage (HV) needle cathode, a cooled nozzle and a grounded skimmer.  
\label{fig:NeedleSkimmer}}
\end{figure}

For the investigation and manipulation of cold metastable noble gas atoms, a long-lived metastable triplet 
state has to be populated.
Excitation to the metastable state can be achieved by collisions with electrons in an electric 
discharge \cite{Gay:96}.
Different types of discharge sources, such as needle-type cathode DC discharge 
sources \cite{Kawanaka:93, Rooijakkers:95}, hollow cathode DC discharge sources
\cite{Swansson:04}, or rf-driven discharge sources \cite{Carnal:91, Chen:01} are used routinely.
A quantitative comparison of different types of dc sources can be found 
in \citet{Lu:01}, \citet{Swansson:04} and \citet{Palmer:04}.
Fig.~\ref{fig:NeedleSkimmer} shows a typical example of a needle-type discharge source. The discharge
runs between the cathode and an anode close to it or through a nozzle of typically 
$\text{0.3 - 2}\mbox{ mm}$ diameter to an external anode, in some cases serving as a skimmer in addition
 (see Fig.~\ref{fig:NeedleSkimmer}).
The source is cooled by liquid nitrogen \cite{Kawanaka:93} or in some cases even by liquid 
helium \cite{Aspect:88, Carnal:91, Woestenenk:01, Swansson:04} for a reduction of the initial mean velocity of the resulting atomic beam.

Only a fraction of $10^{-5}$ to $10^{-4}$ \cite{Metcalf:99, Stas:06} of the atoms leaving the discharge 
are in the metastable state.
For this reason, efficient removal of the unwanted load of ground-state atoms is required. A skimmer
transmits only a small fraction of the solid angle of the atomic beam and serves as a first differential 
pumping stage. The huge remaining gas load is pumped away by oil-diffusion or turbo-molecular pumps backed 
by rotary vane vacuum pumps.
After the skimmer, one or more additional differential pumping stages with turbo-molecular pumps are 
implemented before the atom beam reaches the low pressure region of the main vacuum chamber.
Maximum beam intensities of $10^{15}$ metastable atoms $sr^{-1} s^{-1}$ can be achieved \cite{Lu:01}.

\subsection{Intense and slow beams of metastable atoms}
\label{sec:beams}

Due to the low fraction of metastable atoms and the long distance   
between the source and the main vacuum chamber, an increase in luminance of the atomic beam is desirable.
This is achieved by transverse collimation of the beam of metastable atoms using two-dimensional (2D) 
transverse radiative pressure forces \cite{Aspect:90, Shimizu:90, Shimizu:87, Vansteenkiste:91,Morita:91,Hoogerland:96,Rooijakkers:96,Metcalf:99,Rasel:99, Partlow:04}, focussing in a hexapole~\cite{Woestenenk:01}
or 2D magneto-optical compression \cite{Scholz:94,Schiffer:97,Labeyrie:99}. 
For maximum efficiency, the collimation zone, which has a typical length of several to a few tens of 
$\text{}\mbox{cm}$, has to be implemented as close as possible to the source, typically directly after the skimmer.
With collimation, an increase of the atom flux up to a factor of 150 has been reported \cite{Hoogerland:96b}.

In order to prevent ground-state atoms from entering the main ultrahigh vacuum (UHV) section, the axis of the collimation zone 
can be tilted with respect to the axis of the Zeeman slower~\cite{Scholz:94, Vansteenkiste:91}. The collimated metastable atoms follow this bend and propagate through an aperture,
but the ground-state atoms continue along a straight path which is then blocked. An 
alternative way to 
reduce the gas load in the UHV chamber is to use a shutter that blocks the atomic beam.

Atoms emerging from a liquid nitrogen cooled discharge,  
with a typical mean velocity of $\text{900 - 1300}\mbox{ m/s}$ in 
the case of helium and $\text{300 - 700}\mbox{ m/s}$ in the case of neon and the heavier rare gases, are too fast to be trapped directly in an optical or magnetic trap or even a MOT.
They are therefore slowed by spontaneous light forces \cite{Metcalf:99,Phillips:98} induced by a 
counter-propagating laser beam.
To remain resonant during the deceleration process, the change in Doppler shift in the atoms' rest 
frame is compensated by a position dependent Zeeman shift 
of the corresponding cycling transition~\cite{Metcalf:99,Phillips:82}.
The respective Zeeman slowing sections typically have a length between $\text{40} \mbox{ cm}$ for krypton and more than two meters for helium, although even for helium, 
a liquid helium source and careful 
optimisation of the coil design permitted a length of less than one meter \cite{Dedman:04}.
After the slowing section, the mean velocity of the atoms is reduced to a few tens of $\text{}\mbox{m/s}$~\cite{Morita:91}. An alternative slowing (and collimation) technique for He* was proposed and demonstrated in the group of Metcalf~\cite{Cashen:01, Cashen:03, Partlow:04}. Applying a bichromatic force at high power around 1083~nm a record-high capture angle from the discharge source of $\sim$0.18 radians was demonstrated over the unprecedentedly small distance  of 5~cm, albeit with considerable losses. 
To date, only slowing by 325 m/s has been demonstrated. 
It remains to be seen whether an atomic beam of He* atoms can be slowed down efficiently over a much larger velocity range and down to a few tens of m/s, suitable for trapping in a MOT.

Due to the importance of loading a large number of atoms into a MOT, the production of slow and dense beams of metastable noble gas 
atoms has been optimized thoroughly in recent years \cite{Milic:01,Tempelaars:02,Swansson:07,Kuppens:02,
Rooijakkers:97a,Tol:99,Rauner:98,Morita:91}. 
Experiments with metastable atoms are clearly at or near the state-of-the-art 
of high flux beam preparation techniques. 
Beams of metastable noble gas atoms with up to $2 \times 10^{12}$ 
atoms $s^{-1}$ \cite{Hoogerland:96b} 
and intensities on 
the order of $10^{10}$ atoms s$^{-1}$ mm$^{-2}$ \cite{Rooijakkers:96} have been achieved.

The celebrated experiments on velocity selective coherent population trapping
using metastable helium were also performed, in their first generation, as
transverse beam cooling experiments \cite{Aspect:88, Hack:00}.
In these experiments velocity widths below the single photon recoil are possible. 
The technique has been generalized to two and three dimensions
and these experiments as well as the many associated theoretical issues have been ably 
reviewed in \citet{Cohen:98,Bardou:02}.
We will therefore not discuss velocity selective population trapping further in this review 
except to point out that for the purposes of the production of intense atomic beams,
minimizing the transverse velocity of a beam is not equivalent to maximizing the flux 
through a given area downstream from the source \cite{Aspect:90}. 
Maximum flux is generally attained by maximizing the range of transverse velocities 
which can be brought sufficiently close to zero.
Thus, sub-Doppler or sub-recoil cooling techniques are less important
than having sufficient laser power, curved wavefronts and long interaction regions
to optimize beam flux as described in the references at the beginning of this section.

\subsection{Magneto-optical trapping }
\label{sec:mot}

For the creation of cold and ultracold samples of gas-phase atoms,
laser cooling and trapping techniques  \cite{Metcalf:99} have of course been remarkably successful.
In particular, the magneto-optical trap (MOT) \cite{Raab:87,Metcalf:89} has turned out to be the workhorse for almost all of these experiments.

The principle of the MOT is based on cooling via radiative pressure forces of laser light. 
Superimposed with a magnetic field, the MOT combines both the cooling effect of a three dimensional optical molasses 
with confinement using the spatial dependence of the radiative pressure force due to Zeeman shifts. The MOT consists of three pairs of circularly polarized laser beams for the three dimensions and two coils with anti-parallel currents, which produce a quadrupole magnetic field. 
For red detuning (laser frequency below the atomic 
resonance frequency), both atom trapping and cooling are achieved by radiation pressure forces at the same 
time.

The MOT has been studied both experimentally and theoretically in depth for alkali atoms 
\cite{Metcalf:99,Lindquist:92,Walker:90,Townsend:95}.
However, different limitations arise in the case of metastable noble gas atoms: the high internal
energy leads to Penning ionization (Eq.~\ref{eq:PI}), a two-body loss process which is
particularly rapid in the presence of near-resonant light.
Several observations of the loss processes are discussed in Sec.~\ref{sec:cold_collisions}.
Here we simply emphasize that this loss process puts specific contraints on the design 
and operation of a MOT for metastable noble gas atoms. 
The MOT has to be operated under conditions of high loading and low two-body loss 
rates \cite{Rooijakkers:97a,Swansson:06,Kuppens:02}. 
For optimizing loading, the incoming flux of trappable atoms has to be maximized 
(see Sec. \ref{sec:beams}) and large MOT beams 
are applied for achieving a large capture region. For minimizing losses on the other hand, 
the MOT is operated 
in a low-density regime and with low excitation rates to higher states during loading. For 
increasing the final 
atom density, at the end of the loading phase, the atom cloud may be compressed to a high-density 
non-equilibrium situation. 
With this procedure, MOTs of metastable noble gas atoms can be filled with up to $10^{10}$ atoms at 
a peak density of about $10^{10}$ atoms cm$^{-3}$ and a temperature 
of around $\text{1}\mbox{ mK}$ within a loading phase of a few 100 ms 
to 5 s~\cite{Tol:99,Kuppens:02,Zinner:03}.

The innovations in the operation of MOTs for metastable noble gas atoms include 
the use of four beams rather than the usual six \cite{Shimizu:91},
the application of laser cooling light resonant with transitions to a higher lying 
state \cite{Koelemeij:03,Koelemeij:04a,Tychkov:04}, in some cases applying Stark slowing instead of Zeeman slowing~\cite{Schumann:99,Jung:03}, the
simultaneous application of two-color laser fields for improved MOT 
performance \cite{Rooijakkers:95,Rooijakkers:97b,Kumakura:92}, as well as the simultaneous trapping 
of multiple-bosonic mixtures \cite{Feldker:11,Schuetz:11}, Bose-Fermi 
mixtures \cite{Stas:04,Schuetz:11,Feldker:11}, and metastable-alkali mixtures \cite{Byron:10a,Byron:10b,Sukenik:02,Busch:06}.

\subsection{Magnetic trapping}
\label{sec:mt}

After capturing and cooling atoms in a MOT, it is necessary to
implement evaporative cooling to further increase the phase space density
of the gas (see Sec.~\ref{sec:bectech}).
At the same time, a non-dissipative trapping 
potential is needed to keep the atoms at high phase space density. 
For metastable noble gas atoms, in most
cases, this is based on the Zeeman shift experienced in an inhomogeneous magnetic 
field \cite{Nowak:00,Herschbach:03a}. 
During transfer of the atoms to the magnetic trap, additional laser fields are applied temporarily,
in order to optically pump the atoms into the desired Zeeman substate.

Static magnetic traps, such as the Ioffe-Pritchard trap \cite{Ketterle:99} 
(see Fig.~\ref{ioffepritchardtrap}) and its 
variation, the ``cloverleaf" trap \cite{Mewes:96} are used almost exclusively (for an exception 
see \citet{Doret:09}). Due to the above described requirements for optimized beam slowing and MOT loading, 
the layout of the magnetic trapping coils has to be rather evolved. 
Configurations 
with the field coils inside or outside the vacuum chamber have been used. In any case the required 
currents are large, although 
the high magnetic moment of the metastable triplet states helps to reach strong confinement. 
Specific numbers vary 
for each implementation, but typical currents are in the range a few hundred Amperes 
(electrical power: 1 - 10 kW) giving axial and vibrational 
frequencies in the ranges of a few $\text{hundreds of}\mbox{ Hz}$ and a few $\text{tens of}\mbox{ Hz}$, respectively.
After loading atoms in a magnetic trap, it is also often desirable to further laser cool the gas.
This additional cooling is achieved with two, red detuned, circularly polarized laser beams, 
propagating in opposite directions along the magnetic field axis. 
This configuration preserves the magnetic sublevel of the atom. 
Along the radial directions cooling relies on reabsorption of scattered light by the optically
thick cloud \cite{Schmidt:03,Spoden:05,Tychkov:06}. 

Fig.~\ref{ioffepritchardtrap} shows an example for a Ioffe-Pritchard trap as used for trapping of 
metastable neon atoms \cite{Zinner:03,Spoden:05}. 
On the left, the schematic view is shown: the dipole coils give the 
axial and the Ioffe bars give the radial confinement. 
The additional offset coils are used to define the magnetic field 
bias at the trap center for selecting the trap symmetry and, for non-zero bias field, providing a quantization axis which prevents atom losses 
due to depolarization and Majorana spin flips. 
The photograph on the right shows a view of the actual trap as installed inside a vacuum chamber.
\begin{figure}[hbtp]
\centerline{
\includegraphics[width=0.5\linewidth]{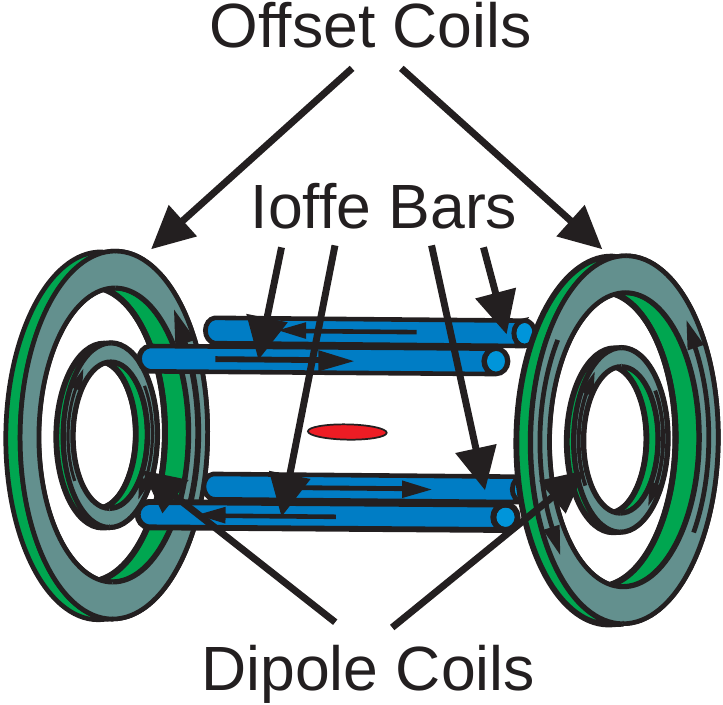}
\includegraphics[width=0.5\linewidth]{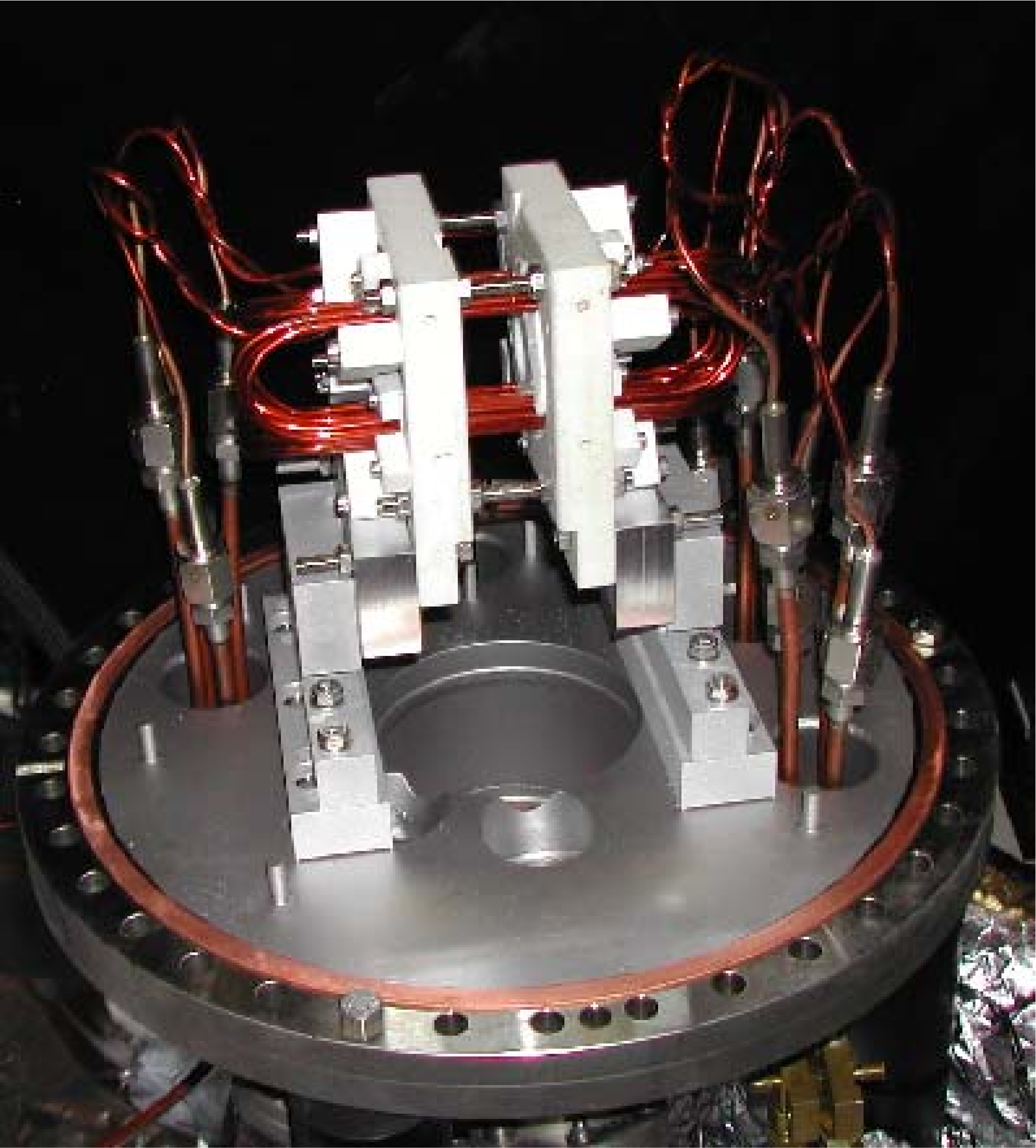}
}
\caption{Left: Schematic view of a Ioffe-Pritchard trap configuration. Right: Photograph of an actual magnetic trap. The current carrying coils and bars are created from hollow copper tubes mounted on two ceramic plates. Between the plates, the Ioffe bars are visible. The trap coils are water cooled through the hollow copper tubes. The whole assembly has a linear dimension of about 15 cm. 
\label{ioffepritchardtrap}}
\end{figure}

\subsection{Optical dipole trapping}
\label{sec:odt}

A second variant of an almost conservative trapping potential is based on the position-dependent energy shift experienced by atoms in inhomogenous light fields \cite{Grimm:00,Dalibard:85}.  
This energy shift, 
the so-called dynamic Stark shift or AC Stark shift, is used for the realization of trapping potentials of flexible 
geometries.
In addition, trapping of atoms in internal states that do not experience a sufficient magnetic energy shift, 
such as neon atoms in the $^3$P$_0$ state, or atoms that experience an anti-trapping magnetic energy shift, such as atoms in high-field seeking states can be achieved as well \cite{VDDiss:08,VanDrunen:11,Dall:10,Partridge:10}.

\begin{figure}[hbtp]
\centerline{
\includegraphics[width=0.8\linewidth]{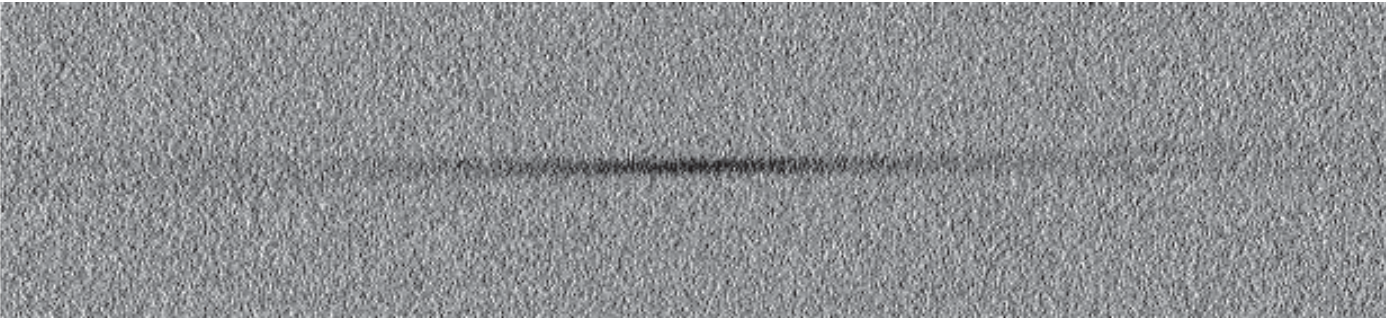}}
\caption{Absorption image of a cold sample of neon atoms trapped in an optical dipole trap formed by two intersecting 
red-detuned Gaussian laser beams at a wavelength of $\text{1064}\mbox{ nm}$. The small intersection angle results in a 
strongly elongated crossed dipole trap.
\label{fig:prod_odt}}
\end{figure}

The most common realizations are based on focused, red-detuned Gaussian laser beams. These can be used in the form 
of a single beam as trapping or guiding potential \cite{Dall:10,Partridge:10}. Superimposing two Gaussian beams under a finite angle and 
preventing interference effects through an appropriate choice of orthogonal laser polarizations 
or different detunings leads to a straightforward extension of the single-beam trap to a crossed dipole trap 
(Fig.~\ref{fig:prod_odt}). 
The main advantage of this configuration is the stronger confinement along the non-radial 
dimension with an improved performance, e.g. for evaporative cooling.
Optical dipole potentials have been applied recently to the investigation of the collisional properties of neon atoms in the $^3$P$_0$ state \cite{VDDiss:08,VanDrunen:11} and the demonstration of Bose-Einstein condensates of spin mixtures of metastable helium atoms \cite{Partridge:10}.

On the other hand, the explicit application of interference effects between multiple laser beams leads to dipole potentials with periodically varying potential surfaces. 
These so called optical lattices \cite{Jessen:96} can be operated in one, 
two, or three dimensions with 
structure sizes of the individual potential wells on the order of the wavelength of 
the light producing the interference pattern. 
In the lattice, atoms spend considerable time trapped in individual potential wells of the lattice.
At low filling factors, they therefore encounter each other less often and it is 
possible to observe a reduction of the Penning ionization rates when metastable atoms
are loaded into an optical lattice  
 \cite{Kunugita:97,Lawall:98}.

\subsection{Evaporative cooling and quantum degeneracy}
\label{sec:bectech}

As phase space densities increase, laser cooling techniques cease to be effective 
in further cooling and compressing a cold gas, and thus quantum degeneracy has never
been achieved with laser cooling alone.
To cool further it is necessary to turn to evaporative cooling techniques \cite{Hess:86,Luiten:96,Ketterle:96,Nguyen:05}. 
Evaporative cooling consists in having the high energy tail of the 
distribution of a trapped gas escape and then allowing the remaining atoms to reequilibrate.
The new equilibrium temperature is lower and under the right circumstances,
the phase space density increases. 
In magnetic traps, evaporation is induced with an "rf-knife" which expels atoms by
causing transitions to untrapped states, while in optical traps, the trapping potential 
is simply lowered \cite{Barrett:01}.

An important condition for successful evaporation is that the ratio of good (elastic) to bad (inelastic) collisions remains high enough.
Thus, a high elastic collision rate is desireable, and in the cases that have
been studied so far, the elastic cross sections are indeed quite high. 
Experiments to determine elastic scattering cross sections are described in Secs.~\ref{sec:cold_collisions} and \ref{sec:ScatteringLength}.
Inelastic loss processes present a more serious problem in metastable gases:
Penning ionization is the most rapid loss process and
evaporative cooling is only possible if it can be suppressed.
In Sec.~\ref{sec:cold_collisions} we describe in detail how Penning ionization can 
be suppressed in spin polarized samples. 
In helium, the suppression factor is of order $10^4$, 
and this has permitted efficient thermalization and evaporative 
cooling~\cite{Browaeys:01,Herschbach:03a,Nguyen:05,Tol:04}.
The supression factor is of order 100 in neon~\cite{Spoden:05, Schuetz:11} and 
has thus far prevented the attainment degeneracy with this atom. 
For the even heavier noble gas atoms no suppression of Penning ionization has been reported.

For a dilute gas of bosons, the achievement of quantum degeneracy 
corresponds to Bose-Einstein condensation (BEC) which was experimentally observed for the first 
time in 1995 for ensembles of alkali atoms \cite{Cornell:02,Ketterle:02}.
BEC was observed in metastable helium in 2001 \cite{Robert:01,Pereira:01a,Westbrook:02},
and since then three more experiments have reported this 
achievement \cite{Tychkov:06,Dall:07a,Doret:09}.
The experiment of \citet{Doret:09} is notable because it uses no laser cooling at all.
Metastable helium atoms are loaded into a magnetic trap and cooled by collisions 
with ground state helium at a temperature of 200 mK.
This sample is the starting point for purely evaporative cooling down to BEC.
This experiment is first example of achievement of BEC by buffer gas cooling \cite{Doyle:95}. 
Another experiment \cite{Tychkov:06} has produced condensates with 
more than $10^7$ condensed atoms.

In Fig.~\ref{fig:prod_bec}, we show an absorption image of a BEC after its release and expansion
from a magnetic trap.
Techniques of optical detection are discussed briefly in Sec.~\ref{sec:detection}.
The various length scales in the image 
can be used to extract parameters such as the temperature and the
chemical potential of the sample.  
Another signature of BEC is the change in ellipticity 
of an expanding BEC which has been created in an asymmetric trapping potential 
(Fig.~\ref{fig:prod_becaspect}). 
In contrast to alkali atoms, electronic detection  is also available for the investigation of 
condensates of metastable atoms (see Sec.~\ref{sec:detection}). 
Time resolved detection of the arrival of metastable atoms
gives data similar to that of an image integrated in one direction (Fig. \ref{fig:prod_becmcp}). 
A BEC signature which is unique to metastable atoms is the observation of an abrupt increase in the ionization rate as the gas passes through the BEC transition \cite{Seidelin:03b,Tychkov:06}.
This type of observation is described in more detail in Section \ref{sec:OtherExperiments}.

Simple evaporative cooling cannot be used to produce a degenerate Fermi gas,
since the elastic collision cross section between identical low energy fermions  vanishes at
low temperature.
Some form of ``sympathetic cooling", in which one species cools another, must be applied. 
In the case of helium, sympathetic cooling of $^3$He* atoms (the fermionic isotope)
has been implemented by making them interact with an evaporatively cooled 
sample of the bosonic isotope $^4$He*.
The two isotopes were held simultaneously in the same magnetic 
trap~\cite{Mcnamara:06,Vassen:07}. 
In Sec.~\ref{sec:correlations} we describe experiments on such a mixture
to compare directly the quantum statistical behavior of bosons and fermions.

\subsection{Detection of metastable noble gas atoms}
\label{sec:detection}

\subsubsection{Optical detection}

Almost all cold atom experiments have relied on atom-laser interactions to detect the atoms. 
In the most common technique a near resonant ``probe" laser beam traverses a sample of cold 
atoms which in turn scatter the light. One can then either collect the scattered light and form a 
positive image of the sample (``fluorescence imaging"), or use the unscattered light in which case the sample appears as 
a shadow in the beam (``absorption imaging"). A third possibility is to tune the probe laser far enough from resonance 
that dispersive effects dominate. The atomic sample acts as a transparent medium with a real 
index of refraction. Phase contrast techniques then allow one to form an image. Imaging methods 
involve many subtleties and tradeoffs, and some of them are reviewed in \citet{Ketterle:99}.

One can also glean information from non-imaging techniques using laser interactions.
As an example, a laser standing wave can form a diffraction grating for matter waves
and thus act as a sort of spectrometer.
This technique is often referred to as ``Bragg spectroscopy" \cite{Stenger:99}.
Optical detection of the atoms however is still used in such experiments although the ability to
image the atoms plays only a minor role.

\begin{figure}[hbtp]
\centerline{
\includegraphics[width=0.6\linewidth]{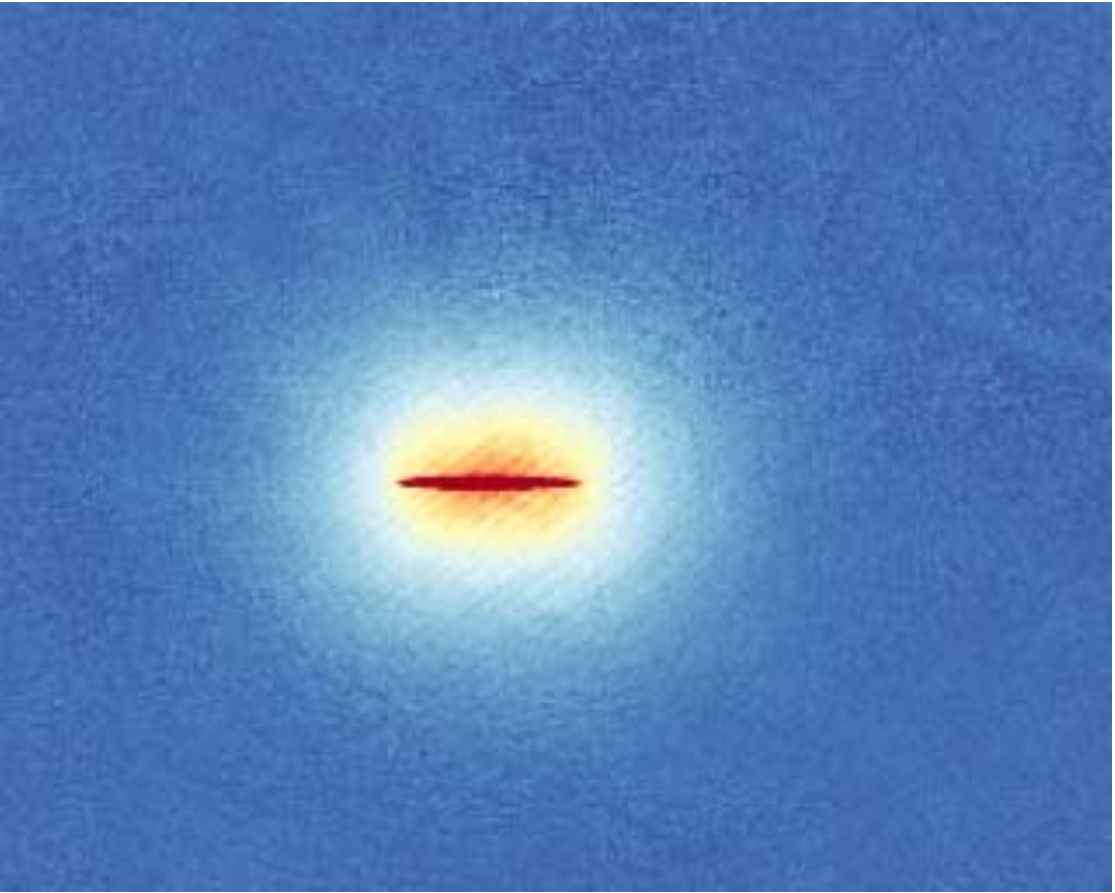}}
\caption{Observation of Bose-Einstein condensation of metastable $^4$He. The figure shows an 
image of an expanding cloud of ultracold atoms after expansion from a magnetic trap. 
The elongated core shows the BEC whereas 
the circular cloud with larger diameter corresponds to the thermal cloud of not-condensed atoms.
The image was acquired with an InGaAs CCD camera. 
\label{fig:prod_bec}}
\end{figure}

All these techniques are in principle also available when using metastable atoms.
To date only imaging techniques have been used with metastable atoms,
and we will not attempt to review the use of optical probing of cold metastable
samples since most examples do not differ substantially from their use
in other atomic species.
We will only comment on one case, metastable helium, which
presents some particularities because of its small
mass, small natural linewidth, 
and primary transition wavelength (see Table \ref{tab:energies}).
This $\lambda=1083\,{\rm nm}$ wavelength is poorly adapted to conventional CCD detectors based
on silicon.
The best measured quantum efficiencies are of order 1\%~\cite{Tychkov:08}.
CCD detectors based on InGaAs technology are becoming available.
These detectors have reported much higher quantum efficiencies,
and first results have become available and are very promising 
(see Fig.~\ref{fig:prod_bec} for a sample image).

The optical detection process depends on the scattering of nearly resonant
light by the atoms, and of course the more light is scattered, 
the greater the signal available for detection.
The scattering however is accompanied by the acceleration of the atoms
along the laser propagation direction due to radiation pressure.
At some point the Doppler shift associated with the increased velocity becomes
comparable to the natural linewidth $\Gamma/(2\pi)$.
Unless the laser's frequency is swept to compensate this
Doppler shift, the light ceases to be resonant and the scattering rate drops.
The typical number of scattered photons is given by $\Gamma/ k v_{\rm rec}$,
where  $v_{\rm rec}=\hbar k/m$ is the velocity transfer associated with the absorption
or emission of one photon of wavevector $k$ by an atom of mass $m$.
For He* this quantity is about 20 for the 2 $^3$S - 2 $^3$P transition at $\lambda=1083\,{\rm nm}$.
For comparison, the same quantity for Ne* atoms and the 2$p^5$ 3s $^3P_2$ - 2$p^5$ 3p $^3 D_3$ 
transition at $\lambda=640\,{\rm nm}$ is about 170
while for Rb atoms and the
5S-5P transition at $\lambda=780\,{\rm nm}$, $\Gamma/ k v_{\rm rec}$ is about 800.
This effect severely limits the available signal for He*, and this
limitation is compounded by a very poor quantum efficiency when using
Si based CCD cameras.

Acceleration of the atoms is not the only difficulty related
to the large recoil velocity in He*.
Successive absorption and emission processes also increase
the three-dimensional momentum spread due to fluctuations in the direction of
absorbed and scattered photons.
Cold atoms thus necessarily heat during optical imaging.
For long exposures this heating can result in a loss of signal, but an additional difficulty is that the atoms will
also move in space during imaging, and smear out their positions so as
to mimic a loss in optical resolution.

\begin{figure}[hbtp]
\centerline{
\includegraphics[width=0.8\linewidth]{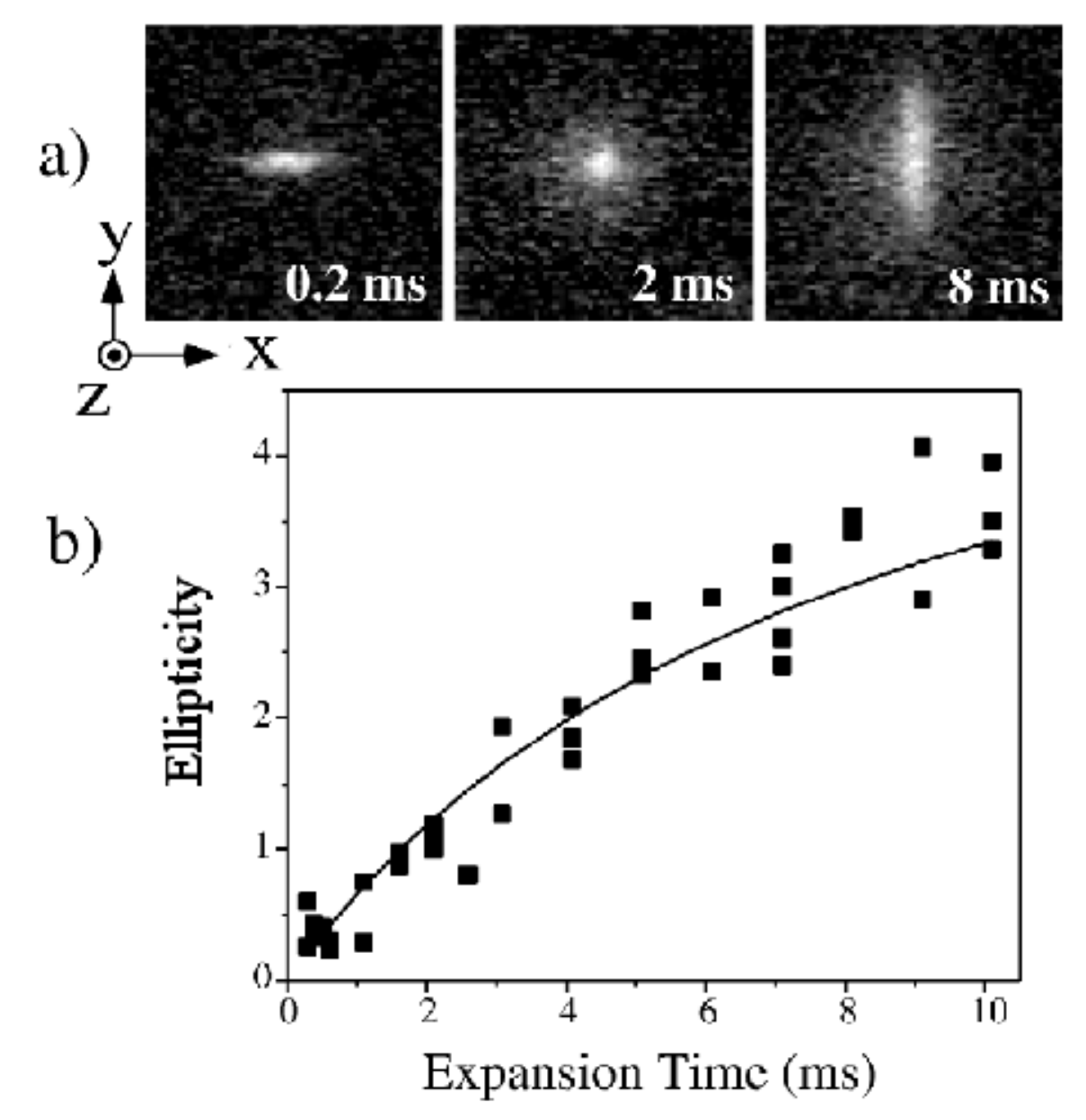}}
\caption{Direct evidence of the occurence of a BEC of $^4$He~\cite{Pereira:01a}: (a) images of a BEC for the given expansion times; (b) ellipticity extracted from the images. 
Due to the asymmetric potential surface of the magnetic 
trap used for condensation, the ellipticity of the expanding cloud of condensed atoms changes
from below 1 to above 1 for increasing expansion times. 
This behavior contrasts to that of a thermal cloud whose ellipticity approaches unity
for long expansion times.
\label{fig:prod_becaspect}}
\end{figure}

\begin{figure}[hbtp]
\centerline{
\includegraphics[width=0.8\linewidth]{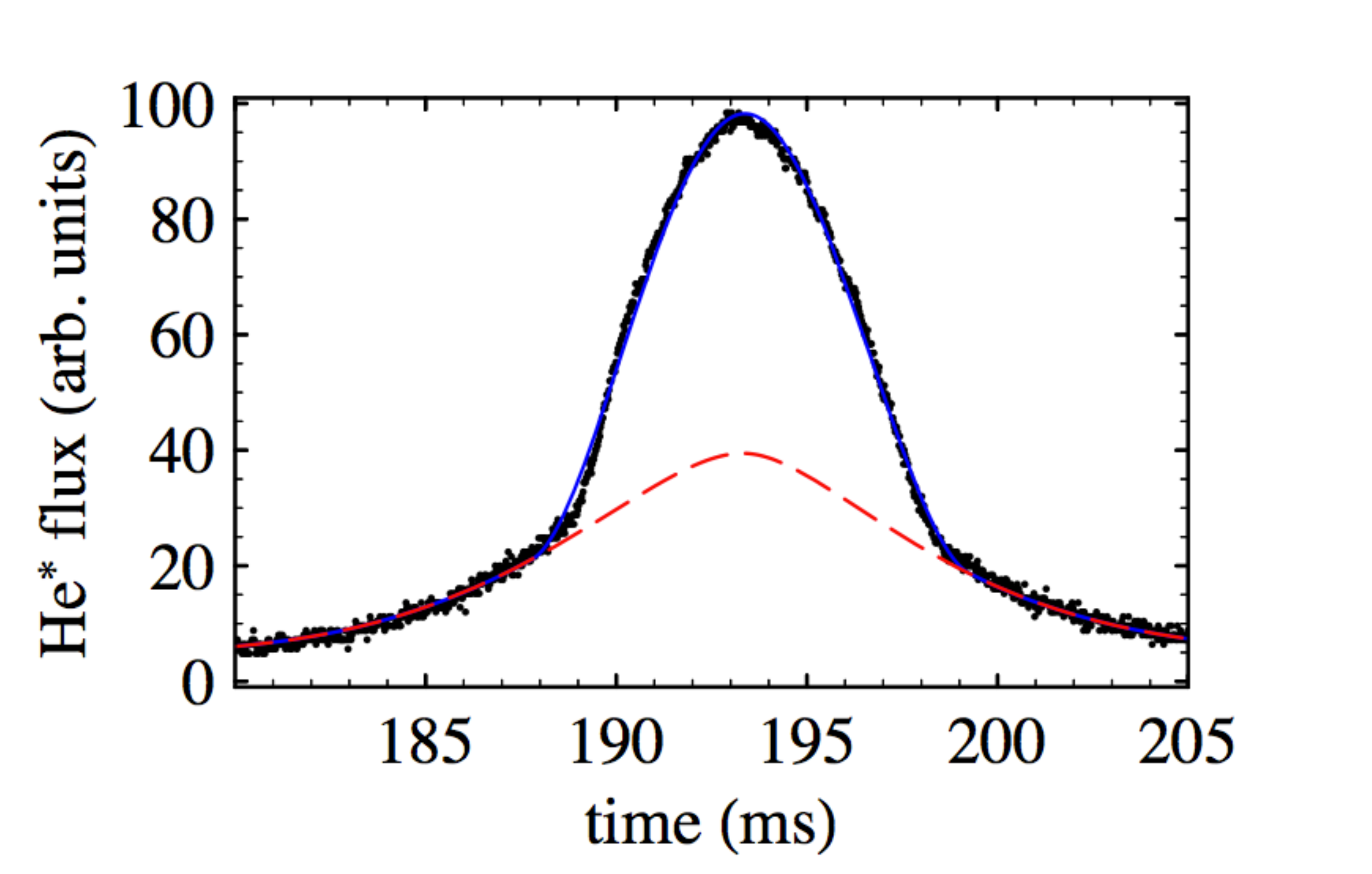}}
\caption{Detection of Bose-Einstein condensation of metastable helium via the electron 
current induced in a micro-channel plate (MCP) detector. 
The internal energy of the metastable atoms 
expels electrons when hitting the surface of the detector mounted below the trap after 
falling under gravity. 
The electrons are amplified in the MCP and the temporal distribution of the detected 
current can be used to extract the parameters of the BEC (narrow distribution, fit in solid blue) 
and the  thermal atom fraction (wide distribution, fit in dashed red).
\label{fig:prod_becmcp}}
\end{figure}

One example of circumventing the large recoil issue is found in the work of \citet{Pereira:01a}.
These workers used two counter-propagating beams for imaging to ensure a zero net radiation pressure on the atoms.
Indeed they tuned the lasers so as to have a net cooling effect.
As shown in Fig.~\ref{fig:prod_becaspect}, this technique realised good quality
absorption images. 
A similar technique was used in the work of \citet{Tol:05},
with the difference that the lasers were tuned on resonance and had an 
intensity far below saturation.
This technique provides a reliable determination of the number of trapped atoms.

An alternative approach for imaging might be to use light resonant
with the 2 $^3$S - 3 $^3$P transition at
$\lambda=389\,{\rm nm}$.
Although the figure of merit  $\Gamma / k v_{\rm rec}$
is a factor of 8 smaller than for the 2 $^3$S - 2 $^3$P transition, the increased
quantum efficiency of Si detectors may compensate the loss
of photons.
\citet{Koelemeij:03} created a MOT using the 2 $^3$S - 3 $^3$P
but they did not attempt to perform absorption imaging 
at the 389 nm wavelength.

\subsubsection{Direct detection using electron multipliers}
\label{sec:ElectronicDetection}
The large internal energy stored in a metastable noble gas atom 
(see Table~\ref{tab:energies}) has naturally led to the
introduction of techniques exploiting this energy and which are not available when using ground state atoms.
Metastable atoms can ionize other atoms or eject electrons from a solid.
These processes therefore lead to the emission of charged particles which can be electronically multiplied and easily detected.

When a metastable atom comes into contact with a metal surface, an electron can be ejected.  Electron ejection probabilities are difficult to measure absolutely and \citet{Siska:93}, \citet{Hotop:96} and \citet{Harada:97} have reviewed some of these measurements, which we summarize here.
For He* electron ejection probabilities in the range of 50-70\% have been reported for various metals (gold, stainless steel).
In Ne* the probability is in the range 30-50\%.
For the other metastable noble gas atoms, this probability can be much smaller.
Roughly speaking, the probability goes down as the metastable state energy decreases,
resulting in values on the order of 1\% for Kr* and Xe*.
The electron ejection process however, is complex and does not depend exclusively on the metastable energy \cite{Hotop:96}.

Even with small electron yields, electron multiplier techniques are an attractive alternative to the use of optical detection methods, because of their low background and fast (sub ns) response.
They can also have sufficient gain to be sensitive to individual atoms. 
A simple detector is the discrete dynode electron multiplier, essentially a photomultiplier without a photocathode.
Laser cooled He* \cite{Aspect:88,Bardou:92}, Ne* \cite{Shimizu:89}, Ar* \cite{Faulstich:92} and Kr*
 \cite{Katori:94,Kunugita:97} have been detected in this way.
\citet{Katori:94} estimated a quantum efficiency
of 2.7\% for Kr* on their detector, consistent with data in the reviews above.

Channel electron multipliers (or ``Channeltrons") operate on a slightly different principle. Instead of having discrete dynodes, the multiplication structure is a coiled, highly resistive tube along which a potential is applied. Electrons travel down the tube colliding with the wall of the tube creating secondary electrons.
We refer the reader to work such as that of Samson \cite{Samson:00} for a technical
discussion of electron multiplication techniques.
As an example, atom interferometry experiments have used channeltron detection of Ar*  \cite{Rasel:95}.
Metastable Ne atoms in a MOT were also detected with a
channeltron \cite{Kuppens:02}.
Closely related to the channeltron is the microchannel plate or ``MCP" detector.
In such a detector, the tube is scaled down to micron size and instead of a single tube,
one has an array of thousands or millions of tubes.
The electron amplification principle is very similar.
An MCP in which the front face served as a simple ionizing surface was
used to observe the BEC transition in \citet{Robert:01}
as well as in \citet{Tychkov:06}.
Quantum efficiencies for He* on microchannel plates of 10\% have
been reported \cite{Tol:05,Jaskula:10}.
MCP's have also been used to detect laser cooled metastable neon \cite{Shimizu:92,Spoden:05} and xenon \cite{Lawall:98}.

An example of the use of electronic detection to achieve fast, low background single atom 
detection can be found in the work of \citet{Yasuda:96} (see Sec.~\ref{subsec:correlations}). 
In this experiment a set of 4 MCPs was used to observe correlations individual atoms 
in a beam of Ne* atoms released from a MOT.
Atoms fell onto a gold surface and produced electrons.
The MCP's collected electrons from different parts of the surface and coincident atoms were detected with ns timing resolution, allowing a correlation time of 100 ns to be demonstrated.

MCP's come in many shapes and sizes and also have imaging capability \cite{Lapington:04}.
Thus MCP's have been widely used in the cold metastable atom community as an alternative
or a complement to optical imaging.
Imaging is possible by placing a phosphor screen behind the MCP and recording the phosphorescence with a camera \cite{Shimizu:92,Lawall:94a,Lawall:94b,Dall:09,Rauner:98}.
A drawback to the phosphor screen method is that it relies on a video camera to acquire the data, and thus the frame rate, or temporal resolution is necessarily limited by the video frame rate, typically 20 ms.

An MCP imaging technique that is not limited by CCD camera technology involves the use of a resistive anode. The anode collects the charge from the MCP at the four corners of a square.
Analysis of the ratios of the charge at each anode allow one to isolate the position of the source of the charge. Such a detector has been used with He* \cite{Lawall:95,Koolen:02}.
Spatial resolution of several tens of $\mu$m has been reported using this technique with cold metastable atoms.
The timing resolution of the technique can be as small as 1~$\mu$s, although in the experiments of Lawall {\it et al.}, the time resolution was 5~$\mu$s, limiting the rate of detection events.
Thus the detector must be used in situations of low atom flux.

Another imaging technique uses an MCP in connection with a delay line anode
\cite{Jagutzki:02}.
In this technique, the anode is configured as a transmission line with a well defined propagation speed.
The charge travels to opposite ends of the line and high precision time to digital conversion provides the position information.
This technique was used with
He*~\cite{Schellekens:05,Jeltes:07,Perrin:08,Manning:10}.
Pulse widths with such anodes are typically in the few ns range, and thus the detectors can be extremely fast.
Simultaneous, multiple hits can in principle be detected \cite{Jagutzki:02}.
This capability has not yet been used with metastable detection.
In experiments with metastable atoms to date, a typical deadtime of at least 100 ns is imposed to simplify the software reconstruction of events.
The permissible atom flux is thus somewhat higher than for an MCP with a resistive anode.
When the flux of detected atoms approaches $10^6\,\rm{s}^{-1}$ however,
the MCP's, which must be used in pulse counting mode (applying close to the maximum allowable voltage), can exhibit saturation effects \cite{Schellekens:07,Krachmalnicoff:09}.
Thus the maximum flux with an MCP + delay line detector is only about an order of magnitude
higher than for an MCP + resistive anode.

The spatial resolution of a delay-line anode detector depends on the precision of the timing electronics.
The larger the MCP the better the position determination because the
timing accuracy generally does not depend on the size of the MCP.
Using He*, FWHM's of 500 - 700~$\mu$m \cite{Schellekens:05,Schellekens:07} and 100~$\mu$m \cite{Manning:10} have been reported.
Workers using other particles (ions etc) have reported resolutions of a 
few tens of $\mu$m \cite{Jagutzki:02}.

\subsubsection {Detection by use of Penning ionization}
We have already mentioned that a significant, indeed often dominant, decay mechanism 
for cold samples of metastable atoms
involves the production of ions by Penning ionization (see Eq.~\ref{eq:PI}).
In Sec.~\ref{sec:cold_collisions}, we will discuss the physics of these collisions. 
Here we concentrate on the use of the ionization signal as a diagnostic. 
Almost all residual gases can be ionized by Penning ionization 
and one metastable atom can ionize another.
If one collects the ions or the electrons from these ionization processes
on an electron multiplier, one also has a
fast, low noise signal to herald the presence of metastable atoms.
In a low density sample
the ion signal is primarily due to ionization of residual gas.
Thus the signal is proportional to the number of trapped atoms \cite{Nowak:00}.
This signal was used as a real time monitor of the trapped atom number in recent
measurements of metastable state lifetimes~\cite{Dall:08a}.

When the density is high enough, ionizing collisions among the trapped atoms
(two-body collisions) become important.
In this situation the ion signal is proportional to the density squared, integrated
over the trap volume, and leads to a non-exponential decay of the trapped atom number.
For metastable atoms in a MOT, light assisted collisions
between the trapped atoms can have large rate constants and the 
two-body losses are often dominant.
This phenomenon has been observed in every metastable noble gas MOT 
equipped with an ion detector
\cite{Spoden:05,Shimizu:89,Bardou:92,Walhout:94,Kunugita:97,Mastwijk:98,Kumakura:99,Tol:99,Rooijakkers:97a}.
The experiment of Bardou {\it et al.}, demonstrated that the ion rate could
be a probe of the local density of atoms.
A cloud of atoms in  MOT was allowed to expand briefly by turning off the laser beams.
When the beams were turned on again, the recompression of the cloud was 
observed as an increase of the ionization rate over about 1~ms. 

In a magnetically trapped BEC, light assisted collisions are absent, but the high
density nevertheless results in Penning ionization.
Both two-body and three-body collisions can contribute to
ionization signals \cite{Sirjean:02,Tychkov:06}.
As in a MOT, the ionization signal can be used as a real time probe of the local atomic density.
As one crosses the BEC threshhold, the abrupt increase in density
due to condensation results in a corresponding increase in ionization~\cite{Robert:01,Tychkov:06}.
A detailed analysis of such signals can be found in~\citet{Seidelin:03b} and is discussed
further in Sec.~\ref{sec:OtherExperiments}.
In Sec.~\ref{sec:AtomOptics} we describe the use of a Penning ionization signal to 
stabilize an atom laser~\cite{Dall:08b}.

%

%
%
%
%
%
%
%

\section{Cold collisions}
\label{sec:cold_collisions}
When the atomic de Broglie wavelength is comparable to or larger than the range of the interatomic potential between two metastable atoms, 
collisions are defined as cold collisions~\cite{Weiner:99,Julienne:89}. 
In that case only a few partial waves contribute to the collision cross section, 
either elastic or inelastic.  
Quantum threshold rules are then valid for collisions in the absence of light close to an atomic resonance. 
In the presence of near-resonant light, such as in a MOT, the interaction potential becomes long-range due to the 
resonant dipole interaction, and higher-order partial waves need to be included. 
At the temperatures relevant for this review, i.e. 
around or below 1~mK, one is in the cold collision regime. 
\citet{Weiner:99} provide an excellent review of cold collision theory, 
both in the presence of light and in the dark. 
That review also discusses experiments in noble gas MOTs 
on ``optical shielding", in which the control of collision dynamics by near-resonant light
is studied. 
We will not cover this topic again here and will primarily restrict ourselves to research performed after that review was published.

Collisions between metastable atoms are different from those between alkalis because inelastic effects play an even more important role~\cite{Julienne:89}. Penning ionization (Eq.~\ref{eq:PI}) is generally rapid and limits the densities that can be obtained in a MOT.
Only under specific conditions can these Penning ionization losses be suppressed (see Sec.~\ref{sec:PI}). 
Penning ionization thus limits the options for realizing cold and dense gases of metastables and 
hampers realization of quantum degeneracy in most noble gas atoms. 

The interaction that drives the auto-ionizing transitions of
Eq.~(\ref{eq:PI}) is electrostatic and only induces transitions between molecular
states of equal total electronic spin. For all noble gas atoms, the
lefthand side of Eq.~\ref{eq:PI} contains two $s=1$ atoms with total spin
$S=0,1$ or 2, while the righthand side, with
$s=\frac{1}{2}$ (twice) and $s=0$, can only form states with $S=0$ or 1. Clearly, the
total electronic spin can only be conserved if $S=0$ or 1, and a
Penning ionization reaction with $S=2$ would involve a violation of spin
conservation. If spin conservation in collisions holds (Wigners spin-conservation rule) it is only the fully spin-stretched states of the metastable atoms that may show suppression of Penning ionization, as these add up to an entrance channel with one unit of total electron spin higher than the maximum spin of the product states.

For He* this is the full argument as there is no orbital angular momentum to consider. 
The suppression of Penning ionization (first observed by~\citet{Hill:72}) turns out to be 4 orders of magnitude, limited by the weak spin-dipole interaction, discussed in Sec.~\ref{sec:spindipole}. In all other noble gas atoms the metastable state has orbital angular momentum $l=1$: the excitation of one $p$ electron out of the $np^6$ ground state creates a metastable state $np^5(n+1)s$ ($n$=2, 3, 4, 5 for Ne, Ar, Kr, and Xe). The $LS$ coupling scheme is expected to break down for heavier atoms. Only for the fully spin-stretched states, however, is $S$ a good quantum number and spin-conservation in collisions (Eq.~\ref{eq:PI}) will still hold if there are no additional spin-state changing interactions. 
On the other hand, as the core is anisotropic, interactions that depend on the relative orientations of the colliding atoms may induce a spin flip to a state in which ionization can occur. These interactions can be so strong that spin-polarization of a gas may even increase the Penning ionization rate~\cite{Orzel:99}.

In the rest of this section we will further elaborate on elastic and inelastic collision properties, first for helium, for which we will discuss a simple model to understand Penning ionization losses and later for the other noble gas atoms and mixtures. We will primarily focus on comparison with theory and comparison of the observed loss rates for the different noble gas atoms. In Table~\ref{tab:data} we have compiled atomic data and collisional properties of all noble gases as we have found in literature.

\begin{center}
\begin{table*}[ht]
\caption{Atomic Data of some common metastable noble gases.
\\The energy of the metastable state is given with respect to the ground state.
The wavelength $\lambda$ and lifetime $\tau$ refer to the most commonly used 
laser cooling transition.
The Doppler limit is $\hbar/(2\tau k_{\mathrm B})$, the recoil limit is 
$h^2/(2 m \lambda^2 k_{\mathrm B})$. "Lifetime" refers to the lifetime of the metastable state.
Loss rate (Pol. and Unpol.) refer to the two-body inelastic rate constants $\beta (= 2K_{SS})$
for a polarized or unpolarized sample in the absence of any resonant light. Reference for energies and vacuum wavelengths: NIST Atomic Spectra Database 
(http://www.nist.gov/physlab/data/asd.cfm). References for other parameters given: (a) \citet{Dall:08a}; (b) 
\citet{Hodgman:09b}; (c) \citet{Hodgman:09a}; (d) \citet{Zinner:03}; (e) \citet{Katori:93}; (f) \citet{Lefers:02};
(g) \citet{Walhout:95b}; (h) \citet{Lach:01}; (i) \citet{Indelicato:94}; (j) \citet{Tachiev:02};
(k) \citet{Small-Warren:75}; (l) \citet{Moal:06}; (m) \citet{Spoden:05}; (n) \citet{Tychkov:06} (see Sec.~\ref{sec:spindipole}); (o) \citet{Stas:06} (see Fig.~\ref{fig:McNamaraMOT}); (p) \citet{Busch:06}, value obtained in a MOT without extrapolation to vanishing light intensity; (q) \citet{Katori:94}; (r) \citet{Kuppens:02};
(s) \citet{VDDiss:08}; (t) \citet{Orzel:99}; (u) \citet{Katori:95}.
}
\hfill{}
\begin{tabular}{l c c c c c c c } \hline \hline
Atom Species & $^{3}$He*& $^{4}$He*& $^{20}$Ne*& $^{22}$Ne*& $^{40}$Ar*& $^{84}$Kr*& $^{132}$Xe*\\  
 \hline
Abundance& 0.01 & 99.99 & 90.48 & 9.25 & 99.60 & 57.00 & 26.91 \\ 
Metastable State & 2 $^3$S$_1$&2 $^3$S$_1$&$3s[3/2]_{2}(^{3}$P$_{2})$&
$3s[3/2]_{2}(^{3}$P$_{2})$&$4s[3/2]_{2}(^{3}$P$_{2})$&$5s[3/2]_{2}(^{3}$P$_{2})$&$6s[3/2]_{2}(^{3}$P$_{2})$ \\ 
Energy (eV) & 19.820 & 19.820 &16.619&16.619 &11.548&9.915&8.315 \\ 
Laser Cool. Wavelength $\lambda$ (nm)& 1083.46& 1083.33 & 640.40 & 640.40 & 811.75 & 811.51 & 882.18 \\ 
Upper State Lifetime $\tau$ (ns)& 97.89& 97.89 & 19.5 & 19.5 & 30.2 & 28.0 & 33.03 \\ 
Doppler Limit ($\mu$K)& 38.95& 38.95 & 196 & 196 & 140.96 & 133.4 & 115.64 \\ 
Recoil Limit ($\mu$K)& 5.433 & 4.075 & 2.337 & 2.125 & 0.727 & 0.346 & 0.186 \\ 
Exp. Lifetime  (s)& - & 7870(510)$^c$ & 14.73(14)$^d$ & - & $38^{+8}_{-5}$ $^e$ & 28.3(1.8)$^f$, $39^{+5\, e}_{-4}$ & 42.9(9)$^g$ \\  
Theory Lifetime  (s) & 7860$^h$& 7860$^h$ & 22$^i$,17.1$^j$,24$^k$  & 22$^i$,17.1$^j$,24$^k$  
& 56$^i$, 51$^k$ & 85$^i$, 63$^k$& 150$^i$, 96$^k$  \\ 
Scat. Length (nm)& - & 7.512$^l$ & -9.5(2.1)$^m$ & $7.9_{-2.7}^{+4.2\, m}$ & - & -& - \\ 
Pol. Loss Rate ($10^{-14}$cm$^{3}$/s) & - &2(1)$^n$&650(180)$^m$ & 1200(300)$^m$ & - & 40000$^{q,u}$ & 6000$^{g,t}$ \\ 
Unpol. Loss Rate ($10^{-11}$cm$^{3}$/s) & 38(6)$^o$ & 20(4)$^o$  & 50(30)$^r$ & $100_{-50}^{+40\, s}$ & 58(17)$^p$ & 40$^q$ & 6(2)$^g$ \\ 
\hline \hline
\end{tabular}
\hfill{}
\label{tab:data}
\end{table*}
\end{center}

\subsection{Elastic collisions of He}
The $s$-wave scattering length is the most important property in cold collision physics. 
At low enough temperatures, for bosons we only need to take $s$-wave interactions into account while 
for fermions $p$-wave interactions are the lowest allowed by symmetrization requirements. 
The $s$-wave elastic cross section becomes a constant approaching $T$=0, while the $p$-wave
elastic cross section goes to zero as $T^2$. 
As discussed before, for the noble gases only spin-polarized atoms may show sufficient 
suppression of Penning ionization to allow long enough lifetimes to reach BEC. 
For spin-polarized $^4$He* atoms, the interaction potential is $^5\Sigma_g^+$ ($S$=2), 
as both atoms have the maximum $m$=+1.
Since the beginning of the 1990s 
theorists~\cite{Starck:94, Dickinson:04, Przybytek:05} have calculated 
this $^5\Sigma_g^+$ potential to determine the energy of its least bound state as 
well as the scattering length. From these calculations it was predicted~\cite{Starck:94} 
that the $^4$He* scattering length should be large and positive (+8~nm), stimulating 
experimental research towards BEC in metastable helium. Apart from calculating the 
scattering length, in the last ten years experimentalists have 
tried to measure this number, using a BEC, measuring evaporation in a 
magnetic trap, and later spectroscopically by actually measuring 
the energy of the least bound state in the $^5\Sigma_g^+$ potential~\cite{Moal:06}. 
The latter determination is by far the most accurate and agrees very well with 
the latest quantum chemistry calculations~\cite{Przybytek:05, Przybytek:08}. 
These experiments are discussed in Sec.~\ref{sec:Photoassociation}.

\subsection{Ionizing collisions in He}\label{sec:PI}
In an unpolarized gas, such as a MOT, atoms populate all magnetic substates and collisions therefore proceed on many interatomic potentials. 
In this case, Penning ionization is a major loss process.
Also, in a MOT, the presence of light with a frequency close to an atomic resonance complicates the analysis of experimental data on trap losses. 
In this section we restrict ourselves first to collisions in the dark between unpolarized atoms, where a partial wave analysis suffices. 
Many experimental data, however, are for losses in a MOT, 
i.e. in the presence of near-resonant light. 
These results will be discussed in Sec.~\ref{sec:MOTPI}. 
As experimental data and theory are best for helium we will focus on helium here.
In the following, we will not distinguish Penning ionization and associative ionization 
(see Eq.~\ref{eq:PI}).
But before proceeding, we point out that the experiment of \citet{Mastwijk:98} included 
a quadrupole mass spectrum analyser which permitted the identification of
the reaction products of the ionizing collisions.
They showed that about 5\% of the ionizing collisions in a He* MOT
come from the associative ionization process. 
To our knowledge this is the only such experiment using cold atoms. 

The possible values $S=0$, $S=1$ and $S=2$ for collisions between $s=1$ He* atoms correspond to the
singlet, triplet and quintet potentials $^1\Sigma_g^+$, the $^3\Sigma_u^+$ and $^5\Sigma_g^+$, respectively.
As the helium atom, with only two electrons and a nucleus, has a
relatively simple electronic structure, interatomic potentials can be
calculated with high accuracy. \citet{Muller:91} performed \emph{ab-initio} calculations in the Born-Oppenheimer
approximation, where the total electronic spin $S$ is a good quantum
number. They also included Penning ionization assuming a complex potential.

At the mK temperatures of a laser-cooled sample of He* atoms, the
collision process can be described conveniently using the partial
wave method: the ionization cross section, written
as a sum of partial wave contributions,
is dominated by only a few partial waves $\ell$. In a sufficiently cold sample of He* atoms in the dark, the cross section
for Penning ionizing collisions is dominated by the $s$-wave
contribution.
For collisions of He* atoms, partial wave cross sections
$\sigma_\ell^{\text{ion}}$ can be derived from the solution of an
effective one-dimensional potential scattering problem~\cite{Orzel:99,Stas:06}. Restrictions
imposed by symmetry on the partial waves that
contribute to the cross section can be
taken into account thereafter, thus accounting for the different quantum
statistics of $^4$He and $^3$He.
>From a semi-classical point of view, the cold ionizing collision can be
described as a two-stage process: (1) elastic scattering of the atoms by the interaction
potential at large internuclear distance, and (2)
Penning ionization at short distance, when the electron clouds of both collision partners start to
overlap.
These successive processes can be treated separately. 
The ionization cross section for collisions with total electronic spin $S$ can be written
as the product of the probability for the atoms to reach a small internuclear distance, and
the probability for ionization to occur at that place. 
As the total spin $S$ is conserved during
ionization, the latter is very small
for collisions that violate Wigner's spin-conservation rule. 
The calculation becomes very simple because for $S=0,1$ ionization occurs with essentially
unit probability (\citet{Muller:91} report an ionization probability of 0.975), and
the calculation of cross sections is reduced to the determination
of partial wave tunneling probabilities. 
To calculate these, the
interaction potentials of the colliding atoms are needed. 
The energy dependence of these probabilities gives rise to an
energy dependent ionization cross section for $S=1$ and $S=0$, that displays
the quantum threshold behavior of the inelastic collisions.

To compare with experiments, the ion production rate $\text{d}N_{\text{ion}}/\text{d}t$ in a
(magneto-optically) trapped atomic sample
can be expressed in terms of an ionization rate coefficient
$K$ (particle$^{-1}$ cm$^3$/s):
\begin{equation}
\frac{\text{d}N_{\text{ion}}(t)}{\text{d}t} = K \int \! n^2(r,t) \, \text{d}^3 r.
\label{eq:ionrate}
\end{equation}
$K$ depends on the temperature $T$ and
can be calculated using the velocity dependent partial wave ionization cross sections for each of the 
potentials $^{(2S+1)}V(R)$. The contribution of the $^5\Sigma_g^+$ potential can then be neglected
compared to contributions of the $^1\Sigma_g^+$ and $^3\Sigma_u^+$ potentials because it corresponds 
to a fully polarized electronic spin which cannot directly couple to the ionized states.

The interatomic interaction is almost identical in the case
of $^3$He* and $^4$He* and partial wave contributions
are therefore similar~\cite{Stas:06}. The composition of the total ionization
cross section or rate coefficient from these contributions is very
different for the bosonic ($^4$He*) and fermionic isotope ($^3$He*).
Symmetrization requires that a scattering state
describing a colliding pair of identical bosons has even symmetry
under exchange of the atoms, while a state describing identical
fermions has odd symmetry. As a result, partial
waves with improper symmetry do not contribute to the total cross
section or rate coefficient, and are excluded from the summations.
Also, the hyperfine structure of $^3$He* complicates the analysis as
$S$ is not a good quantum number for large internuclear distances,
where atom pairs are characterized by $F$. However, $S$ is a good
quantum number for small internuclear distances, where the molecular
interaction dominates and Wigner's spin-conservation rule applies.
In a laser-cooled sample of He* atoms, collisions occur for all
values of the total atomic angular momentum, $F=0,1,2,3$ in case of
$^3$He* in the $f=\frac{3}{2}$ hyperfine level, and $S=0,1,2$ in case of $^4$He*. The contribution of each
collision channel depends on the distribution of magnetic
substates in the sample, where $m$ is the azimuthal quantum number
of the atom, which can take on values
$m_f=-\frac{3}{2},-\frac{1}{2}, \frac{1}{2},\frac{3}{2}$ in case of
$^3$He* and $m_s=-1,0,1$ in case of $^4$He*.
The unpolarized ionization rate coefficient
$K^{\text{(unpol)}}$ is obtained for a
laser-cooled sample of He* atoms where magnetic substates
are uniformly populated. 
For samples with a temperature around 1~mK, only $s$ and $p$-waves need
be taken into account.

The results of this
theoretical model~\cite{Stas:06} for $^4$He* turn out to agree very well with the 
results of detailed close-coupling theory calculations
\cite{Leo:01}, as well as with a simpler calculation~\cite{Mastwijk:98}; 
at 1~mK $K^{\text{(unpol)}}$=$8.3 \times 10^{-11}$ cm$^{3}$/s which agrees 
with $K^{\text{(unpol)}}$=$8.9 \times 10^{-11}$ cm$^{3}$/s~\cite{Leo:01} and 
$K^{\text{(unpol)}}$=$7.3 \times 10^{-11}$ cm$^{3}$/s~\cite{Mastwijk:98}.

\subsubsection{Measurement of ionizing collisions in a MOT}
\label{sec:MOT}
The theoretical ionization rate constants can be compared to
experimental values extracted from MOT data such as the ion production rate,
number of atoms and cloud size.
For a MOT, the time dependence of the
number of trapped atoms $N$ can be described by the phenomenological
equation \cite{Bardou:92,Stas:06}
\begin{equation}
\frac{\text{d}N(t)}{\text{d}t}=L - \alpha N(t) - \beta \int \! n^2(r,t) \, \text{d}^3 r.
\label{eq:diffnumb}
\end{equation}
The first term governing $N(t)$ is a constant loading rate $L$,
representing the capture of atoms from the decelerated atomic beam
into the MOT. 
The second and third terms are the linear and
quadratic trap loss rates, respectively.
If the loss is exclusively due to ionization, we can make a connection
to Eq.~\ref{eq:ionrate} because $\beta=2K$.
For He* samples in a 1083~nm 
MOT~\cite{Stas:04,Stas:06,Tol:99,Pereira:01b,Browaeys:00b,Kumakura:99,Mcnamara:07} 
or 389~nm MOT~\cite{Koelemeij:03,Koelemeij:04a,Tychkov:04}, 
collisional loss mechanisms give rise to significant trap loss. 
Quadratic trap loss is determined by collisions between trapped He* atoms, while linear
trap loss results from collisions with background gas particles.
In a MOT, the density is generally small enough that three-body
processes can be neglected.

For a Gaussian density distribution
the ion current, measured on an MCP detector,
can be written as~\cite{Tol:99}
\begin{equation}
\varphi =  \frac{V \beta}{4 \sqrt{2}} \, n_0^2 + \varphi_{\text{bgr}}.
\label{eq:phi_dark}
\end{equation}
Here $n_0$ is the central density in the MOT, 
and the effective volume $V$ is defined by $V=(2 \pi)^{3/2} \sigma_\rho^2 \sigma_z$
(where $\sigma_{\rho}$ and $\sigma_z$ are the rms radii of the cloud).

\begin{figure}[t]
\centerline{\includegraphics[width=0.9\linewidth]{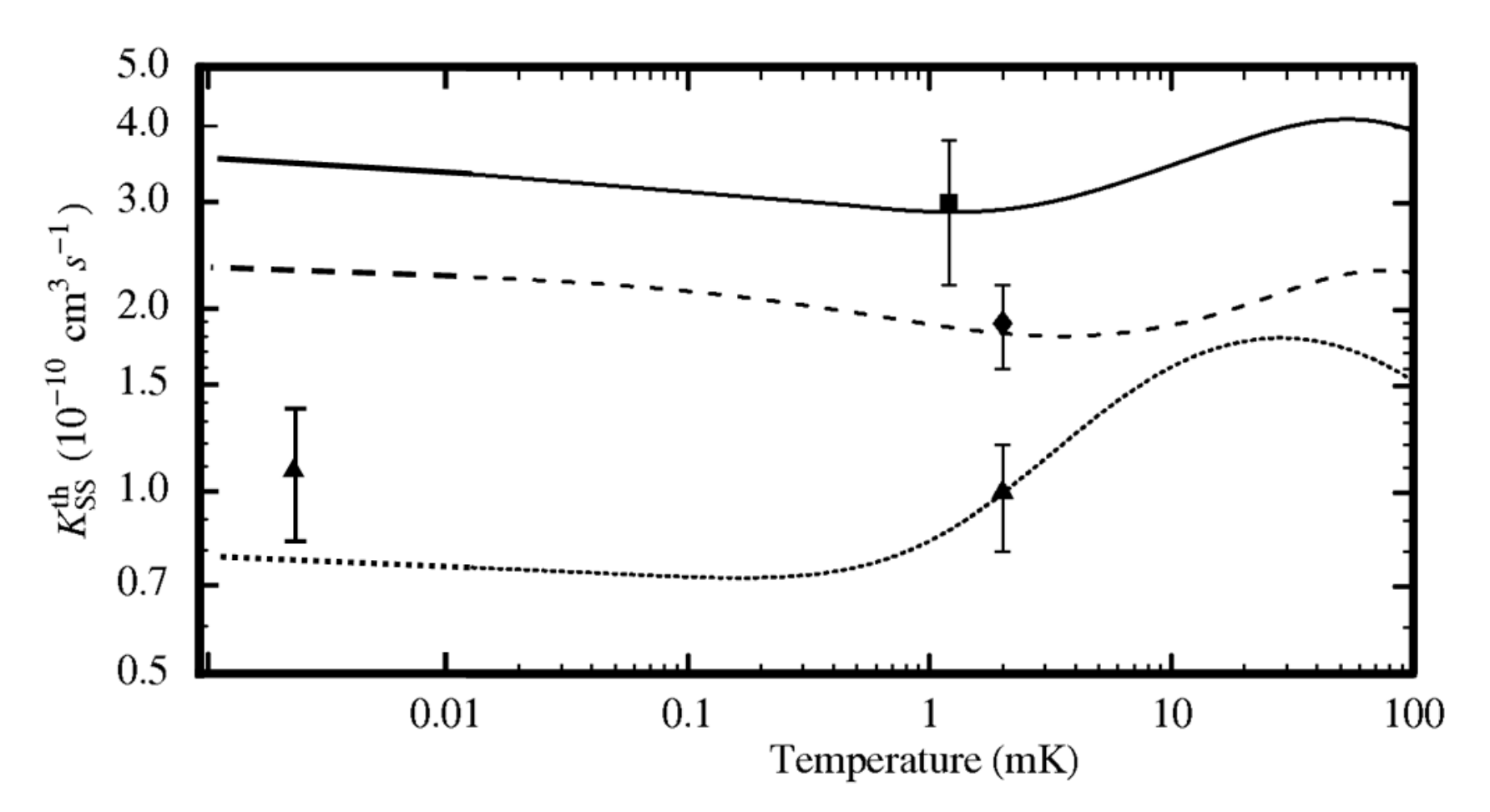}}
\caption{Experimental Penning ionization loss rate coefficients for an unpolarized gas of helium atoms
in the metastable 2 $^3$S$_1$ state, compared to single-channel theory~\cite{Mcnamara:07}.
Dashed curve: theory for $^3$He - $^3$He*, and experimental point at $T$= 2 mK (diamond);
solid curve: theory for $^4$He - $^3$He*, and experimental point at $T$= 1 mK (square);
dotted curve: theory for $^4$He - $^4$He*, and experimental points at $T$= 2 mK and $T$= 3$\mu$K (triangle).
The $T$= 3 $\mu$K point is from \citet{Partridge:10}, deduced from experiments in an optical dipole trap, see Sec.~\ref{sec:spindipole}.
\label{fig:McNamaraMOT}}
\end{figure}

An experimental value for $K_{SS}$ in the absence of trapping light can be derived 
by measuring the ion production rate both in the MOT phase and while the trapping light is turned off briefly.
For calibration one needs an independent measurement of the loss
rate constant in the MOT ($\beta_{MOT}$) which can be deduced from measuring the
decay of the MOT (see Sec.~\ref{sec:MOTPI}). 
This procedure, and an analogous one for the heteronuclear case
of a two-isotope MOT for $^4$He* and $^3$He*~\cite{Mcnamara:07}, provides values of the ionization loss rate constant at the temperature of the MOT.
Inspection of Fig.~\ref{fig:McNamaraMOT} shows that experimental loss rate coefficients are very
well understood from the simplified theory described in the previous section in all cases.

As the theoretical model shows good
agreement with other theoretical work and with the experimental
results, cold
ionizing collisions of He* atoms can be understood as single-channel
processes that are determined by Wigner's spin-conservation rule,
quantum threshold behavior and the symmetrization postulate. Using
the model, the difference between the ionization rate coefficients
for $^3$He* and $^4$He* can be interpreted as an effect of the
different quantum statistical symmetry of the two isotopes and the
presence of a nuclear spin in the case of $^3$He*. As the model is
relatively simple, it is complementary to the more complete (and
precise) close-coupling theory that has been developed for $^4$He*
collisions as well~\cite{Venturi:99,Venturi:00,Leo:01}.

\subsubsection{Two-body and three-body losses for spin-polarized $^4$He*}
\label{sec:spindipole}
The most important mechanism that causes decay of an ultracold gas of spin polarized metastable 
$^4$He atoms in the $m=+1$ state is Penning ionization due to spin relaxation, 
caused by the spin-dipole interaction~\cite{Shlyapnikov:94,Leo:01}. 
At the temperatures and 
magnetic fields relevant for Bose-Einstein condensation experiments it turns out that the rate 
constant for this two-body process is $\sim2 \times$10$^{-14}$ cm$^3$/s, four orders of 
magnitude smaller than the Penning ionization loss rate for an unpolarized gas 
(see Fig.~\ref{fig:McNamaraMOT}). This rate constant was also calculated for higher temperatures 
and magnetic fields~\cite{Fedichev:96}. It was the prediction of this small loss rate constant, 
comparable in magnitude to two-body loss rates in the alkalis, that stimulated experimental 
research towards realization of BEC in several labs at the end of the 1990s.

The large suppression of Penning ionization has been measured in several studies on metastable $^4$He*, 
first by spin-polarizing a cloud of atoms from a MOT, where a suppression of at least 2 orders of magnitude 
was demonstrated~\cite{Herschbach:00a}, later in studies of a spin-polarized cloud in a magnetic 
trap~\cite{Nowak:00}, close to or at 
quantum degeneracy~\cite{Sirjean:02,Seidelin:03b,Seidelin:04,Tychkov:06}, where the calculated 
rate constant~\cite{Shlyapnikov:94,Leo:01} was experimentally confirmed.

Actually, at the high densities near BEC, the two-body losses are comparable in size to the 
three-body losses. One can study the decay
of the condensate atom number as a function of time~\cite{Pereira:01a,Tychkov:06}
or directly measure the ion rate as a function of the atom
number~\cite{Sirjean:02,Seidelin:04}. These experiments result in compatible
values for the two- and three-body loss rate constants with the most accurate values
being $\beta^{(pol)}=2(1) \times 10^{-14}$~cm$^3$/s for the two-body loss rate constant
and $\gamma^{(pol)}=9(3) \times 10^{-27}$~cm$^6$/s for the three-body loss rate constant
~\cite{Tychkov:06} in good agreement with the theoretical
predictions~\cite{Shlyapnikov:94,Leo:01}. The amount of experimental data is limited and it shows a strong correlation between $\beta$ and $\gamma$ values. More measurements at different densities are required to improve the accuracy of the quoted numbers.

The lifetime of a mixture of quantum degenerate gases of $^3$He$^*$ and
$^4$He$^*$ was also measured and is compatible with the theoretical three-body
(boson-boson-fermion) inelastic rate constant estimation of $1.4 \times
10^{-24}$~cm$^6$/s, 2 to 3 orders of magnitude larger than for the fully
polarized bosonic case~\cite{Mcnamara:06}.

In a magnetic trap, only $^4$He* atoms in the $m=+1$ state can be trapped. This automatically 
generates a spin-polarized gas. Recently, He* atoms were also optically trapped~\cite{Partridge:10}. 
In this case also $m=0$ and $m=-1$ atoms as well as mixtures of different $m$-states may be trapped. 
As only a spin-polarized gas of either $m=+1$ or $m=-1$ atoms is stable against Penning ionization, 
all other possibilities are expected to be considerably less stable. Indeed, the authors found that 
a gas of $m=0$ atoms shows a loss rate constant of $6.6 (1.7) \times 10^{-10}$ cm$^{3}$/s, while for 
a mixture of $m=+1$ and $m=-1$ atoms they find $7.4 (1.9) \times 10^{-10}$ cm$^{3}$/s (25\% error). 
In these cases 
Penning ionization can occur in the $^1\Sigma_g^+$ potential. At the temperatures where these 
experiments were performed (a few $\mu$K) Penning ionization in the $^3\Sigma_u^+$ potential is expected 
to contribute negligibly~\cite{Venturi:00}. The measured loss rate constants may be compared to the 
loss rate constants deduced in unpolarized clouds from a MOT as discussed previously and shown in 
Fig.~\ref{fig:McNamaraMOT}. $K^{\text{(unpol)}}$=$1.10 (28) \times 10^{-10}$ cm$^{3}$/s and 
$K^{\text{(unpol)}}$=$1.23 (28) \times 10^{-10}$ cm$^{3}$/s are deduced from these $m=0$ resp. 
$m=\pm1$ results, in fine agreement with the other experiments and theory.

Ionization rates for spin polarized metastable helium have been studied theoretically in
strongly confining traps, both isotropic and anisotropic. 
In isotropic traps, strong confinement reduces the trap lifetime due to spin-dipole interactions \cite{Beams:04}. 
Surprisingly, in anisotropic traps, there are situations in which interference effects involving
the anisotropic trapping potential and the spin-dipole interaction can dramatically
change the trap lifetime due to ionization.  
The work of \citet{Beams:07} reports trap lifetime enhancements of 2 orders of magnitude for
some, specific trap states at specific trap aspect ratios.

\subsubsection{Collisions in the presence of light}
\label{sec:MOTPI}
Although the loss rate constant for Penning ionization
collisions between unpolarized metastable helium atoms is
of order $10^{-10}$~cm$^3$/s (see Sec.~\ref{sec:MOT}), the dominant losses in a MOT are due to
photoassociative collisions. During such a collision, a transition
is made to a quasimolecular state with a potential, due to the
resonant dipole-dipole interaction, scaling as $\pm C_3/R^3$ at
long range, where $R$ is the internuclear distance, and $C_3\simeq\hbar\Gamma(\lambda/2\pi)^3$
is the squared atomic dipole matrix element of the transition,
which has a linewidth $\Gamma/2\pi$, and wavelength
$\lambda$. In the presence of light, red detuned by
an amount $\Delta$, a resonant transition to such
an excited state is possible at the Condon
radius, $R_C$, where the molecular energy compensates the detuning:
\begin{equation}
\label{condonpoint}
R_{\mathrm{C}}=\left(\frac{C_3}{2\pi\hbar |\Delta|}\right)^{1/3}.
\end{equation}
At the Condon radius the Van der Waals interaction between two noble gas
atoms is small compared to the kinetic energy (if the detuning is not very large) and has no
noticeable effect on their relative motion. This contrasts to the
situation in the excited state, for which the interaction is
strongly attractive. Therefore, after excitation the two atoms are
rapidly accelerated towards small internuclear distances, where
couplings exist to loss channels involving autoionization or
fine structure changing mechanisms. Before reaching this region,
however, there is a probability that the molecule decays back to
the lower state by spontaneous emission. This results in two fast
He* atoms colliding, leading either to Penning ionization, or to elastic
scattering, after which the atoms may have sufficient kinetic
energy to escape from the trap, a mechanism called radiative escape~\cite{gallagher:89}.

The loss rate constant can also be modified for blue-detuned light. 
This occurs by excitation of the colliding atom pair to the long-range 
repulsive part of the $C_3/R^3$ potential instead of the attractive part. 
In this case, ``optical shielding" of cold collisions may occur since the atoms 
cannot reach a short internuclear distance where Penning ionization occurs. 
The loss rate constant can be several percent smaller than even in the 
dark~\cite{Walhout:95a,Katori:94,Orzel:98}. ~\citet{Weiner:99} review these 
experiments, which are performed for the heavier noble gas atoms Kr and Xe.
For metastable helium, optical shielding was demonstrated as well, although 
it is far less dramatic~\cite{Herschbach:03b}.

To measure the loss rate constant $\beta_{MOT}$, the time dependence 
of the number of trapped atoms $N$ is determined by the loading of atoms 
into and/or loss of atoms from the trap.
A cloud of $N$ atoms, trapped in a MOT, can be characterized by a
Gaussian density distribution with central density $n_0$, and the
effective volume defined in Sec.~\ref{sec:MOT} for which $V=N/n_0(0)$. 
Eq.~\ref{eq:diffnumb} can then be written as
\begin{equation} \label{generalrateeq}
\frac{\mathrm{d}N(t)}{\mathrm{d}t}=L-\alpha N(t)-\frac{\beta_{MOT} N^2(t)}{2\sqrt{2}V}.
\end{equation}
If all parameters but $N$ (or $L$) are known,
the steady-state number of atoms (or the loading rate)
follows from Eq.~\ref{generalrateeq} by setting $\mathrm{d}N/\mathrm{d}t=0$. Alternatively,
any time dependence of, for instance, the volume may readily be included, and
Eq.~\ref{generalrateeq} can then be used to describe the resulting time dependence
of $N$.
Experimentally, one can deduce the loss rate constant $\beta_{MOT}$
from the ionization signal in a MOT and, for instance, a measurement of the decay of the trap when the loading is stopped. When the central density in the MOT $n_0(0)$ is measured
by absorption imaging and the linear decay rate is negligible or known,
$\beta_{MOT}$ is deduced from a fit of the trap loss rate $\beta_{MOT} n_0(0)$.

In this way several groups have measured loss rate constants in a He* MOT at 1083~nm~\cite{Bardou:92,Mastwijk:98,Kumakura:99,Tol:99,Browaeys:00b,Pereira:01b,Stas:06} as well as 389~nm~\cite{Koelemeij:03, Koelemeij:04a}.
These loss rate constants are at least one order of magnitude larger than for
an unpolarized metastable helium cloud in the dark, ranging from $2 \times10^{-8}$ cm$^{3}$/s to $5 \times10^{-9}$
cm$^{3}$/s for a 1083~nm MOT (detuning -5~MHz resp. -40~MHz) and $2 \times10^{-9}$ cm$^{3}$/s for a 389~nm MOT (detuning -10~MHz).
The numbers for a 1083~nm MOT at large detuning are the same within a factor 2 for both isotopes.

In general, the
loss rate constant increases with decreasing red detuning, until a
certain value for the detuning (around $\sim
-5$~MHz~\cite{Tol:99, Herschbach:00b}). Beyond this detuning, the probability of
decay by spontaneous emission starts to approach unity. Also the
gradient and, therefore, the acceleration on the excited-state
potential decreases with detuning. Consequently, the probability
of reaching the short internuclear distances (in the excited
state), where loss mechanisms reside, goes to zero. Additionally, resonances in the ionization rate may be observed. These are due to
vibrational states in the excited-state potentials and will be discussed in more detail in Sec.~\ref{sec:Photoassociation}.

\subsection{Elastic and inelastic collisions in Ne}
\label{NeonCollisions}
Much of the research on cold collisions in metastable neon (the lightest noble gas atom after helium) is motivated by the quest for 
Bose-Einstein condensation. Initial theoretical 
research indicated that suppression of Penning ionization in the metastable $^3$P$_2$ state, 
for atoms in the $m=+2$ state, might be sufficient~\cite{Doery:98}. 
But since {\it ab-initio} calculations of scattering properties of all metastables except helium are very demanding, these propertieas are best determined experimentally. Accordingly,
groups in Eindhoven~\cite{Kuppens:02} 
and Darmstadt (previously Hannover)~\cite{Zinner:03,Spoden:05,VDDiss:08} designed and built MOTs 
containing more than 10$^9$ atoms of the bosonic isotope $^{20}$Ne, more than 10$^8$ atoms of the other bosonic isotope $^{22}$Ne, and $3 \times 10^6$ atoms of the fermionic isotope $^{21}$Ne.

In a MOT, where the atoms are unpolarized, a Penning ionization loss rate (in the dark) 
$\beta_{unpol}$=$5(3) \times10^{-10}$ cm$^{3}$/s for $^{20}$Ne was  measured~\cite{Kuppens:02} in an analogous way to He*.
In an optical dipole trap, loss measurements on unpolarized atoms in the $^3P_2$ state gave Penning ionization loss 
rates $\beta_{unpol}$=$5_{-3}^{+4} \times10^{-10}$ cm$^{3}$/s for $^{20}$Ne and
$\beta_{unpol}$=$10_{-5}^{+4} \times10^{-10}$ cm$^{3}$/s for $^{22}$Ne~\cite{VDDiss:08}.
These numbers are similar in a He* MOT (see Table~\ref{tab:data}).

The rates of elastic and inelastic collisions of cold spin-polarized neon atoms in the metastable 
$^3$P$_2$ state for $^{20}$Ne and $^{22}$Ne were measured by transfering the atoms, after spin-polarization, 
from a MOT to a magnetic trap. Penning ionization loss rates
$\beta$=$6.5(18) \times10^{-12}$ cm$^{3}$/s for $^{20}$Ne and $\beta$=$1.2(3) \times10^{-11}$ cm$^{3}$/s 
for $^{22}$Ne were obtained. These losses thus indeed 
occur less frequently (a reduction by a factor 77 for $^{20}$Ne and 83 for $^{22}$Ne) than for unpolarized 
atoms. This proves the suppression of Penning ionization due to spin polarization in Ne*, but the suppression factor is about two orders of magnitude smaller than for He*.

>From cross-dimensional relaxation measurements in a magnetic trap, elastic scattering lengths of 
$a$=-180(40)a$_0$ for $^{20}$Ne and $a$=+150$_{-50}^{+80}$a$_0$ for $^{22}$Ne were obtained as well.
These numbers show that concerning the magnitude and the sign of the elastic scattering length, 
$^{22}$Ne would be a good candidate for pursuing BEC in metastable neon.
Accordingly, evaporative cooling of $^{22}$Ne in a magnetic trap has been demonstrated \cite{VDDiss:08}, 
but the small ratio of elastic to inelastic collisions has so far prevented realization of Bose-Einstein condensation in Ne*. 

\subsection{Ionizing collisions in Ar, Kr and Xe}
\label{sec:ArKrXe}
In the 1990s the heavier noble gas atoms were studied in MOT experiments. Detailed studies were performed on ionizing collisions in MOT's of Kr~\cite{Katori:94,Katori:95} and Xe~\cite{Walhout:95a,Suominen:96}, in particular in the presence of near-resonant light, that may cause shielding of collisions. These studies have been reviewed~\cite{Weiner:99} and we will focus here only on the experimental results on rate constants, relevant when comparing noble gas atoms.

For metastable $^{40}$Ar MOTs have been realized~\cite{Katori:93,Sukenik:02} and Penning ionization losses in the presence of the MOT light have been studied~\cite{Busch:06}. The homonuclear loss rate in the presence of MOT light turned out to be surprisingly small: $\beta^{\text{(unpol)}}$=$5.8(1.7) \times 10^{-10}$ cm$^{3}$/s. These studies have so far not been extended to collisions in the dark.

More results were obtained for $^{84}$Kr* and $^{83}$Kr*~\cite{Katori:93,Katori:94,Katori:95}. 
For an unpolarized cloud of $^{84}$Kr* an ionization rate constant in the dark 
$K^{\text{(unpol)}}$=$2 \times 10^{-10}$ cm$^{3}$/s was reported~\cite{Katori:94}, 
again of the same order of magnitude as for Ne* and He*. 
In the presence of near-resonant light this rate increased for negative detuning and decreased 
for small positive detuning (optical shielding) as discussed in Sec.~\ref{sec:MOTPI} for He*. 
Cooling and spin-polarizing the fermionic isotope $^{83}$Kr* demonstrated the effects of quantum statistics on the Penning ionization rate. It was shown~\cite{Katori:95} that the rate constant for ionizing collisions decreased 10\% when the temperature was decreased below the $p$-wave threshold in the case of the fermionic isotope while it remained constant for the bosonic isotope. Only marginal indications of suppression of ionization due to spin conservation in collisions were observed~\cite{Katori:95}.

Experiments in Xe* provide results that are very similar to those in Kr*. Focused primarily on the 
effects of near-resonant light on collision dynamics in a MOT, the $^{132}$Xe loss 
rate constant for ionizing collisions in the absence of light was measured: 
$\beta^{\text{(unpol)}}$=$6(3) \times 10^{-11}$ cm$^{3}$/s~\cite{Walhout:95a}. 
This value agreed very well with a theoretical number of
$\beta^{\text{(unpol)}}$=$6.5 \times 10^{-11}$ cm$^{3}$/s~\cite{Orzel:99}, 
obtained using the same model as discussed in Sec.~\ref{sec:PI}, assuming unit 
ionization probability for atoms that have penetrated the centrifugal barriers. 
For the other even isotopes $^{134}$Xe and $^{136}$Xe identical rates were measured, 
in accordance with theory. Also the fermionic isotopes $^{129}$Xe and $^{131}$Xe showed 
the same loss rate constant $\beta^{\text{(unpol)}}$ (close to the $p$-wave threshold), 
also predicted by the model. Compared to Kr, in Xe a much larger difference in loss rate 
between bosons and fermions was found when comparing the ionization rate for a spin polarized 
gas to that of an unpolarized gas. 
The two rates differed by a factor 3 at temperatures far below the $p$-wave threshold~\cite{Orzel:99}, 
while they were equal at the $p$-wave threshold. Measuring the ratio 
$\beta^{\text{(pol)}}$/$\beta^{\text{(unpol)}}$ for the fermions as a function of 
temperature yielded a factor of 2 decrease, in
perfect agreement with the predictions of the simple one-dimensional single-channel potential 
scattering model, also discussed in Sec.~\ref{sec:PI}.
Interestingly, comparing the ionization rate for a spin-polarized gas to that of an unpolarized gas 
for the bosons $^{132}$Xe*, $^{134}$Xe* and $^{136}$Xe*, it was observed that instead of a suppression 
of Penning ionization an enhancement by 60\% was found. This clearly shows that spin-conservation effects, 
seen in He* and Ne*, are absent in bosonic Xe. This result confirms that anisotropic interactions in the
$np^5(n+1)s$ $^3$P$_2$ states may lift the spin-conservation restriction. It is however expected that 
these anisotropic interactions are strongest for heavier noble gas metastable atoms, explaining why 
for Ne* spin-conservation still holds to some extent.
\subsection{Mixtures}
Mixtures of different species (with at least one metastable noble gas atom) allow an extension of 
research possibilities. One can mix two isotopes of the same element or mix two different chemical species. 
A priori it may seem difficult to simultaneously trap two different chemical species
as most elements are easily ionizable by a metastable noble gas atom.
Therefore Penning ionization may strongly affect the densities and lifetimes that can be realized. 
However, Penning ionization due to optically assisted heteronuclear collisions in a MOT is expected to 
be less important in comparison to the homonuclear case. The interaction is not a long-range resonant dipole 
interaction but a short-range van der Waals interaction. In the dark, an unpolarized mixture is therefore
expected to show a two-body loss rate constant $\sim 10^{-10}$ cm$^{3}$/s, just as in the homonuclear case.

For a spin polarized gas it is difficult to predict whether suppression of Penning ionization may occur. 
Based on the discussion on anisotropic interactions it seems that mixtures containing He* and an 
alkali atom (both in a symmetric $S$-state) may be the most promising.

So far, mixtures with at least one metastable noble gas species have been realized experimentally
for helium, neon and argon. Since the study of collisions in the neon mixtures ($^{20}$Ne/$^{21}$Ne,
$^{20}$Ne/$^{22}$Ne,$^{21}$Ne/$^{22}$Ne) is still in progress~\cite{Feldker:11,Schuetz:11} we will focus
on a qualitative discussion of the mixtures  $^{3}$He/$^{4}$He, $^{4}$He/$^{87}$Rb, and $^{40}$Ar/$^{85}$Rb
in the subsequent subsections.
\subsubsection{$^{3}$He / $^{4}$He}
In Sec.~\ref{sec:MOT} we already discussed some collision properties of a mixture of the bosonic and fermionic isotope of helium in a MOT. Taking into account the quantum statistics in the collisions between unpolarized $^{3}$He and $^{4}$He atoms the Penning ionization losses can be well understood, both in the homonuclear and heteronuclear case, as illustrated in Fig.~\ref{fig:McNamaraMOT}.

It was expected that for a spin polarized mixture of $^{3}$He in the $|f,m_f\rangle=|$3/2,+3/2$\rangle$ 
state and $^{4}$He in the  $|j,m_j\rangle=|$1,+1$\rangle$ state a similar suppression of Penning ionization 
would hold as in the case of $^{4}$He in the  $|j,m_j\rangle=|$1,+1$\rangle$ state alone. Indeed, this was 
observed in a magnetic trap at $\sim$mK temperatures~\cite{Mcnamara:06}.
Cooling towards quantum degeneracy, however, a significant reduction in the lifetime of a condensate was 
observed in the presence of an $\sim\mu$K dense cloud of $^3$He atoms. For a condensate 
of $|$1,+1$\rangle$ $^4$He* atoms the lifetime was a few seconds while it was only $\sim$10 milliseconds in the presence of $|$3/2,+3/2$\rangle$ $^3$He* atoms~\cite{Mcnamara:06}. 
This reduction can be explained assuming large three-body losses for the heteronuclear mixture. 
It turns out that the heteronuclear scattering length is extremely large, 
$a_{34}=27.2\pm0.5$~nm~\cite{Przybytek:08}, calculated by mass scaling of the accurately 
known $^5\Sigma_g^+$ potential. As the three-body losses are proportional to the fourth 
power of the scattering length~\cite{Fedichev:96b} this explains the short lifetime that was 
observed~\cite{Mcnamara:06}.
\subsubsection{$^{4}$He / $^{87}$Rb and $^{40}$Ar / $^{85}$Rb}\label{sec:RbM}
Simultaneous trapping of alkali atoms with noble gas atoms has been studied for two cases. A dual-species MOT for Rb and Ar* was
built~\cite{Sukenik:02} and Penning ionization was observed, both of Rb and Ar*~\cite{Busch:06}. This experiment was the first to identify the molecular ion RbAr$^+$ (associative ionization, see Eq.~\ref{eq:PI}), measured in a time-of-flight mass spectrometer. Two loss rates could be measured in the presence of light: the loss rate of Rb due to the presence of Ar* MOT light and the loss rate of Ar* due to the presence of Rb MOT light, $\beta_{Rb-Ar}^{\text{(unpol)}}$=$3.0(1.3) \times 10^{-11}$ cm$^{3}$/s and $\beta_{Ar-Rb}^{\text{(unpol)}}$=$1.9(0.9) \times 10^{-11}$ cm$^{3}$/s, respectively. These are much lower (by about two orders of magnitude) than loss rates in a single-isotope noble gas MOT.

Recently studies started on simultaneous trapping of Rb and He*~\cite{Byron:10a,Byron:10b}. Here, for the loss rates in the presence of light, analogous results were reported as in the Rb - Ar* case: the total two-body loss rate $\beta_{Rb-He}^{\text{(unpol)}}$=$6(2) \times 10^{-10}$ cm$^{3}$/s is relatively small~\cite{Byron:10a}. To investigate the possibilities of a dual-species BEC, both Rb and He* were spin polarized and the ion production was measured to see whether suppression of Penning ionization occurs. A suppression of Rb$^+$ ion production of at least a factor of 100 was observed, only limited by the detection sensitivity. 
This suppression is the largest measured for any noble gas boson except $^4$He*, and is very promising for future progress towards a dual-species BEC.
\subsection{Feshbach resonances}
The possibility to tune the scattering properties between atoms has pushed the research with 
ultracold and degenerate alkali atoms. For the metastable noble gases helium has been subject 
of a recent study. Spin-polarized $^4$He* has only very limited possibilities to tune the 
scattering length as $^4$He has no hyperfine structure. This only allows resonances due to 
the weak magnetic dipole interaction which are expected to be very narrow. For efficient 
coupling and experimentally accessible resonance widths at least one of the two collision 
partners should be $^3$He*. As the fully spin-polarized mixtures do not allow efficient mixing 
at least one of the two atoms should be in a state with magnetic quantum number $|m|<m_{max}$. 
This induces admixtures of molecular states with singlet and/or triplet character increasing the 
Penning Ionization rate. 
Moreover, as the singlet and triplet potentials are not as accurately known 
as the quintet potential, this also leads to reduced accuracy in calculating the Feshbach 
resonances using knowledge of the molecular potentials. However, a detailed \emph{ab-initio} 
study of the possibilities to magnetically tune the scattering length in all isotopic mixtures 
of He* was performed recently~\cite{Goosen:10}. One promising very broad Feshbach resonance 
was found for the case of a mixture of $^4$He* and $^3$He* atoms. For the single-isotope case 
narrow resonances are expected to exist, however, as the admixture of singlet and triplet 
character will be substantial in this case, accurate predictions could not be made.

In a second study the possibilities of optical Feshbach resonances were investigated for the case of spin-polarized $^4$He*~\cite{Koelemeij:04b}. It was found that a substantial modification of the scattering length can be realized experimentally applying laser radiation near the long-range states discussed in Sec.~\ref{sec:Photoassociation}, however only for a time of less than a millisecond, as strong losses and heating of trapped atoms are anticipated at this time scale.

\section{Photoassociation}
\label{sec:Photoassociation}

\subsection{General features for metastable atoms}
Photoassociation processes involve two colliding atoms that absorb a photon to
form an excited molecule. 
Photoassociation has been used for a long time in the field of
molecular physics. It attracted renewed interest when cooling methods
appeared (see \citet{Jones:06} and references therein). 
Long-range molecules can be created
from ultracold atoms and 
photoassociation resonances also provide
a tool to modify the atomic collisional properties via optical Feshbach resonances.
Finally, photoassociation spectra can be used to determine elastic scattering lengths
as we will discuss in Sec.~\ref{sec:ScatteringLength}.

Photoassociation has been succesfully applied to cold and ultracold
gases of metastable helium providing spectroscopical data of unprecedented
accuracy for many different studies. 
Photoassociation with metastable atoms has some special features. 
First, noble gases are almost chemically inert and it
is quite counter-intuitive to be able to form dimers of such species. 
Moreover, each atom in the dimer has a high internal electronic energy. 
Thus the molecules also carry this energy.  
Nevertheless, photoassociation of $^4$He* atoms has been demonstrated by different groups
\cite{Kim:04,Herschbach:00b,Zwan:06,Pieksma:02}. 
These molecules are of special interest because of the simplicity of the helium atom. 
The interaction potential between two colliding metastable atoms can be 
calculated \textsl{ab-initio} with a good accuracy \cite{Przybytek:05}; 
this makes the comparison between experimental and calculated values a stimulating challenge.

\subsection{One-photon photoassociation of metastable helium}

The relevant molecular levels of helium are shown in Fig.~\ref{fig:potentialhelium}. 
Starting from two atoms in 2$^3$S$_1$ state, a
resonant laser can be tuned to excite the pair of atoms to a molecular state.
Two photoassociation spectroscopy experiments reaching the molecular $J=0$ and $J=2$
levels have been performed.

The first one  \cite{Herschbach:00b}  used a MOT at a
temperature $\sim$1 mK and with atomic densities $\sim5\times10^9$
cm$^{-3}$. A probe laser was scanned over a frequency range of 20 GHz below the atomic
2 $^3$S$_1$ - 2 $^3$P$_2$ transition. 
Photoassociation spectra were recorded by measuring the variation
of the ion rate induced by the photoassociation beam. 
The molecules are in an excited
state and decay faster by Penning ionization than by radiative decay. 
This is especially true for
molecular state with total spin 0 or 1: these are ionized with a probability close to unity.
The autoionization effect is reduced for molecules with a total spin 2 due to spin conservation
rules, as in the atomic case. 
However, the spin-orbit interaction causes spin mixing and the spin 2
molecules can also create ions. 
Three vibrational series were identified  and many lines could be interpreted, in spite of the fact
that the short range part of the interaction potential is not well-known \cite{Herschbach:00b}.

Other photoassociation studies were later performed \cite{Kim:04}, with two main differences with
the previous experiments :  first, an ultracold gas in a magnetic trap was used,  just above the
BEC threshold at a few $\mu$K, thus enhancing the photoassociation rate because of higher densities; second, spectra were recorded through the temperature changes of the cloud. The improved accuracy was used to determine the scattering length (see Sec.~\ref{sec:ScatteringLength}).

\subsection{Formation of long-range helium molecules}

The long-range part of the interaction potentials between two metastable helium
atoms is very well-known, in contrast with its short range part. Then, it was
possible and interesting to search for purely long-range potentials. One such potential could
be found within the manifold of molecular potentials linking to the
2 $^3$S$_1$ - 2 $^3$P$_0$ asymptote: it is a $0_u^+$ potential, with a steep inner wall
at 150 $a_0$ and a very weak attractive part at long distance. This potential
well can support five molecular bound states that can be excited by
photoassociation.

The excited dimers in the  $0_u^+$ potential are giant molecules with classical inner turning
points at about 150 $a_0$ and outer turning points as large as 1150 $a_0$. As a consequence of
this extremely large size, the autoionization process is blocked because the two atoms can not
get close enough to Penning ionize \cite{Leonard:03}. These molecules therefore decay by fluorescence.

A spectrocopic study of the giant photoassociated molecules provided a value of the  binding energy
of each of the five vibrationnal bound states of the $0_u^+$ potential with an accuracy of 0.5 MHz.
The results are in perfect agreement with calculations \cite{Leonard:03,Leonard:04,Cocks:10a}.
The accuracy of the measurements was sufficient to clearly show the role of retardation effects in
the interaction between the two atom carried by the electromagnetic wave  \cite{Leonard:04}.


\section{Scattering length measurements}
\label{sec:ScatteringLength}
In cold atom physics, the $s$-wave scattering length
characterizes the strength of the interaction between the atoms. It is a crucial
parameter for describing the physical properties of a condensate. In the case of
metastable helium, several groups have determined its value using different methods
with increasing accuracy over the years. In the case of neon, collisional properties
were used.

\subsection{Determinations from collisional properties}\label{a_collision}
The first idea is to deduce the scattering length $a$ from the measurement of
the chemical potential of the Bose-Einstein condensate. Within the Thomas-Fermi
approximation, the chemical potential $\mu$ is  estimated from the size of the
condensate and the number of Bose-condensed atoms~\cite{Dalfovo:99}. 
The large uncertainty in the number of
atoms gives a large error bar for the scattering length. For helium, the reported values
for $a$ were $20 \pm 10$ nm \cite{Robert:01a} and $16 \pm 8$ nm
\cite{Pereira:01a}. 
Another estimation, based on the temperature dependence 
of the collisional cross-section in a
thermal cloud, led to $10 \pm 5$ nm \cite{Tol:04}.

The work of \citet{Seidelin:04} attempted to improve on the initial estimates
which used measurements of the chemical potential. 
Instead of directly measuring the number of atoms in the condensate,
the authors compared Penning ionization rates in an almost
pure condensate and in a cloud at the BEC transition temperature.
Observation of the ionization rate was also used to place the cloud 
close the BEC transition (see Sec.~\ref{sec:OtherExperiments}). 
To a good approximation, the value of the critical temperature in a weakly 
interacting gas depends only on the trap geometry and the atom number \cite{Stringari:03}, 
thus the ionization production rate and detection efficiency could be calibrated.
This calibration in turn gave the atom number, and the condensate expansion could
again be used to find the scattering length.
The authors found $a=11.3^{+2.5}_{-1.0}$~nm.

For neon, studies of thermalizing collisions were used to determine the magnitude
and sign of the scattering lengths 
of $^{20}$Ne ($a=-9.5 (2.1)$~nm) and $^{22}$Ne ($a=7.9^{+4.2}_{-2.7}$~nm)(see Sec.~\ref{NeonCollisions}).

\subsection{Determinations using photoassociation}

\begin{figure}[h]
\centerline{
\includegraphics[width=0.8\linewidth]{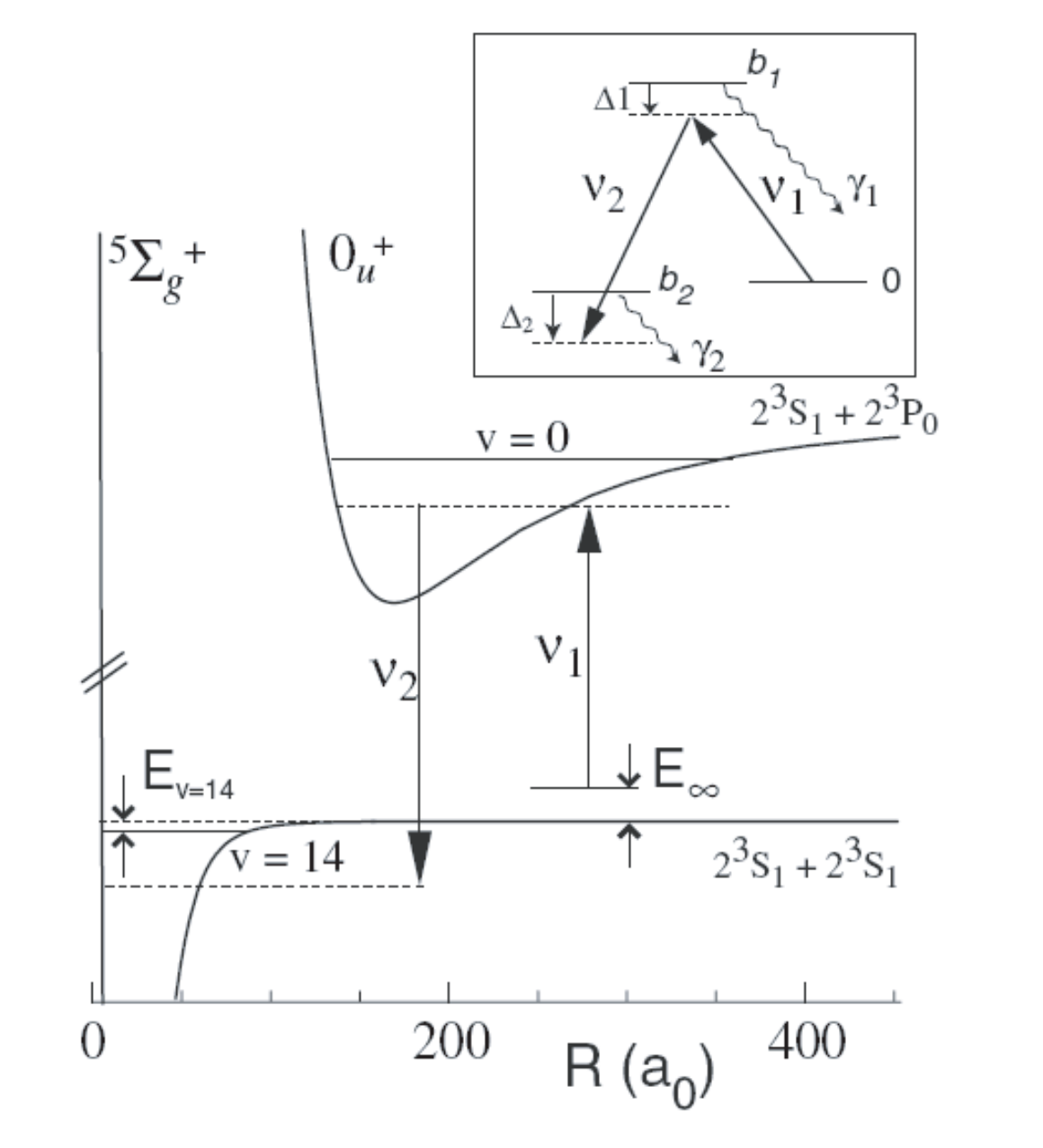}}
\caption{Relevant molecular potentials for photoassociation of spin-polarized metastable $^4$He
atoms (not to scale). The $0_u^+$ potential is purely long-range and very shallow. The arrows represent
the two laser frequencies involved in the two-photon photoassociation experiment (from \citet{Moal:06}).}
\label{fig:potentialhelium}
\end{figure}

 The basic idea in this method is to deduce the energy of the least bound-state  $v=14$ in the  $^5\Sigma_g^+$
 molecular potential (see Fig.~\ref{fig:potentialhelium}), from which the value of the scattering
 length can be derived. 
 Because it is essentially a frequency measurement, its potential accuracy is much
 higher than those directly exploiting collisional effects.  
 The first such measurement using He* was based on frequency shifts 
 in a one-photon experiment, a second one used two-photon photoassociation.

In the one-photon photoassociation experiment, light-induced frequency shifts of the photoassociation
spectra \cite{Kim:04} were found to be linearly dependent on the intensity of the photoassociation light beam.
These light shifts depend on the scattering length $a$ because the laser couples the excited vibrational
states in the $0_u^+$ potential to the continuum of free unbound scattering states above the dissociation
limit, as well as to the bound states in the $^5\Sigma_g^+$ molecular  potential.  Consequently,
the resulting shifts of the excited molecular levels depend on the value of the binding energy of the
least-bound state and also on the Franck-Condon overlap between the excited and the ground states.
Both these parameters depend on the scattering length $a$.  A reliable determination of $a$ resulted
from the measurement of the shifts and of a theoretical analysis \cite{Portier:06,Cocks:09,Cocks:10b}. Uncertainties
related to the laser intensity at the atom cloud were eliminated by comparing results for three different
excited vibrational states in the $0_u^+$ potential. $a=7.2\; \pm \; 0.6$ nm was deduced,
significantly smaller than the determinations discussed in Sec.~\ref{a_collision}.
Recently this method to detect the photoassociation resonance was extended to observe 
collective dipole oscillations as a result of momentum transfer of the photoassociation laser to
the cloud. This simple technique also allowed extraction of the scattering length with similar accuracy: 
$a=7.4\; \pm \; 0.4$ nm~\cite{Moal:08}.

\begin{figure}[h]
\centerline{
\includegraphics[width=0.8\linewidth]{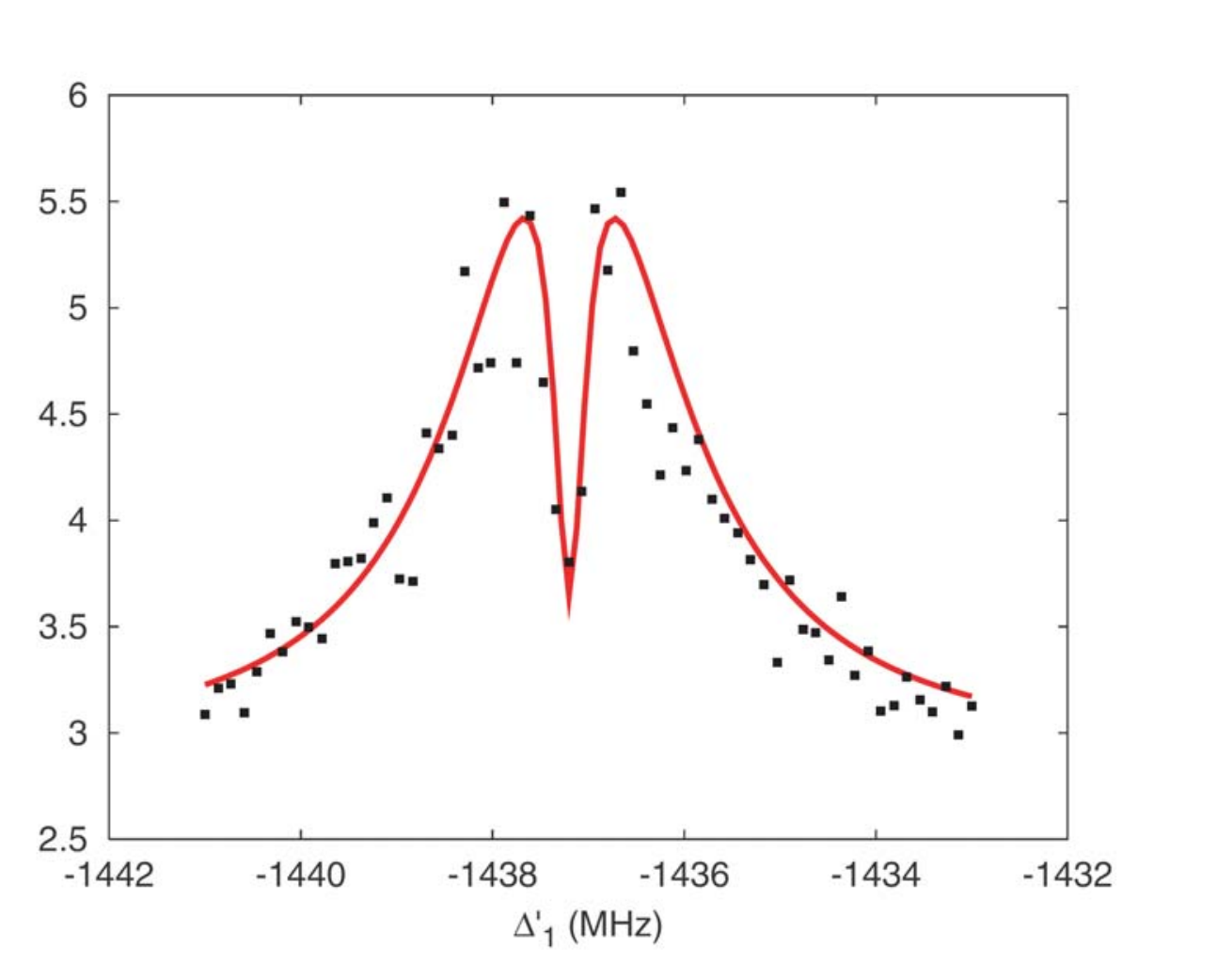}}
\caption{Atom-molecule dark resonance in a two-photon photoassociation
experiment with $2 ^3S_1$ metastable helium atoms (from ~\citet{Moal:06}. The resonance is detected by
measuring the temperaure increase of the cloud (vertical axis: temperature in $\mu$K unit).}
\label{fig:darkresonance}
\end{figure}

In a one-photon photoassociation process, the created molecules have a lifetime
limited by radiative decay, creating a pair of free atoms of high kinetic
energy. However the molecules can also decay to a bound state of the ground state
molecular potential. Interestingly, this decay can be stimulated using Raman transitions in a two-photon
photoassociation process. This led to a new determination of $a$, which
confirmed the first one \cite{Moal:06} and improved its accuracy. The principle
of the measurement is illustrated in Fig.~\ref{fig:potentialhelium}, showing
the two laser beams simultaneously illuminating the sample. Here the binding
energy of the least-bound state $v=14$ in the  $^5\Sigma_g^+$ molecular
potential is directly deduced from the position of the resonance given by the
Raman resonance condition $E_{\infty}+ h \nu_1 - h \nu_2=E_{\nu=14}$. A dark
resonance spectrum of narrow width was recorded when scanning one of the two
lasers, the other one being kept at a fixed frequency.  Such bound-free dark
resonance signals are shown in Fig. \ref{fig:darkresonance}. The dependence
on the relative kinetic energy $E_{\infty}$ was eliminated by extrapolating  to
zero temperature.

The deduced binding energy is $E_{\nu=14}=-91.35\; \pm 0.06$~MHz. The final
determination of the scattering length then relied on the precise interaction
potentials calculated \textsl{ab-initio} \cite{Przybytek:05}. This led to the
very accurate value $a=7.512 \; \pm \; 0.005$ nm \cite{Moal:06}.

\section{Spectroscopic measurements of atomic properties}
\label{sec:spectroscopy}

%
%
%
%
%
%
%

\subsection{Metastable state lifetimes}
\label{sec:Lifetimes}

As noted earlier in this review, the long lifetimes of the noble gas metastable triplet states enable these species 
to act as effective ground states for atom optics experiments \cite{Baldwin:05}.   
The metastable behaviour of these states arises from the doubly-forbidden nature of the decay process to the $^{1}$S$_{0}$ 
ground state for each species:  both electric dipole and spin-flip (triplet-singlet) transitions are forbidden.  
As shown in Table~\ref{tab:data}, the lifetimes of the metastable states range from $\sim$15 s to $\sim$8000 s.

For the heavier noble gases (Ne, Ar , Kr and Xe) the dominant decay process for the metastable n$s[3/2]_{2}(^{3}$P$_{2})$ 
states is \textit{via} a magnetic quadrupole (M2) transition.  The He \ptwo state also decays to the ground state 
principally \textit{via} an M2 transition, but its \textit{lifetime} is dominated by electric dipole (E1) decay to the 
metastable \met state.  By contrast, the lifetime of the metastable \met state itself is determined solely by decay to the ground state \textit{via} a magnetic dipole (M1) transition.

In all the noble gases, the $^{3}$P$_{1}$ transition to the ground state is not forbidden by dipole selection rules 
(i.e. it can decay \textit{via} an electric dipole transition), but the decay time is nevertheless relatively long 
since the transition is still spin-flip forbidden.  Conversely, decay of the $^{3}$P$_{0}$ state to the ground state 
is strictly forbidden to all orders of the multipolar expansion since there is no change in the total quantum 
number J = 0.

\begin{figure}
\centerline{\includegraphics[width=0.7\linewidth]{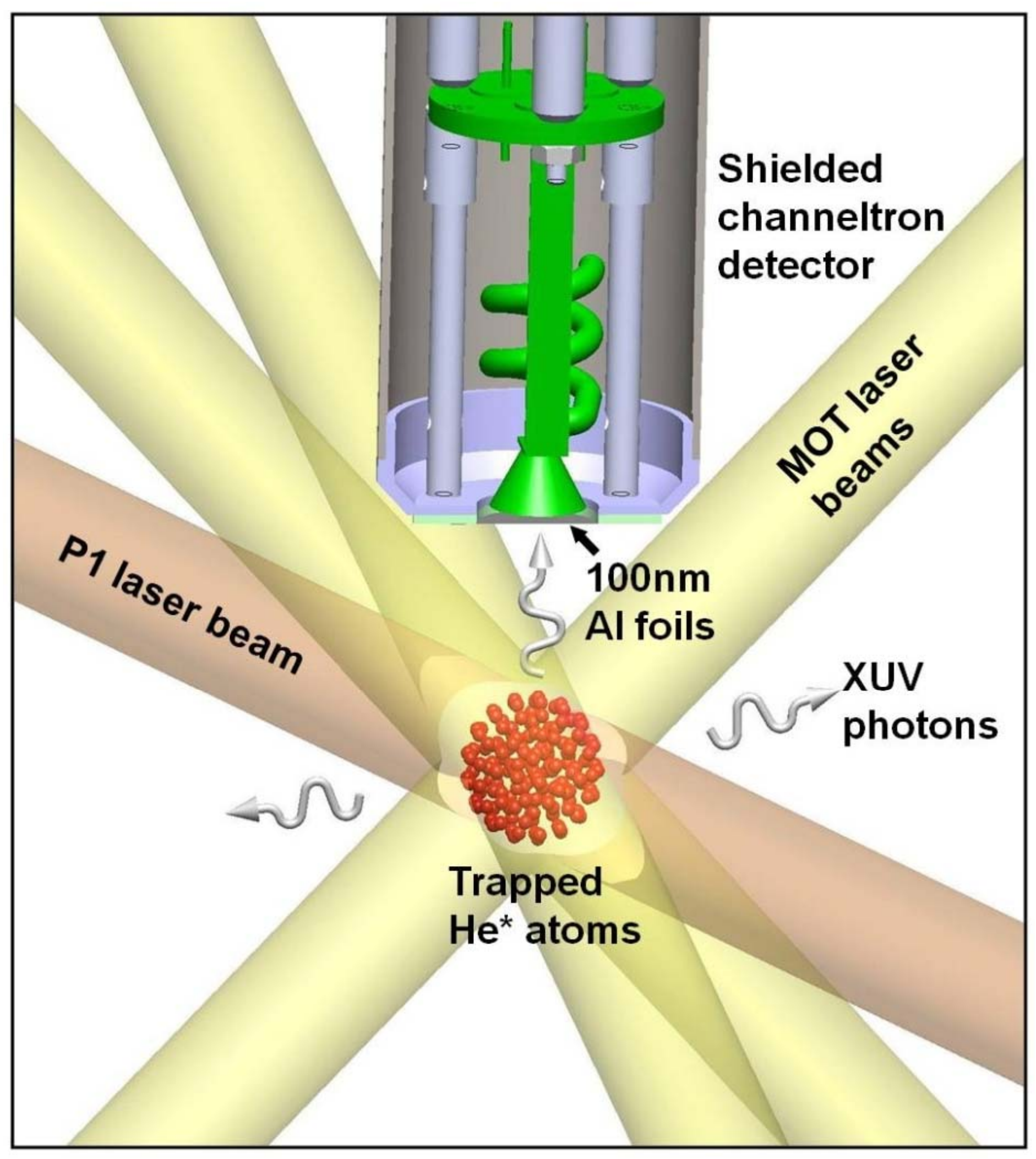}}
\caption{Experimental schematic showing the trapped atomic ensemble whose decay rate can be determined directly from the 
trap number loss rate, or by measuring the emitted XUV photons incident on a shielded channeltron detector 
(adapted from~\citet{Hodgman:09b}).}
\label{experiment}
\end{figure}

The long lifetime of metastable atoms enables them to be laser cooled and trapped in ultra-high vacuum environments 
for extended periods (comparable to their decay time).  This allows the metastable state lifetime to be measured 
either directly by determining the decay rate of the atomic ensemble ~\cite{Katori:93,Zinner:03,Dall:08a}, or by 
measuring the XUV photon emission rate and calibrating that against another known XUV emission 
rate~\cite{Walhout:95a,Lefers:02,Hodgman:09b,Hodgman:09a}. A schematic of such an experiment utilising both techniques 
is shown in Fig.~\ref{experiment}. However, measurement of the decay times of the metastable states is not simply 
motivated by their usefulness in atom optics.   For example, the two-photon transition from the 
Xe $6s[3/2]_{2}$ state 
to the $6s[3/2]_{0}$ state is also potentially useful as an atomic clock transition 
~\cite{Walhout:95a}, and the long life time of the He 2 $^3$S$_1$ state is important as an energy 
reservoir in electron collision-dominated plasmas for which this state has a large scattering 
cross-section~\cite{Uhlmann:05}.

Perhaps most importantly, the metastable state lifetimes are of significant interest as a test-bed for quantum 
electrodynamics (QED) - one of the most robust and long standing theories in modern physics.  
\begin{figure}[hb]
\centerline{\includegraphics[width=0.7\linewidth]{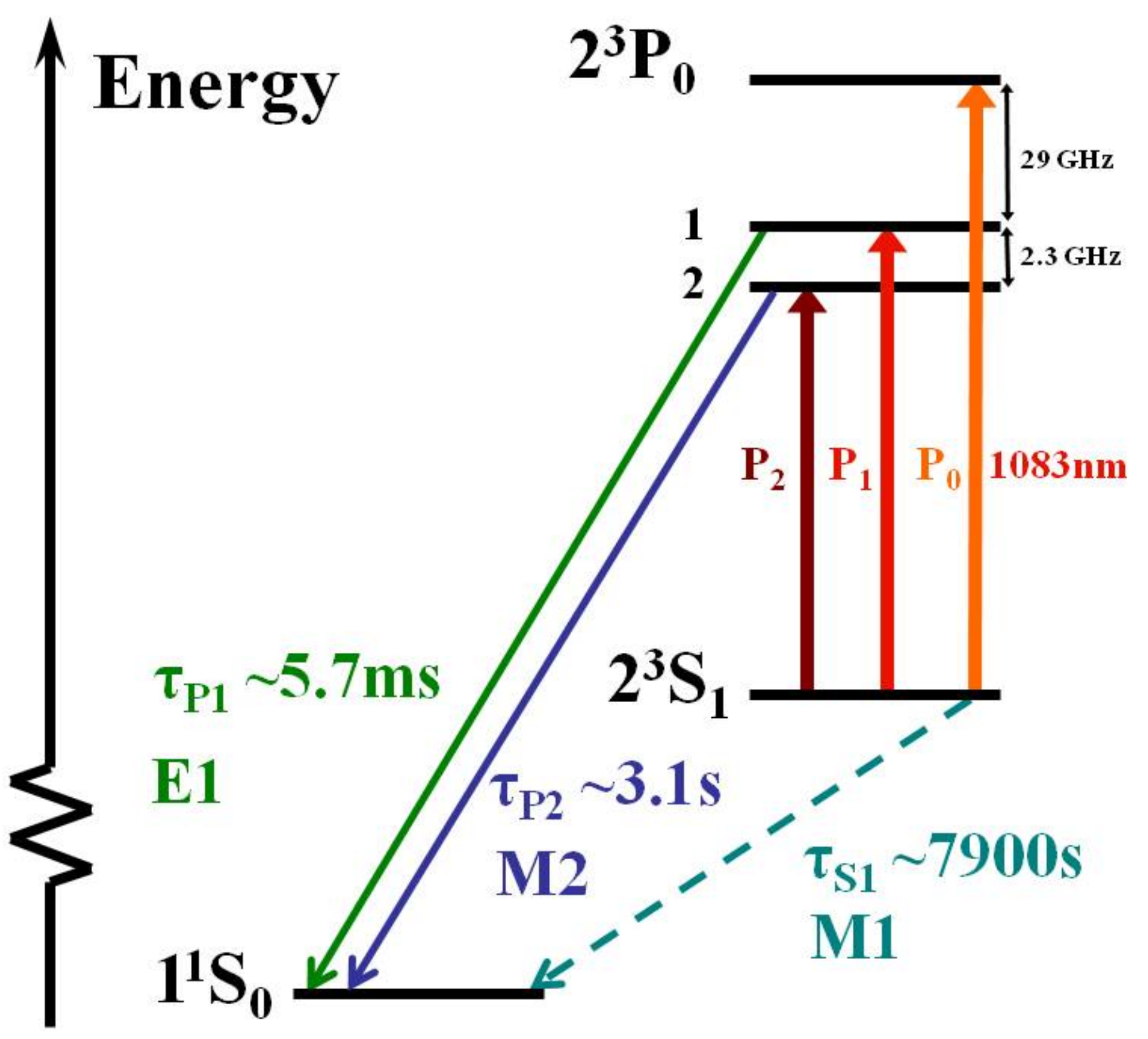}}
\caption{Triplet states of $^4$He showing XUV decay times to the ground state \textit{via} the mechanisms indicated 
(adapted from ~\citet{Hodgman:09a}).}
\label{transitions}
\end{figure}
However, QED calculations for atomic processes such as transition lifetimes are not as well-determined, with 
accuracies often no better than the percent level even in helium, the simplest multielectron atom.  Surprisingly, 
the theoretical accuracy again matches very closely the experimental uncertainty in measuring the transition rates, 
even though the measurement accuracy is up to nine orders of magnitude worse than for the atomic energy level separations.

For the heavier noble gases, there are further complications due to the widely acknowledged uncertainties arising from 
the theoretical treatment of relativistic effects and electron correlations, to which the significant discrepancies 
between theory and experiment (see Table~\ref{tab:data}) are attributed.  In Ne, the lightest of these, the relativistic 
effects are relatively minor, making Ne a good test-bed for studying electron correlations. However, while the agreement 
with theory is indeed somewhat better, the experimental value is still significantly less than 
predicted~\cite{Small-Warren:75,Indelicato:94,Tachiev:02}.  For the heavier species (Ar, Kr and Xe) the discrepancy 
increases with Z (as does the lifetime), possibly due to inadequate treatment of relativistic effects.  
This disagreement between theory and experiment for the metastable lifetimes of the heavier noble gases has 
yet to be resolved.  Indeed, a similar level of disagreement occurs for the decay of the metastable \p state 
in Sr~\cite{Yasuda:04}.

Theoretical attention has therefore focused more closely on He, the simplest multi-electron atom.  
Being the lightest of the noble gases it is also less susceptible to computational uncertainty contributions from 
relativistic effects and electron correlations.  

Until very recently, no published measurements had been made for transition rates to the ground state from the He \p manifold.  Fig.~\ref{transitions} summarises the atomic energy level scheme for He, and shows the primary decay mechanism to the ground state for each of the four lowest triplet states.  Once again, decay of the 2 $^{3}$P$_{0}$ state to the ground state is strictly forbidden.
Early attempts to measure the fastest (2 $^{3}$P$_{1}$) decay rate in a helium discharge~\cite{Tang:72} remained unpublished, 
as were more recent attempts at the Institut d'Optique in Orsay using a He* trap~\cite{Poupard:00}.  

However, in the first of a series of experiments at the Australian National University in Canberra, the UHV capabilities 
of a well-characterised He BEC apparatus~\cite{Dall:07a} were exploited to produce the first published measurements of 
the He \pone transition rate to the ground state~\cite{Dall:08a}.  The initial experiment employed the Penning ionization 
of background gas atoms \textit{via} (He*) single-body collisions as a diagnostic.  In this way, a direct measurement 
could be made of the trap loss arising from decay to the ground state by atoms continually cycled into the \pone state 
from the metastable state \textit{via} optical pumping with 1083~nm laser light.

This enabled the \pone transition rate to the ground state to be determined with an uncertainty of 4.4\% (see Table~\ref{tab:energies}).  
The technique did not require any additional absolute measurements, only knowledge of the optical excitation fraction 
into the \pone state.  The result was in excellent agreement with previous QED 
calculations~\cite{Drake:71,Johnson:95,Lach:01} and anchored the isoelectronic sequence for this transition at low Z.

Furthermore, the \pone XUV flux can be used to accurately calibrate the decay of other transitions.  
Using a shielded channeltron detector, the transition rate for the \ptwo decay to the ground state was determined relative to the 
(now known) count rate of the XUV photons emitted \textit{via} \pone decay to the ground state, with an absolute 
uncertainty of 5\% ~\cite{Hodgman:09b}.   This was again in excellent agreement with the same QED frameworks as 
before~\cite{Drake:69,Johnson:95,Lach:01}, and was able to discount several others.  An upper bound was also placed 
on the $2^{3}$P$_{0}$ decay rate of 0.01 s$^{-1}$ as determined by the channeltron background count rate.

Finally, the lifetime of the He 2 $^3$S$_1$ metastable state itself could be measured with improved accuracy using 
the same relative XUV flux technique.  Earlier measurements using an electric discharge in a highly perturbed environment yielded error bars of at least 30\%~\cite{Woodworth:75,Moos:73}, compared with the most recent experimental uncertainty of 6.5\% ~\cite{Hodgman:09a}.  It should be noted that the recent experiment covered a range of more than six orders of magnitude in XUV count rates.  Again, the agreement with QED predictions was excellent. Summarizing, experiment and theory are in good agreement for all 3 decay rates at a level of accuracy of $\sim5$\%. The results are listed in Table~\ref{tab:energies}.

The value determined for the He metastable lifetime - 7870~(510)s ~\cite{Hodgman:09a} - is 
the longest of any excited neutral atomic species yet measured.  Once again, this experimental determination anchors the 
isoelectronic sequence for the helium-like metastable lifetime at low Z and again confirms the previous three most 
consistent theories ~\cite{Drake:71,Johnson:95,Lach:01}.  The excellent agreement between the calculated and recently 
measured decay rates to the ground state for the four lowest triplet states in He is another validation of the 
robustness of QED theory.

\begin{table}[ht]
\caption{Lifetime of $^4$He states for decay to the ground state, deduced from MOT measurements (for references see Table~\ref{tab:data})}
\begin{tabular}{l c c c} \hline \hline
$^4$He State &	\pone &	\ptwo & 2 $^3$S$_1$\\ \hline 
Exp. Lifetime (s) & 5.66(25) $\times 10^{-3\,a}$ & 3.09(15)$^b$ & 7870(510)$^c$ \\  
Theory Lifetime (s) & 5.63 $\times 10^{-3\,h}$ & 3.06$^h$ & 7860$^h$  \\ 
\hline \hline
\end{tabular}
\label{tab:energies}
\end{table}

%
%
%
%

\subsection{Precision spectroscopy of atomic structure}
\subsubsection{Precision spectroscopy on triplet levels of He*}
\label{sec:precisionspectroscopy} As pointed out in the introduction, helium has long been a favorite testing 
ground for fundamental two-electron QED theory and for new techniques in atomic physics, both experimental 
and theoretical.  In this context, the spectroscopic measurements and theoretical determinations of the energies of low-lying triplet states have pushed the limits of precision. The readily 
accessible 2 $^3$P and 3 $^3$P manifolds have been the focus of much experimental activity, particularly 
following the introduction of optical frequency combs (see \citet{Maddaloni:09} for a recent review 
and \citet{Consolino:11}).

Among all triplet level transitions, that between the 2 $^3$S and 2 $^3$P
states in $^4$He at 1083 nm is by far the most experimentally studied. 
In a series of experiments 
\cite{Frieze:81,Minardi:99,Castilleja:00,Storry:00,Giusfredi:05,Zelevinsky:05,Borbely:09,Smiciklas:10},
1083 nm light was used to access the $^4$He 2 $^3$P fine structure, with the prospect of
obtaining an accurate value for the fine structure constant $\alpha$. 
Assuming the QED theory of fine structure energies 
of an atomic system to be correct, a determination
of $\alpha$ is possible by frequency measurements of these fine structure splittings, which are proportional to
$\alpha^2$.
Although the possibility of using the helium 2 $^3$P states to determine $\alpha$ was pointed out as early as 
1969 \cite{Hughes:69}, serious experimental work only began in the 1980s. 
With the development of laser 
spectroscopy and improved calculations, helium was recognized as the best atom for an $\alpha$ 
determination which could compete with other methods. These methods use many 
different physical systems and energy scales \cite{Mohr:08}, and thus a method based purely on atomic 
spectroscopy is of great interest.

Different high-precision spectroscopy approaches ranging  from optically-pumped magnetic resonance microwave 
spectroscopy \cite{George:01,Borbely:09} to heterodyne frequency differences of the 1083~nm transitions 
\cite{Giusfredi:05,Zelevinsky:05,Smiciklas:10}
have produced sub-kHz accuracy 2 $^3$P fine structure measurements with a remarkable agreement among them.
In particular the recently reported 9~ppb accurate value for the largest interval
$\Delta \nu_{2 ^3\text{P}_{0-2}}$= 31~908~131.25 (30)~kHz, would lead to 
an uncertainty of 4.5~ppb in the inferred value of $\alpha$ 
if the theory were exact~\cite{Smiciklas:10}.

Unfortunately, theory of  2 $^3$P fine structure \cite{Drake:02,Pachucki:09,Pachucki:10} 
has not experienced an 
accuracy improvement comparable to that of the experiments. 
Moreover, larger discrepancies between 
theory and measurements for the fine structure energies have prevented, up to now, an $\alpha$ value from helium competitive 
with determinations from other physical systems \cite{Mohr:08}.  
In fact, in the past, fine structure measurements were used to test 
the fine structure QED theory
in He, rather than to determine $\alpha$. 
However, the most recently published theoretical results \cite{Pachucki:10} have almost resolved the discrepancies between measurements and
theory for all three fine structure intervals, although the theoretical uncertainty
is still almost one order of magnitude worse than the experimental one. 

\begin{figure}
\centerline{\includegraphics[width=1.0\linewidth]{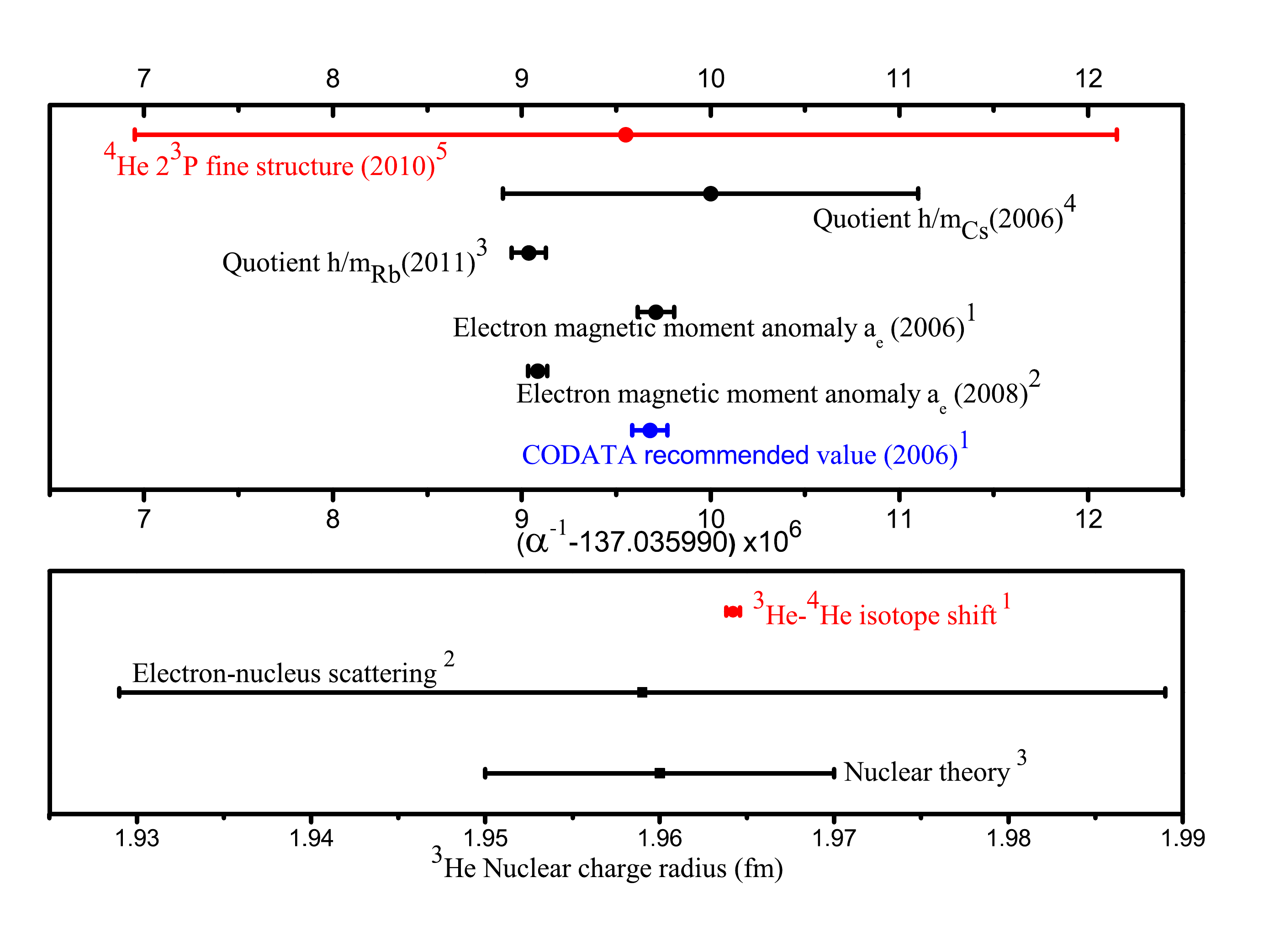}}
\caption{Upper graph: Comparison of most updated $\alpha$ determinations ($^1$\citet{Mohr:08}
and references therein, $^2$\citet{Hanneke:08}, $^3$\citet{Bouchendira:11}, $^4$\citet{Gerginov:06})
with the recent $\alpha$ value determined by $^4$He 2 $^3$P$_0$ - 2 $^3$P$_2$ fine structure
($^5$\citet{Smiciklas:10}). Lower graph: Comparison of the $^3$He nuclear charge radius
measured from the $^3$He - $^4$He isotope shift ($^1$\citet{Zhao:91,Shiner:95,Marin:95,Morton:06})
and by electron-nucleus scattering ($^2$\citet{Amroun:94}) and calculated by nuclear theory
($^3$\citet{Pieper:01}).   \label{fig:alpha-NCR}}
\end{figure}

Combining the above cited 
$\Delta \nu_{2 ^3\text{P}_{0-2}}$ measurement~\cite{Smiciklas:10} and QED corrections from the recent theory
\cite{Pachucki:10}, an $\alpha$ value from the He FS with an uncertainty of 31~ppb is determined.  
Such an uncertainty is mainly due to uncalculated  high-order  QED terms of the fine structure splittings. In the upper 
graph of Fig.~\ref{fig:alpha-NCR},
the agreement between this determination and the most recent $\alpha$ values from other
physical systems is shown.
The weight of this result in a new adjustment of $\alpha$ is still very weak, 
as the accuracy from helium spectroscopy is more than one order of 
magnitude lower than its competitors. 
Finally, we would like to mention some similar spectroscopic measurements of the 3 $^3$P
fine structure by using the 2 $^3$S$ \rightarrow $3 $^3$P transitions at 389 nm \cite{Pavone:94,Mueller:05}.
These measurements provided an independent test of He fine structure QED theory, albeit with lower accuracy.

Precise frequency measurements of larger energy intervals, like 2 $^3$S$ \rightarrow $2 $^3$P~\cite{Shiner:94,Cancio:04,Cancio:06}  provide a unique opportunity to test two-electron Lamb shift calculations, which, of
course, are absent in one-electron atoms. The most
precise Lamb-shift contribution to a transition frequency in a simple atomic
system, including hydrogen and deuterium, was determined by the optical frequency comb assisted
frequency measurements of the 1083~nm $^4$He transitions
\cite{Cancio:04,Cancio:06}. Here,
2 $^3$S$_1\rightarrow$ 2 $^3$P$_{2,1,0}$ transitions were probed by saturated
fluorescence spectroscopy in absence of external magnetic fields, where the
frequency of the exciting 1083 nm laser was measured against a
Quartz-GPS-disciplined optical frequency comb. The 2 $^3$S - 2 $^3$P centroid frequency was measured with
an accuracy of about 2~kHz (8$\times$10$^{-12}$), representing the best known optical frequency difference in 
helium \cite{Morton:05}. Theoretical calculations of such frequencies \cite{Yerokhin:10} are in reasonable agreement  with measurements, taking into account the larger uncertainty of theoretical energies (1-10~MHz).  
In fact, a QED test at the accuracy of the measurement is very challenging due to the 
difficulty of calculating all high-order QED contributions for S- and P-states. 

In addition to testing QED, the frequency measurements at 
1083 nm contribute to improving the 2 $^3$P level 
ionization energy, assuming the 2 $^3$S ionization energy is well known. 
Unfortunately, the final 2 $^3$P accuracy in the ionization energy
is still limited to 60~kHz as a result of the uncertainty in the ionization energy of the 2 $^3$S state.
Surprisingly, the widely cited value for the 2 $^3$S ionization energy is not a conventional
experimental determination based on extrapolation of a Rydberg series, but
instead a hybrid result obtained by combining an accurate measurement of the
2 $^3$S $\rightarrow$ 3 $^3$D transition \cite{Dorrer:97} with QED and
relativistic theoretical corrections for 3 $^3$D state \cite{Morton:05}. As
proposed recently \cite{Eyler:08}, highly accurate measurements of
transitions from the 2 $^3$S state to high-n Rydberg states can help to improve
our knowledge of the ionization energy both by increasing the experimental accuracy and by using Rydberg levels
with negligible QED and relativistic corrections. 
In particular, an optical frequency comb assisted measurement of the two-step transition 2 $^3$S $\rightarrow $ 3 $^3$P $\rightarrow $ 40 $^3$S
in a cooled and trapped ensemble of He$^*$ atoms,
may take the 2 $^3$S accuracy in the ionization energy to the kHz level.

Another important fundamental physics parameter which can be extracted from precision
spectroscopy of helium triplet levels is the determination of the nuclear charge radius
(r$_c$) of helium isotopes. 
This, and the nuclear mass, are the two basic atomic physics observables
defining the structure of an atomic nucleus, and hence provide a link to the 
nuclear theory of helium. 
Nuclear mass and volume differences between atomic isotopes 
determine the isotope shift of spectral lines. 
Precise isotope shift frequency measurements, 
together with a calculation of the nuclear mass contribution, extract the nuclear volume contribution, and thus the 
difference of the square charge radius of both isotopes \cite{Morton:06}. Since the r$_c$ of 
the $\alpha$-particle has been measured independently \cite{Borie:78, Sick:08}, measurements of the isotope shift of other helium 
isotopes, with respect to $^4$He, give a determination of r$_c$ of these isotopes. In a light atom such as helium, the difference in nuclear mass contributes
more than 99.99$\%$ to the isotope shift. 
However, the isotope shift is calculated with an uncertainty of a few tens of 
ppb \cite{Morton:06}. 
By making measurements of comparable accuracy, uncertainties in r$_c$  of 
1$\times$10$^{-3}$ fm can be achieved,
limited only by the accuracy of the  r$_c$  of the $\alpha$-particle. 

The above method was applied for the first time to determine the $^3$He r$_c$ by measuring the isotope shift for the transitions at 1083~nm \cite{Zhao:91} and at 389~nm \cite{Marin:95}. 
More accurate results were obtained by using isotope shift measurements
at 1083~nm by \citet{Shiner:95}. 
The determined $^3$He r$_c$ from these
measurements are in good agreement even though isotope shifts of different
helium transitions were used \cite{Morton:06}.
The agreement provides validity to the method. 
In the lower graph
of Fig. \ref{fig:alpha-NCR}, the weighted mean of r$_c$ for $^3$He, determined
by $^3$He/$^4$He isotope shift measurements is compared with the calculated value from nuclear theory
\cite{Pieper:01}, and with the electron-nucleus
scattering measurement \cite{Amroun:94}. The potential of this method is demonstrated by the
good agreement and the more than an order of magnitude higher accuracy. In fact, precise measurements of the size of halo
nuclei of rare $^6$He and $^8$He isotopes were performed by using isotope shift measurements at 389~nm in 2 $^3$S cooled and trapped in a MOT \cite{Wang:04,Mueller:07}. 
Moreover, another strength of this method is that, unilke electron-nucleus scattering, the resulting
charge radius is independent of the theoretical model for the nucleus. 

\subsubsection{Precision spectroscopy on heavier metastable noble gases}
Precision spectroscopy involving metastable levels of noble gases other than helium was mainly devoted 
to measuring isotope shifts and hyperfine structure in the case of fermionic isotopes 
\cite{Feldker:11, Klein:96, Blaum:08, Cannon:90, Walhout:93}. These measurements have not allowed precision 
levels as in helium due to the fact that metrology tools such as the optical frequency comb have not yet been used. 

Particularly interesting from the point of view of this review are the spectroscopic measurements in 
Ne$^*$ and Xe$^*$, performed in magneto-optical traps.  For neon \cite{Feldker:11}, the isotope shift of 
the $^3$P$_2 \rightarrow$ $^3$D$_3$ transition at 640.2 nm for all combinations of the three stable isotopes 
was measured, as well as the hyperfine structure of the  $^3$D$_3$ level of $^{21}$Ne.  The potential 
of two different spectroscopic techniques, absorption imaging and MOT fluorescence, used to realize these measurements was demonstrated, also for the very rare isotope $^{21}$Ne.  The realized accuracy 
improves previous results by about one order of magnitude. 
Similar improvement was reported for isotope shift measurements on the $^3$P$_2\rightarrow$ $^3$D$_3$ transition 
at 882 nm for all possible combinations of the nine stable xenon isotopes, including the most rare 
ones ($^{124}$Xe and $^{126}$Xe) \cite{Walhout:93}. In addition, hyperfine structure measurements in the upper level ($^3$D$_3$) were performed for the two stable fermions ($^{129}$Xe and $^{131}$Xe). As for neon, MOT fluorescence spectroscopy was used. 

Although magneto-optical trapping of Ar$^*$ and Kr$^*$ has been demonstrated (see Sec.~\ref{sec:ArKrXe}), measurements of properties other than the metastable
lifetime have not been performed with such samples. 
  

\section{Atom optics experiments}
\label{sec:AtomOptics}
\subsection{Interferometry} 

Shortly after the development of laser cooling techniques, and before the advent of BEC, atom interferometry 
developed as an important application of cold atoms. These developments are extensively described in a recent 
review \cite{Cronin:09}. 
That review also discusses interferometry experiments using conventional beams of metastable atoms.
This topic is beyond the scope of the present review; we will restrict ourselves to
a brief description of some interferometry experiments done with cold metastable atoms. 

The first interferometry experiment using laser cooled metastable atoms was reported in 1992 \cite{Shimizu:92}. 
Neon atoms from a MOT passed through a double slit and propagated to an MCP. High-contrast 
interference fringes were observed on a phosphor screen. 
The experiment introduced an important technique to enhance the spatial 
coherence of the source.
In neon, it is possible to optically pump atoms from the $^3{\rm P}_2$ state to the $^3{\rm P}_0$ with 
50 \% efficiency.
The optically pumped atoms have no magnetic moment and are insensitive to the
lasers creating the MOT; they therefore fall out and create a beam.
The source size is approximately the pumping laser waist 
(smaller than 20~$\mu$m in this case), and much smaller than the size of the MOT itself.
This technique is crucial for obtaining good fringe contrast with reasonable propagation distances.
The same experimental technique was subsequently used by the same group to
realize an amplitude hologram \cite{Morinaga:96}, a phase hologram \cite{Fujita:00} and a reflecting 
hologram \cite{Shimizu:02} for neon atoms.

\subsection{Metastable Helium Atom Laser}
\label{sec:atomlaser_introduction}
Since the observation of BEC, the ability to coherently outcouple atoms to
form a coherent beam of matter waves, or ``atom
laser"~\cite{Mewes:97} has attracted much attention,
in part because of its potential applications in atom interferometry.
Coherent matter wave optics is reviewed in \citet{Bongs:04}.
Like its optical counterpart, the atom laser is
interesting from both a fundamental and applied point of view.  
>From a fundamental perspective, atom lasers can be used to study atom-atom
interactions \cite{Doring:08}
as well as coherent properties \cite{Bloch:00} of matter waves. 
Atomic scattering is responsible for non-linearities 
in atom laser formation, which can generate
non-classical matter waves such as entangled beams \cite{Dall:09}.
Such beams are of interest for
tests of quantum mechanics \cite{Reid:09}, and for performing Heisenberg-limited
interferometry \cite{Dowling:98}.
>From an applied perspective, the atom
laser has the potential to revolutionise future atom interferometric
sensors~\cite{Cronin:09}, in which a high flux of collimated atoms is required. 
Ultimately the performance of such sensors will
depend on the signal-to-noise ratio with which atoms in the atom laser beam can
be detected. 
Thus the unique detection possibilities offered by metastable atoms
(see Sec.~\ref{sec:detection}) may prove important for future atom laser applications. 

\subsubsection{Atom laser spatial profiles}

\begin{figure}[h]
\centerline{\includegraphics[width=0.8\linewidth]{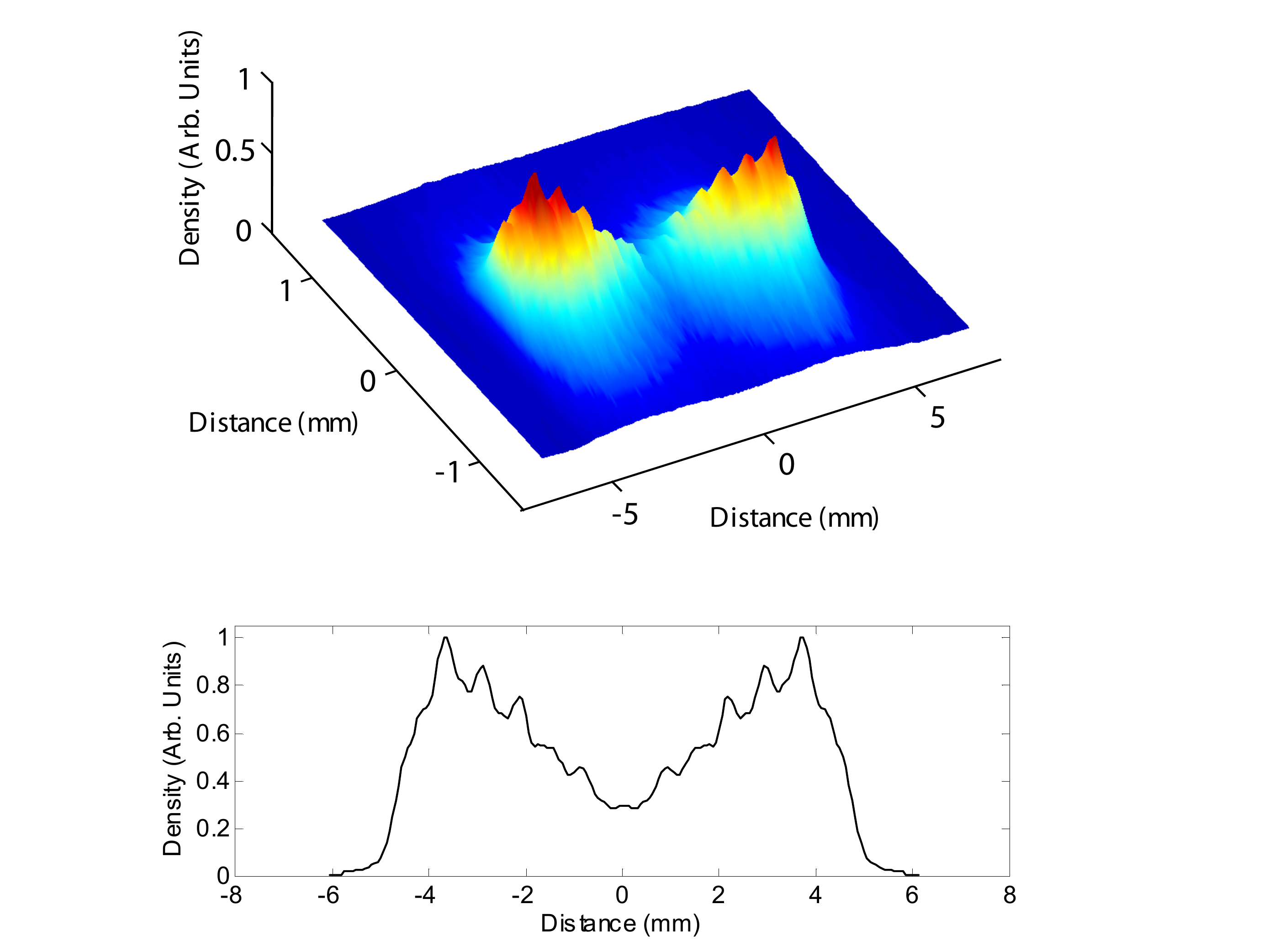}}
\caption{Image (upper plot) and cross-section (lower plot) showing
interference fringes of a metastable $^4$He atom laser~\cite{Dall:07b}. 
Reproduced with permission from The Optical Society of America (OSA).}
\label{AL_interference}
\end{figure}

Because of their low mass and large $s$-wave scattering length,
trapped He* atoms experience only a small gravitational displacement
(relative to the cloud size) away from
the magnetic trap minima, compared to an atom such as Rb. 
For such a system, rf output-coupling leads to
output-coupling surfaces that are oblate spheres rather than planes, as is the
case of previously studied atom lasers \cite{Bloch:99}. 
Atoms which are out-coupled above the trap centre
experience an upward force and therefore travel upwards and then
drop back through the condensate.  
The resulting
transverse atom laser profiles exhibit a central shadowed region
(Fig.~\ref{AL_interference}), cast by the condensate, since atoms passing
back through the condensate are pushed off axis due to the strong mean
field repulsion.

Besides these large-scale classical effects, it has been predicted that
interference fringes should be present on an atom laser beam. Atoms starting
from rest at different transverse locations within the outcoupling surface can
end up at a later time with different velocities at the same transverse
position, leading to interference~\cite{Busch:02}. In the
case of  He* these interference fringes are readily observable in the spatial
profile of the atom laser (see Fig.~\ref{AL_interference})\cite{Dall:07b}.

For the purposes of atom interferomety, the complicated structure exhibited by the
helium atom laser is likey to be a hindrance.
In principle atom laser beams can be guided in the lowest mode of a confining
potential in much the same way that optical fibres guide laser light. 
Recent experiments with $^{87}$Rb
atoms out-coupled from BECs into optical waveguides \cite{Guerin:06,
Couvert:08, Gattobigio:09} have achieved up to 85\% single mode occupancy. The
mode population in these experiments was inferred by observing the propagation
of atoms along the waveguide via absorption imaging, which also permits the
transverse energy of the guided atoms to be determined and compared with that
expected for various transverse mode combinations.

\begin{figure}
\centerline{\includegraphics[width=0.8\linewidth]{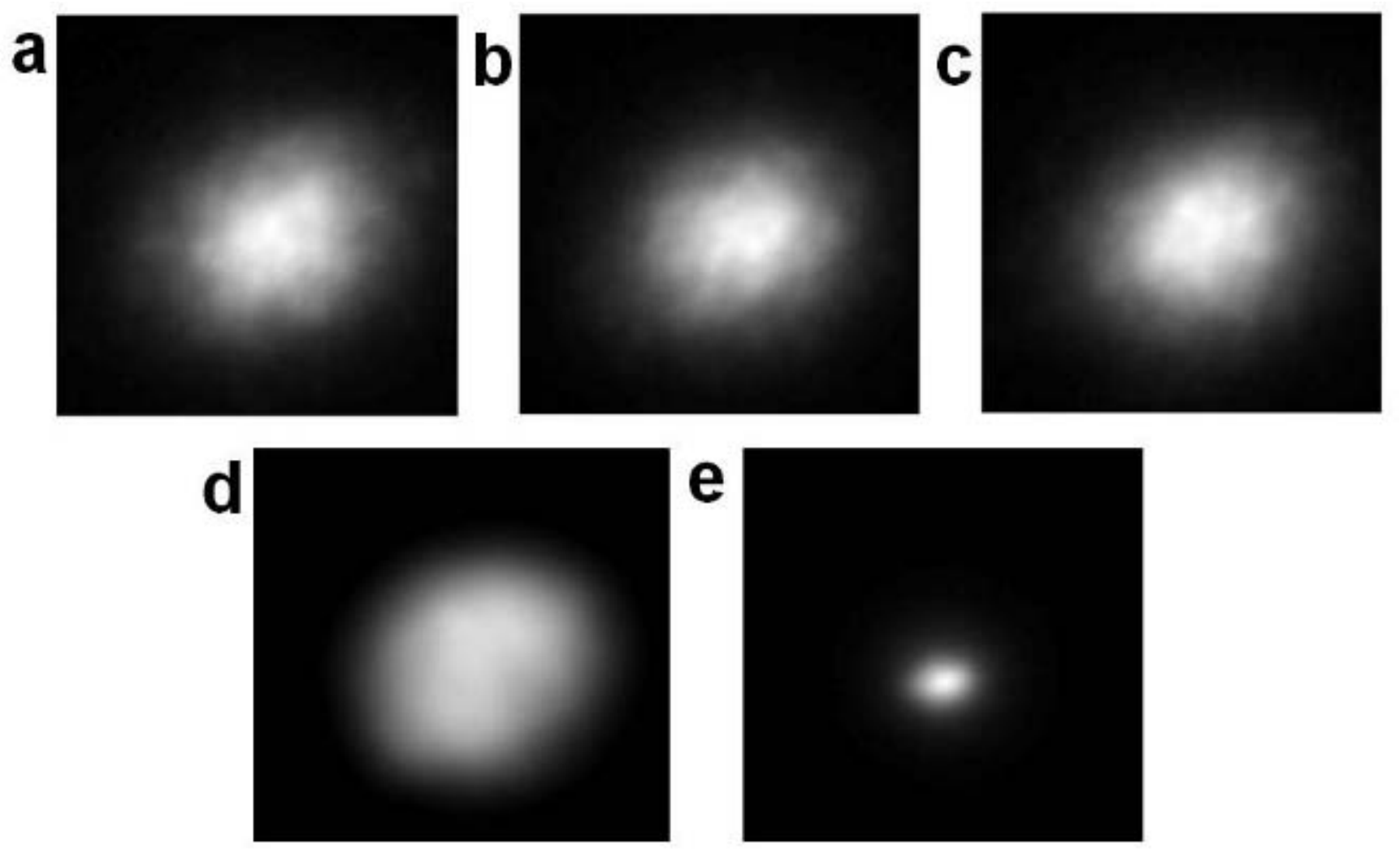}}
\caption{Multi-mode speckle images: (a-c) Successive experimental realizations 
produced by guiding only the thermal component of the trapped atoms, showing 
interference between the thermal modes (“speckle”).  Each panel is a single 
experimental run, and the pattern is seen to change between images due to the 
somewhat random nature of the speckle.  (d)  Average of twenty runs of the 
experiment,  (e)  Image of a predominantly single-mode profile.  
All 5 images show a 3mm window~\cite{Dall:11}.} 
\label{guiding}
\end{figure}

Recent experiments with an optically guided He* BEC have allowed direct imaging of the
transverse mode structure of the guided matter waves (see
Fig.~\ref{guiding})~\cite{Dall:10}. The laser beam used to confine the BEC is vertical, 
so that by adiabatically reducing the intensity of the
optical trap atoms are pulled out of the trap and into the waveguide by
gravity.  Since this process is adiabatic, or nearly so,
an almost 100\% single mode guided beam should be possible.
This idea is confirmed by simulations.
Guiding the atom laser beam results in a smooth Gaussian mode profile (see
Fig.~\ref{guiding}e), avoiding the formation of structure that is often present
in atom laser beams (see Fig.~\ref{AL_interference}).  In the case of He* 65\%
of the atoms have been shown to be guided in the fundamental mode, while the
coherence of the guided atom laser was demonstrated by the high visibility
interference pattern generated from a transmission diffraction grating.

In some very recent experiments~\cite{Dall:11}, atom guiding for several 
lowest-order modes was achieved by loading thermal atoms into the guide in a controlled fashion.  
Interference between the modes created atomic speckle which was imaged for the first 
time (Fig.~\ref{guiding}a-c).  In addition, measurement of the second-order correlation 
function (Section VIII) demonstrated atom bunching associated with the speckle pattern, 
while no atom bunching was observed for single-mode guiding (Fig.~\ref{guiding}e), as expected for a 
coherent atomic wavefront. 

\subsubsection{Feedback Control of an Atom Laser Beam}
\begin{figure}
  \centerline{\includegraphics[width=0.8\linewidth]{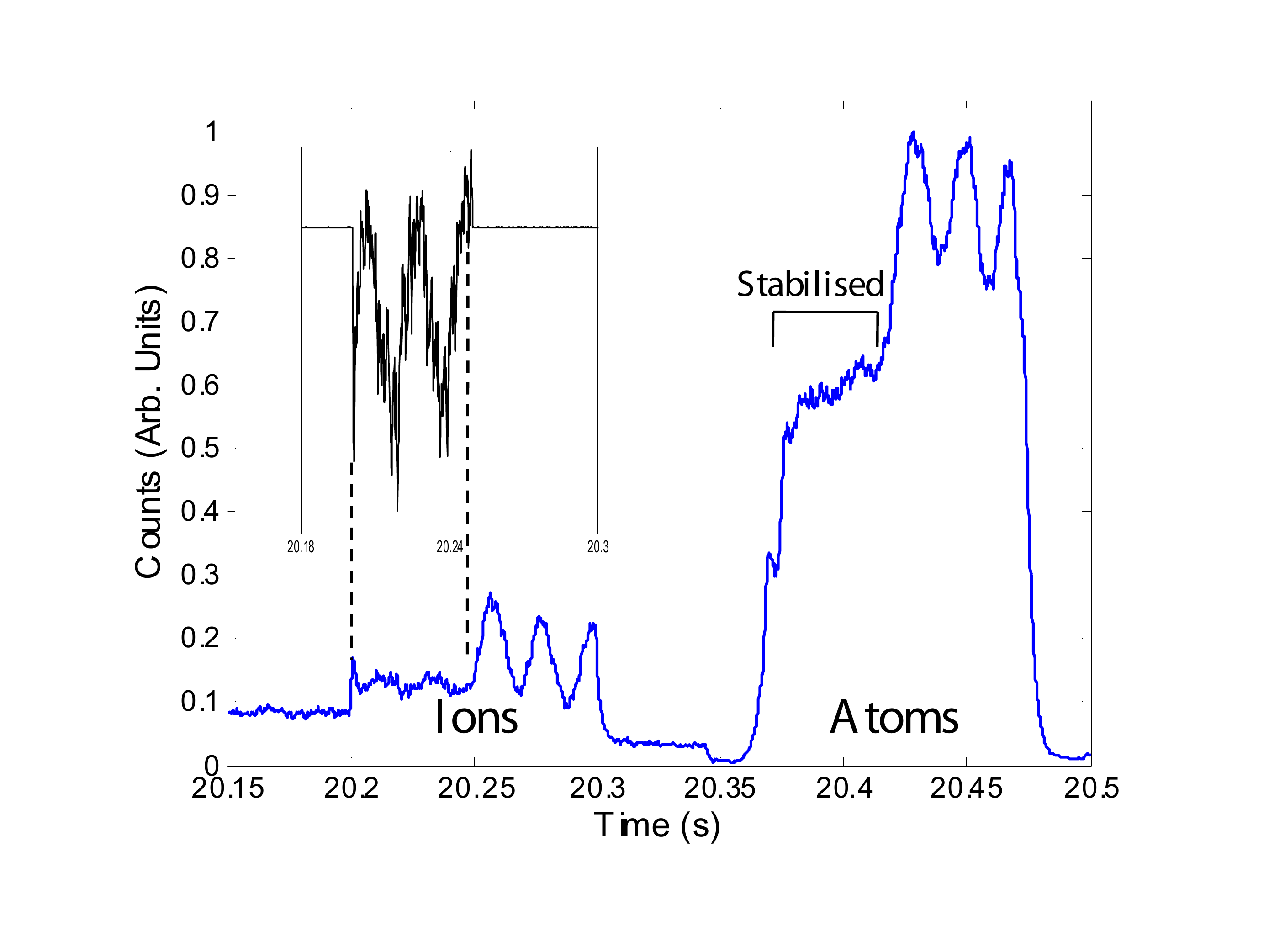}}
  \caption{Electron multiplier signal demonstrating stabilisation of the atom laser beam.
  The electron multiplier detects both the ions, which arrive first, and the beam of atoms, 
  which arrives after a 150 ms time of flight.
  Feedback control is only implemented for the first half of the atom laser signal.
  The inset shows the closed-loop error signal~\cite{Dall:08b}. 
  Reproduced with permission from The
  Optical Society (OSA).}\label{locking}
\end{figure}

Many precision applications of the optical laser involve active control, in
which an error signal is used in a feedback loop to control the laser output
\cite{Drever:83}. Similarly, the success of the atom laser as a practical
instrument may well depend on feedback control. 
For a beam of matter waves, flight times from the source to the detector are typically of
the order of many milliseconds, rather than the nanosecond times
possible with light waves. 
This difference renders feedback for most atom lasers less useful. 
In the case of helium however, there is a way
around this problem by probing the atom laser beam at the source
rather than monitoring the atoms in the beam itself. 
An error signal can be derived from the ion signal which accompanies the outcoupling. 
This error signal can be fed back to the rf-outcoupling frequency. 
When this scheme was implemented~\cite{Dall:08b}, 
a significant reduction in the intensity fluctuations of the atom laser beam was 
observed as shown in Fig.~\ref{locking}.
Since presumably, the feedback mechanism is acting on the position in the BEC
where atoms are being outcoupled, the beam energy and its spatial structure may be
stabilized as well.

\section{Pair correlation experiments}
\label{sec:correlations}
The particle counting techniques that naturally accompany the use of metastable
atoms have enabled and encouraged explorations of particle correlation effects
in cold gases (see Sec.~\ref{sec:ElectronicDetection}). Analogous experiments using optical imaging techniques have also
been carried out \cite{Folling:05,Greiner:05,rom:06},  but in the
following sections we will concentrate on results obtained with metastable
atoms.

Correlation measurements involve measuring at least two-particles and answering
the question, given the detection of one particle at point $r_1$ and time
$t_1$, what is the probability of finding a second one at point $r_2$ and time
$t_2$. The correlation is generally expressed in terms of second quantized field
operators by the second-order correlation function
\cite{Naraschewski:99,Gomes:06}:
\begin{equation}
g^{(2)}(r_1,t_1;r_2,t_2)=\\
{\left\langle\psi^\dagger(r_2,t_2)\psi^\dagger(r_1,t_1) \psi(r_1,t_1)\psi(r_2,t_2) \right\rangle
\over
{\left\langle \psi^\dagger(r_1,t_1)\psi(r_1,t_1)\right\rangle
 \langle\psi^\dagger(r_2,t_2)\psi(r_2,t_2) \rangle}
}.\label{eq:g2definition}
\end{equation}

If this function is different from unity, particles are somehow correlated. In the first sub-section we will describe experiments emphasizing the role of the quantum statistical properties of atomic clouds through the Hanbury Brown and Twiss effect. In the second sub-section this same function will allow us to charaterize atomic sources produced by a four-wave mixing process which have non-classical properties and potential applications in quantum information processing.

\subsection{Correlation effects in equilibrium ensembles}
\label{subsec:correlations}
Correlation measurements go to the heart of many quantum effects because, as has
been shown in the field of quantum optics \cite{Glauber:06}, it is in the two
particle correlation effects that a second quantized field becomes truly necessary 
to adequately treat a system of particles or photons. 
Thus one often speaks of the
beginning of modern quantum optics as coinciding with the experiments of
Hanbury Brown and Twiss [HBT] \cite{Hanburybrown:56a} and their elucidation by
Glauber \cite{Glauber:63a,Glauber:65}. 
We refer the reader to several works which give introductions and
simple explanations of the HBT effect both for photons and for particles
\cite{Baym:98, Loudon:00, Westbrook:09}.

\begin{figure}[t] 
   \centerline{\includegraphics[width=0.5\linewidth]{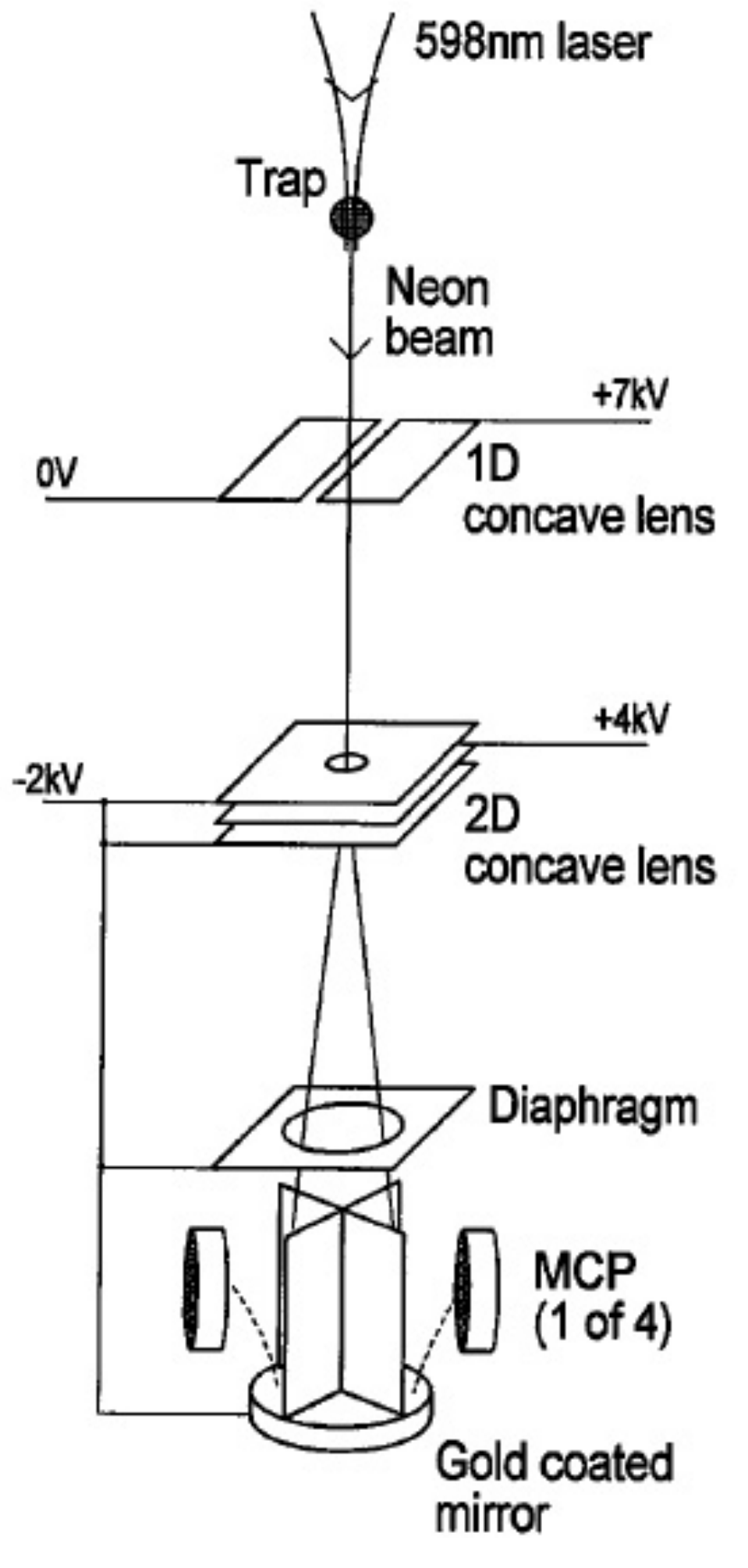}}
   \caption{Schematic diagram of the atom correlation experiment of \citet{Yasuda:96}.
   Metastable neon atoms are ejected from a MOT with a focused laser which
   pumps atoms into an untrapped state. Below the trap is a metallic plate
   (gold coated mirror) which can emit an electron when struck by a metastable atom.
   Four Microchannel plates collect the ejected electrons from different parts of the plate.
   The Hanbury Brown Twiss effect corresponds to an enhanced probability of detecting
   coincident electrons with zero delay, compared to the coincidence rate with a delay greater
   than the correlation time, about 100 ns. The electrodes between the source and the detector
   act as lenses and allow one to modifiy the source size as seen from the detector and thus
   modify the transverse coherence length of the beam.}
   \label{fig:yasuda}
\end{figure}

In the field of atom optics, the first experimental work on HBT was done using
metastable neon atoms \cite{Yasuda:96}. Fig. \ref{fig:yasuda} shows a schematic
diagram of their experiment. The experiment was a tour de force, because the
correlation length (or time) for thermal atoms in a MOT
is very short and, since an atom cloud in a MOT is far from quantum
degeneracy, the probability of finding two particles within a coherence
volume is very small. It was necessary to acquire data for a time on the
order of 50 hours,  and even with this amount of data, the signal to noise
ratio was low, but the experimental result definitively exhibited the bunching 
behaviour expected for the HBT effect of thermal bosons.

The situation became more favorable with the advent of evaporation techniques
to achieve quantum degeneracy (Sec.~\ref{sec:prod}). Quantum
degeneracy corresponds to one particle per coherence volume, rendering the
probability of finding two particles in the same coherence volume much higher.
Thus, with the demonstration of BEC of He*, a more
powerful version of the Yasuda and Shimizu experiment became an attractive
possibilty. In 2005,  \citet{Schellekens:05} succeeded in observing HBT
correlations with He*.

\begin{figure}[h] 
   \centerline{\includegraphics[width=0.8\linewidth]{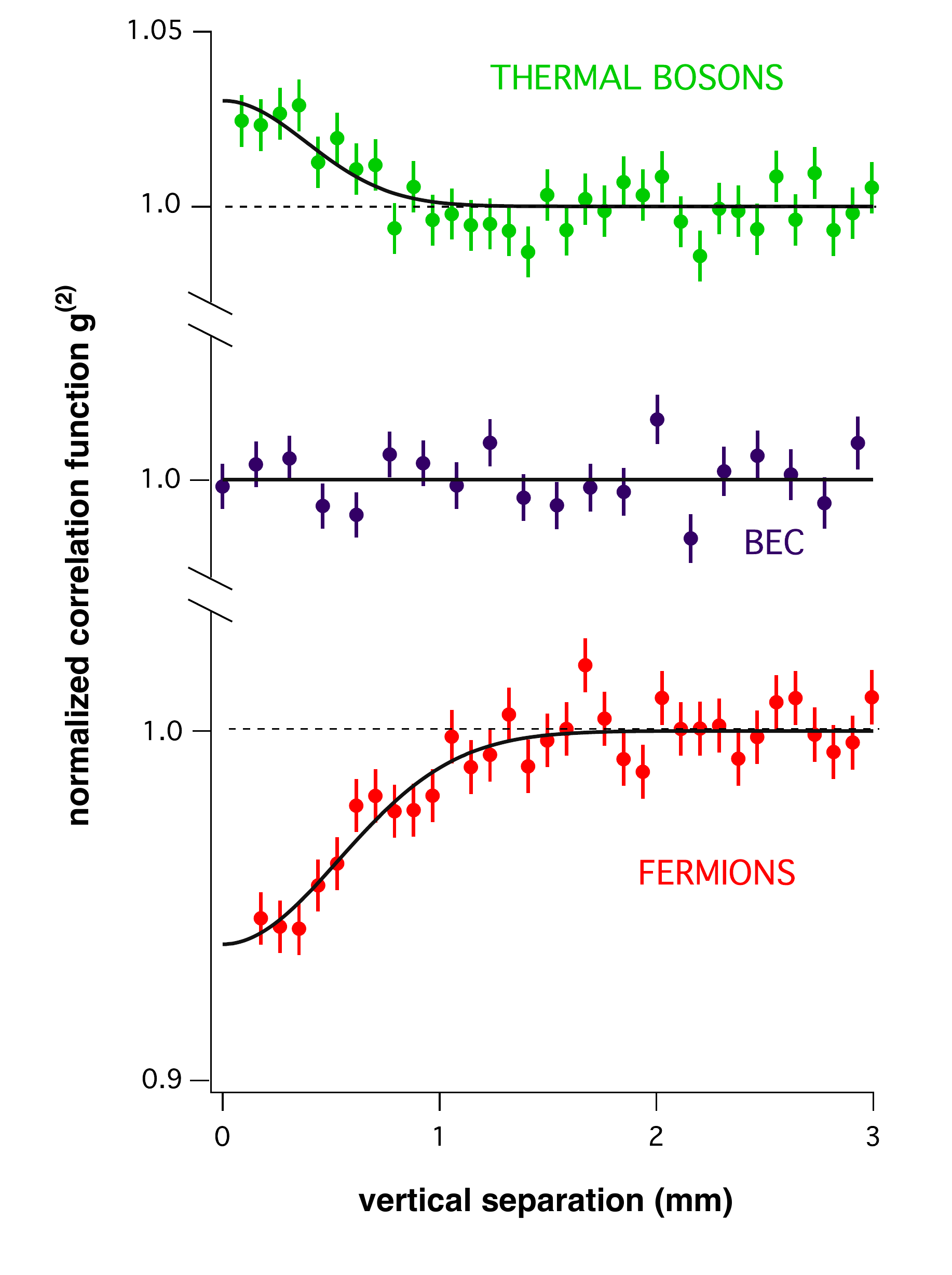}}
   \caption{Data on bunching and anti-bunching as shown by the normalized second-order
   correlation function Eq.~\ref{eq:g2definition}. The vertical scale of each plot is the same.
   The upper plot shows bunching of a thermal gas of bosons ($^4$He). The lower plot
   shows anti-bunching of a thermal gas of fermions ($^3$He). The horizontal axis shows
   the spatial separation, but it can be converted into a temporal separation by
   multiplying by the mean velocity of the atoms (about 3 m/s). The middle plot
   shows the flat correlation function of atoms from a BEC. The upper and lower
   plots are derived from the data of \citet{Jeltes:07}. The middle plot is derived
   from \citet{Schellekens:05}.
   }
   \label{fig:fermibosebec}
\end{figure}

This work reported measurements both of a thermal gas slightly above the BEC
threshhold, and of a degenerate gas. As shown in the top two plots of
Fig.~\ref{fig:fermibosebec}, the thermal gas shows a HBT effect, while a BEC
does not. The absence of HBT correlations in the BEC is an indication of the
fact that, like the intensity of a laser, density fluctuations are suppressed.
The positions of the particles in an ideal BEC are entirely uncorrelated with
each other. In contrast to the work of \citet{Yasuda:96}, which used a
continuous beam of atoms, the inherently cyclical nature of evaporative cooling
experiments imposed a pulsed mode of operation on the work of \citet{Schellekens:05}: 
an entire trapped sample was released and allowed to ballistically
expand. Another difference between the two experiments is the observed
correlation time which was 3 orders of magnitude larger than in the Ne*
experiment. The difference is in part due to the smaller mass of helium, but
also to the fact that in a ballistically expanding cloud, the slow and fast
atoms separate during propagation leading to a local momentum spread which is
smaller than that of the initial source. The spatial correlation length at the
detector is inversely proportional to the momentum spread, and thus the
correlation length increases as the cloud propagates. The correlation time is
given by the longitudinal correlation length divided by the mean velocity of
the atoms \cite{Gomes:06}. 

Recently a new thermal gas -- BEC comparison was
published~\cite{Manning:10}. In that work, atoms were rf-outcoupled from a trap
in roughly 30 small bunches, each containing only a fraction of the total atom
number. This technique permitted the use of a higher atom number in the BEC
without saturating the detector. The data showed that the second-order coherence
properties of both a thermal cloud and of a BEC are not perturbed by the
operation of an rf-outcoupler. 
This work has been extended to measurement of the three-body correlation function~\cite{hodgman:11}. It has demonstrated the long-range coherence of the BEC for correlation functions to third order, which supports the prediction that like coherent light, a BEC possesses long-range coherence to all orders. 

The achievement of quantum degeneracy of
a gas of metastable $^3$He, the fermionic isotope of this atom, was
reported in \citet{Mcnamara:06}.
In the fermion case, the exclusion principle, or alternatively the anti-symmetry of fermion
wavefunctions under exchange of particles, leads to an antibunching effect
rather than a bunching effect as with bosons. Antibunching has no classical
wave analog and thus by observing the HBT effect with fermions, one is truly
entering the domain of what might be called quantum atom optics in which we can
have a non-classical interference effect. 
A clear antibunching signal using $^3$He* was reported in 2007 \cite{Jeltes:07}. 
This work also
repeated the experiment for a thermal bose gas ($^4$He*) of very nearly the same
temperature and density. The comparison, shown in Fig.~\ref{fig:fermibosebec}
shows the dramatic effect of the different quantum statistics.

\subsection{Four-wave mixing of matter waves}
After the experiment of Hanbury Brown and Twiss, the field of quantum optics
developed further with the availability of non-classical photonic sources
\cite{Scully:97}. One well-known example is the source produced in spontaneous
parametric down conversion where photons are created in pairs through a
non-linear process \cite{Burnham:70}. Strongly correlated states are now at the
heart of quantum information processing and of future interferometers~\cite{giovannetti:04,giovannetti:11}. Quantum
{\it atom} optics is only at its early stages but is progressing rapidly. 
The fact that the atomic non-linearity is intrinsically present due to atomic
interactions and could be several orders of magnitude larger than optical
non-linearity \cite{Molmer:08} has attracted considerable interest. 
The search for efficient non-classical atomic sources is therefore both natural and desirable.
There have been many proposals concerning atom pairs, especially the production and observation
of individual entangled pairs of atoms through atomic collisions or the breakup of diatomic molecules
\cite{Band:00,Pu:00,Duan:00,Opatrny:01,Naidon:06,Zin:05,Kheruntsyan:02,Zin:06,Savage:06,Norrie:06,Deuar:07}. As emphasized in ~\citet{Duan:00}, pair production can be studied in two limits. If many atoms are created in a single pair of modes, stimulated emission of atoms is important and one speaks of two-mode squeezing in analogy with ~\citet{Heidmann:87}. The opposite limit, in which the occupation number of the modes is much less than unity, corresponds to the spontaneous production of individual, entangled atom pairs, either in spin or momentum states in analogy with \citet{Ou:88},\citet{Shih:88} and \citet{Rarity:90}.

\subsubsection{Pair production in the spontaneous regime}
The correlation among scattered atoms has been studied experimentally in the spontaneous limit in the breakup of K$_2$ molecules~\cite{Greiner:05}, using the technique of noise correlation in absorption images~\cite{Grondalski:99, Altman:04}, and in the collision between two Bose-Einstein condensates of metastable helium atoms~\cite{Perrin:07,Perrin:07b} using a 3D single atom detector \cite{Schellekens:05}. The performance of such a detector (see Sec.~\ref{sec:detection}) has enabled a careful characterization of the pair production mechanism.

The collision between two BECs produces scattered particles by elastic
collision and can be viewed as a spontaneous four-wave mixing process. This can
be shown using the Hamiltonian governing the system,
$$\hat H=\int d{\bf r}\;\hat\Psi^\dagger({\bf r})\left( -\frac{\hbar^2}{2m}\Delta+V({\bf r})+g
\hat\Psi^\dagger({\bf r})\hat\Psi({\bf r})  \right)\hat\Psi({\bf r})$$
with $V$ the trapping potential and $g$ the interaction coupling constant.
In a simple picture one can write $\hat\Psi({\bf r})$ as
$$\hat\Psi({\bf r})=\phi_{\bf Q}({\bf r})+\phi_{\bf -Q}({\bf r})+\hat\delta({\bf r})$$
with $\phi_{\pm\bf Q}$ representing the two coherent colliding condensates of relative momentum
$\pm \hbar {\bf Q}$, and $\hat\delta$ the scattered field.
When the depletion of the condensates can be ignored, the Hamiltonian contains a term of the
form $g\phi_{\bf Q}\phi_{-\bf Q}\hat\delta^\dagger\hat\delta^\dagger+h.c.$ similar to the one
found in spontaneous parametric down conversion or molecular dissociation. One then expects that the scattered field has also similar
quantum properties. 

\begin{figure}
\centerline{\includegraphics[width=\linewidth]{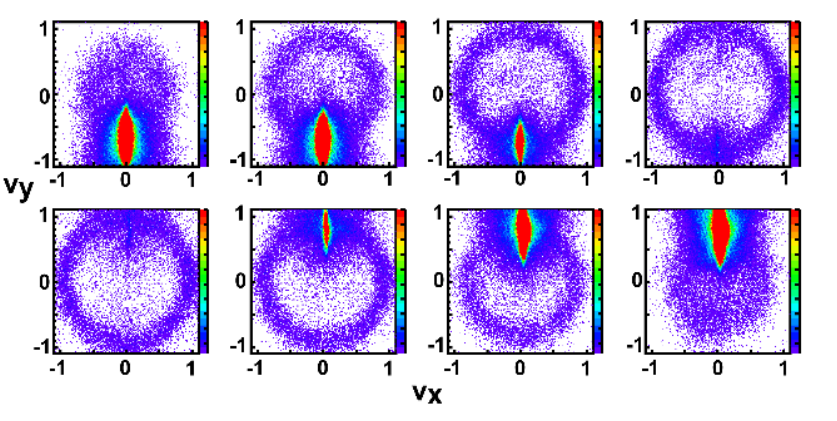}}
\caption{(Color online) 
Slices in vertical velocity $v_z$ of the spherical shell of atoms in velocity space 
(in units of $v_{rec}$) for a collision of two BECs. The data is similar to that of
~\citet{Perrin:07}.
All plots use the same linear false color scale: 3-blue, 100-red.
The scattering halo is the circular shell which intersects the BECs.}
\label{BEC_collision_fig2}
\end{figure}
\begin{figure}[htb]
\centerline{\includegraphics[width=1.0\linewidth]{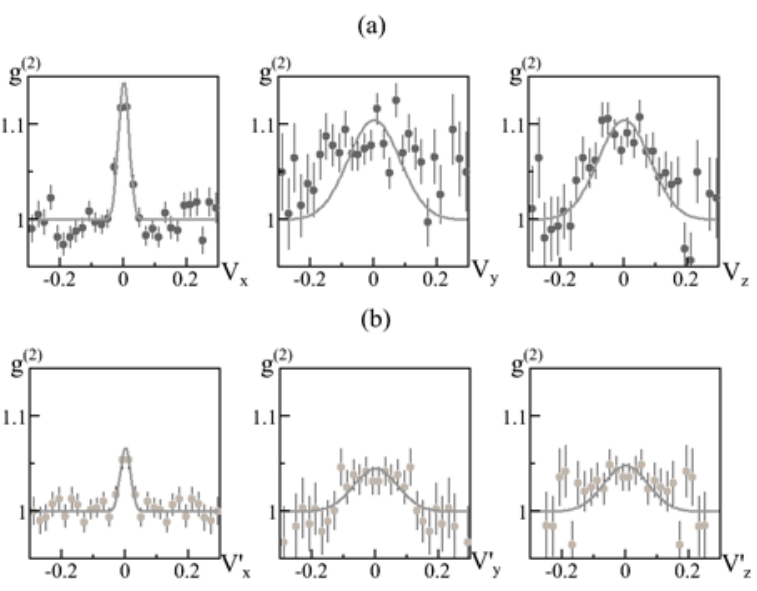}}
\caption{Back to back (panel a) and collinear (panel b) correlation peaks 
observed in the collision of two BEC's. (a)
Projection of the normalized two-body correlation function along the different
axes of the experiment and around $\mathbf{v+v'}=\mathbf{0}$. The projection
consists in averaging the correlation in the two other directions over a surface
equal to the products of the corresponding correlation lengths. The peak is the
signature for correlated atoms with opposite velocities. (b) Like (a) but
for $\mathbf{v-v'}=\mathbf{0}$. This peak is due to the Hanbury Brown and Twiss
bunching effect. In all the graphs, velocities are expressed in units of the recoil
velocity. Reprinted from ~\citet{Perrin:07b}.}
\label{BEC_collision_fig3}
\end{figure}

In the experiment of \citet{Perrin:07b}, two stimulated Raman transitions
transfer the atoms from a condensate confined in a magnetic trap (magnetic
sub-state $m=1$) to the magnetic insensitive state $m=0$. Since the laser beams of the Raman transitions are different, the momentum they transfer to the atoms is also different. The two ``daughter'' condensates have a relative velocity of $2 v_{rec}\approx
18.4~$cm/s, at least 8 times larger than the speed of sound in the initial
condensate. Since the collisions are elastic, the scattered particles are
expected to lie approximately \cite{Krachmalnicoff:10} on a spherical shell
in velocity space of radius $v_{rec}$, as shown in
Fig.~\ref{BEC_collision_fig2}. Since atoms are expected to be scattered in pairs
with opposite velocities (in the center of mass frame), the normalized two-body
correlation function $g^{(2)}({\bf v,v'})$ (equivalent of Eq.~\ref{eq:g2definition} in velocity space)
peaks at $\bf v'=-v$; this is demonstrated in Fig.~\ref{BEC_collision_fig3}(a).

In addition to the correlations of opposite momenta, 
the panel (b) of Fig.~\ref{BEC_collision_fig2} shows the correlation for $\bf v'\approx v$. Momentum conservation forbids two atoms to be scattered with the same velocity,
but, since the experiment is performed with bosonic atoms, bunching between
pairs of atoms is expected in analogy with the Hanbury Brown Twiss effect (see
section \ref{subsec:correlations})~\cite{Molmer:08}. The local correlation can hence be explained
by a four-body process. As seen in Fig.~\ref{BEC_collision_fig3}, the
correlation function is anisotropic with a width along $x$ much shorter than
along the $yz$ plane. A perturbative approach~\cite{Chwedenczuk:08} confirms
that the widths of the opposite and collinear correlation are mainly controlled
by the spatial size of the BECs as in the Hanbury Brown Twiss
experiment~\cite{Gomes:06}. Since the trap is anisotropic with a long axis
along $x$, the correlation along that axis is the shortest as expected.
Numerical calculations are in good agreement with the experiment. The local correlation
function can also be used to define a mode volume; this leads to a mode
population of the order of 0.1, indicating the spontaneous nature of the
collision. To fully understand the results, the BEC
collision is also simulated using a fully quantum, first-principles numerical
calculation based on the positive-$P$ representation method~\cite{Deuar:07,Perrin:08}.
The good agreement between these calculations and the experimental data shows that 
the physical picture is correct and exemplifies the power of this method.

One expects that a zone in momentum space centered at $\bf k$ should have exactly the same atom number as the corresponding zone centered at $-\bf k$. 
Sub-shot noise number differences have indeed been observed  in such a  BEC collision experiment~\cite{Jaskula:10}. 
Although the measured noise reduction (0.5~dB) was modest, it has been shown to be completely dominated by the finite detection efficiency of the detector (see Sec.\ref{sec:detection}), demonstrating that the collision between two BECs indeed produces a good non-classical atomic source.
The result is not entirely trivial because the presence of correlations in 
opposite momenta does not guarantee a sub-Poissonian number difference 
(see \citet{buchmann:10} for an example).

\subsubsection{Paired atom laser beams in the stimulated regime}
Stimulated four-wave mixing in a trapped
BEC was first demonstrated in 1999 \cite{Deng:99}, using three matter waves to generate a fourth. More recently, twin atomic beams were created using a similar process
\cite{Vogels:02}.  
Although phase coherence of these matter waves was
demonstrated correlation properties were not observed. 

In the case of a He* atom laser, pairs of beams can be produced simply by the
process of rf outcoupling from a He* BEC \cite{Dall:09}.
Unlike the previous methods, which required pairs of atoms traveling at high
kinetic energies as a source, this process involves scattering between atoms in
the same zero momentum state to states with non-zero momentum. At the heart of
the method are the different scattering lengths that are available for the
different magnetic sublevels of He* \cite{Leo:01}.  As atoms are outcoupled in a different magnetic sublevel, their interaction
energy abruptly changes, driving collisional processes consisting in atom pairs moving in opposite directions.

\begin{figure}[h]
\centerline{\includegraphics[width=0.8\linewidth]{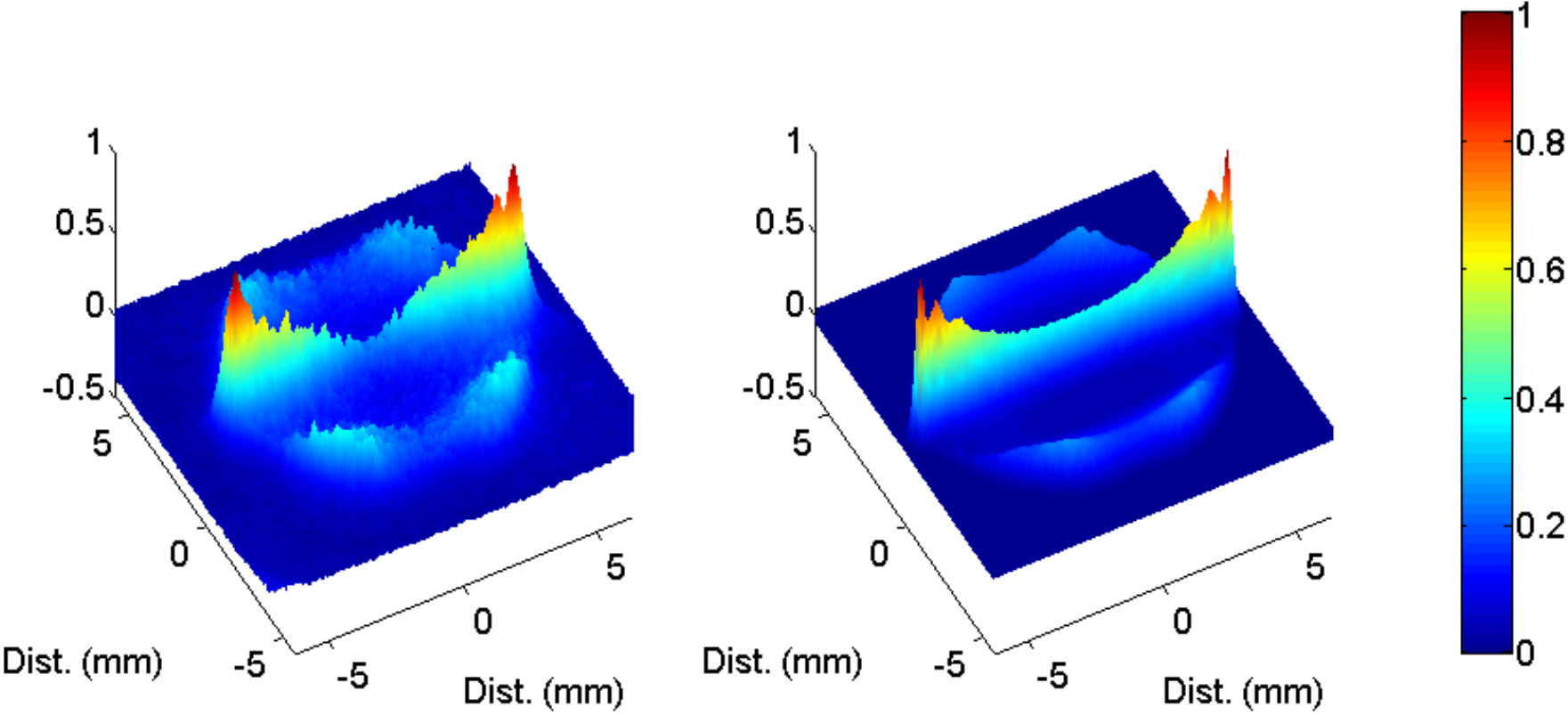}}
\caption{Spatial profiles of the He* atom laser when the conditions for
stimulated four wave mixing are met (experimental image - left, theoretical simulation -
right). 
Four extra peaks, resulting from four wave mixing, are observed around the usual He*
profile. 
Adapted from \citet{Dall:09}.} \label{FWM_al}
\end{figure}

Stimulated four-wave mixing occurs if the outcoupling surface is chosen to be an ellipsoidal shell, for example by detuning the outcoupling from the center of the condensate by $2\, \text{kHz}$. This causes a range of densities in the trapped
and untrapped fields to exist over the surface. Atoms accelerated in the mean-field and trap potentials by initial four-wave mixing scattering are amplified, allowing resonant stimulated scattering into higher and higher final momentum
modes.  This sweeps out the halo shown in Fig.~\ref{FWM_al}, and finally reaches the momentum corresponding to the well defined peaks (shown on the outskirts of the halo in Fig.~\ref{FWM_al}) which then become heavily
populated.

Quantum and semiclassical models indicate that these peaks are formed from scattering of pairs of atoms in a BEC, and should be therefore entangled upon formation.  It remains to be seen whether correlation and useful entanglement remain after the
outcoupling process.










\section{Other experiments using cold, metastable noble gases}
\label{sec:OtherExperiments}
In this last section we discuss several other experiments that have been performed 
in recent years using cold noble gases.

\subsection{Reflection of slow metastable atoms from surfaces}
\label{sec:Reflection}

Using cooling techniques, atoms can be rendered sufficiently slow that they probe the weak, 
long-range interactions with surfaces corresponding to the Casimir-Polder force \cite{Casimir:48}.  
Since the potential in this regime is \emph{attractive}, the reflection of atoms from a surface 
is remarkable. 
The phenomenon is a wave effect, analogous to the partial reflection of an electromagnetic wave 
at a dielectric interface, and is generally referred to as ``quantum reflection".
The low noise electronic detection methods, discussed in Sec.~\ref{sec:prod}, have rendered 
metastable atoms useful for such experiments because reflection coefficients, and thus the 
available signal, can be very small.
In 2001, \citet{Shimizu:01} reported the observation of specular reflection of Ne* atoms from 
both a silicon and a glass surface.
A grazing incidence geometry permitted incident velocities as low as 1~mm/s normal to the surface. 
For Ne*, this velocity corresponds to a deBroglie wavelength of approximately 20~$\mu$m.
For incident normal velocities between 1~mm/s and 35~mm/s, 
the observed reflectivities varied from above 0.3 to below $10^{-3}$.
A careful study of the velocity dependence showed that, within the estimated uncertainties, 
the data agreed well with the Casimir-Polder theory.
This work was followed up by a study of reflection and diffraction from a Si surface with a 
periodic structure of parallel ridges with heights of a few $\mu$m \cite{Shimizu:02b}. 
Using the same atomic source geometry, the authors demonstrated somewhat higher reflectivities 
at 1~mm/s incident velocity and \emph{much} higher reflectivity at 30~mm/s.
The group later experimented with reflection of He* from a flat silicon surface 
 \cite{Oberst:05}. 
The lowest incident normal velocity was higher in this case (30~mm/s), but at this velocity a 
reflectivity above 10\% was nevertheless observed.
The authors also compared the velocity dependence with that of Ne* and showed that it scaled
with the mass and polarizability of the atoms. 

Interaction of He* with surfaces was also studied theoretically. 
\citet{Yan:98} made detailed calculations of the polarizability and gave interaction
potentials for a perfectly conducting or dielectric surface.
\citet{Marani:00} used these results to analyze how the atom surface interaction would
modify the behavior of atom diffraction effects at an evanescent wave atomic mirror.
\citet{Halliwell:11} determined the reflection probability for He* 
atoms inside hollow optical 
fibres to determine the contribution to the atomic transmission efficiency.

\subsection{Birth and death of a Bose-Einstein condensate}
\label{BECdecay}

As pointed out in
Sec.~\ref{sec:intro} and \ref{sec:cold_collisions}, inelastic collisions with a metastable atom
usually lead to production of ions and electrons. Since charged particles are
efficiently detected with an MCP detector, the corresponding ionization
signal can be used to monitor the metastable atomic cloud in real time. 
We have already seen how such a signal can be used in a feedback loop (Sec.~\ref{sec:AtomOptics}).
The ion signal can also serve to monitor condensation dynamics.
At low density, the ion rate is due to Penning collisions with background gas
and thus allows one to monitor the metastable atom number. 
At higher density, reached for clouds close to quantum degeneracy, 2- and 3-body inelastic collisions 
dominate the ion rate which in turn gives information about the
atomic density. 
Because the N-body correlation functions are different
in a thermal cloud and a condensate (see Sec.~\ref{sec:correlations}),
inelastic N-body collisions occur with a rate constant $N!$ smaller for a
condensate than for a thermal sample~\cite{Kagan:85, Burt:97}. 
Nevertheless, in contrast to a homogeneous gas, 
the ion rate of a trapped condensate is still significantly higher than for a 
thermal sample in the same trap because of its
higher density.

\begin{figure}
\centerline{\includegraphics[width=1.0\linewidth]{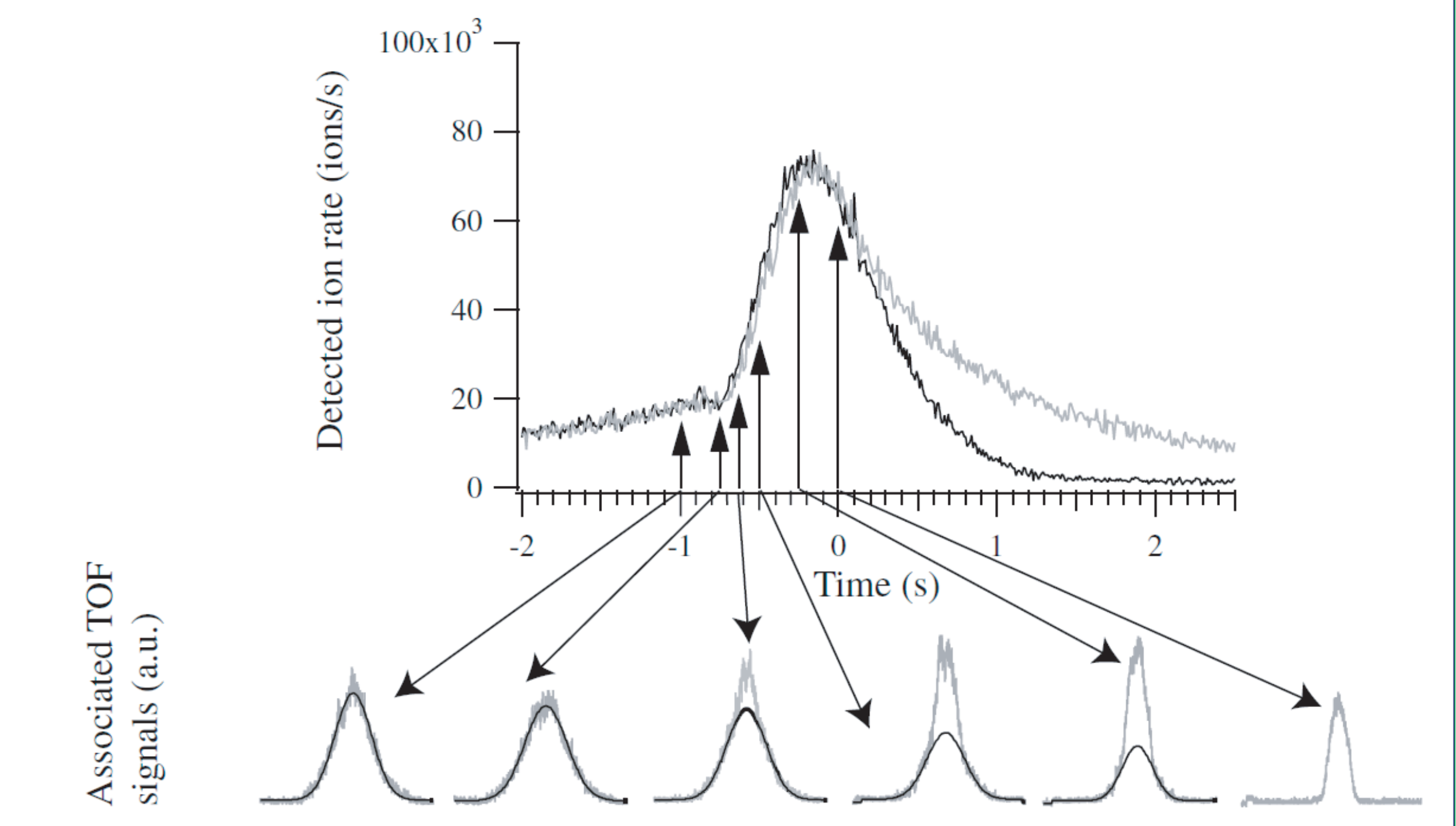}}
\caption{Ion signal during the last stage of RF-evaporation. Its sharp increase at
$t_{th}=-0.7$~s indicates that the cloud crosses the Bose-Einstein threshold.
This is confirmed by switching off the trap at various times and measuring the
time-of-flight signal. For $t>t_{th}$ a double structure is clearly visible indicating
the presence of a Bose-Einstein condensate. The upper lighter curve corresponds to a
situation where the RF-shield is always on whereas it is off after BEC formation for
the lower darker curve. The difference indicates that without an RF-shield, the condensate heats
up rapidly (see text). Figure extracted from ~\citet{Seidelin:03b}.}
\label{fig1_sec6A}
\end{figure}

In a trap therefore, one can observe a gas crossing the threshold for Bose-Einstein 
condensation as a sudden increase in the ion production
rate~\cite{Robert:01,Seidelin:03b,Tychkov:06}.
This phenomenon is illustrated in
Fig.~\ref{fig1_sec6A}. 
This signal is a good indicator to pinpoint the BEC threshold
and reproducibly place a gas close to that threshold.
In addition, since for a given trap and  temperature, the threshold corresponds to a 
well defined number of atoms, locating the BEC threshold can be used to
calibrate the number of atoms.
This idea was used in an experiment to measure 
the scattering length~\cite{Seidelin:04}, as discussed in Sec.~\ref{sec:ScatteringLength}.
Figure~\ref{fig1_sec6A} also shows that the ion rate permits monitoring of
condensate decay. 
Such a study was also carried out in \citet{Tychkov:06}, 
and modeled in
\citet{Zin:03},~\citet{Gardiner:97} and \citet{Soding:99}. 
It has been shown that under
the assumption of rapid thermalization, a transfer of atoms from the condensate
to the thermal component should occur, enhancing the condensate decay.

\subsection{Hydrodynamic regime close to the bosonic degenerate regime}
\begin{figure}
\centerline{\includegraphics[width=0.8\linewidth]{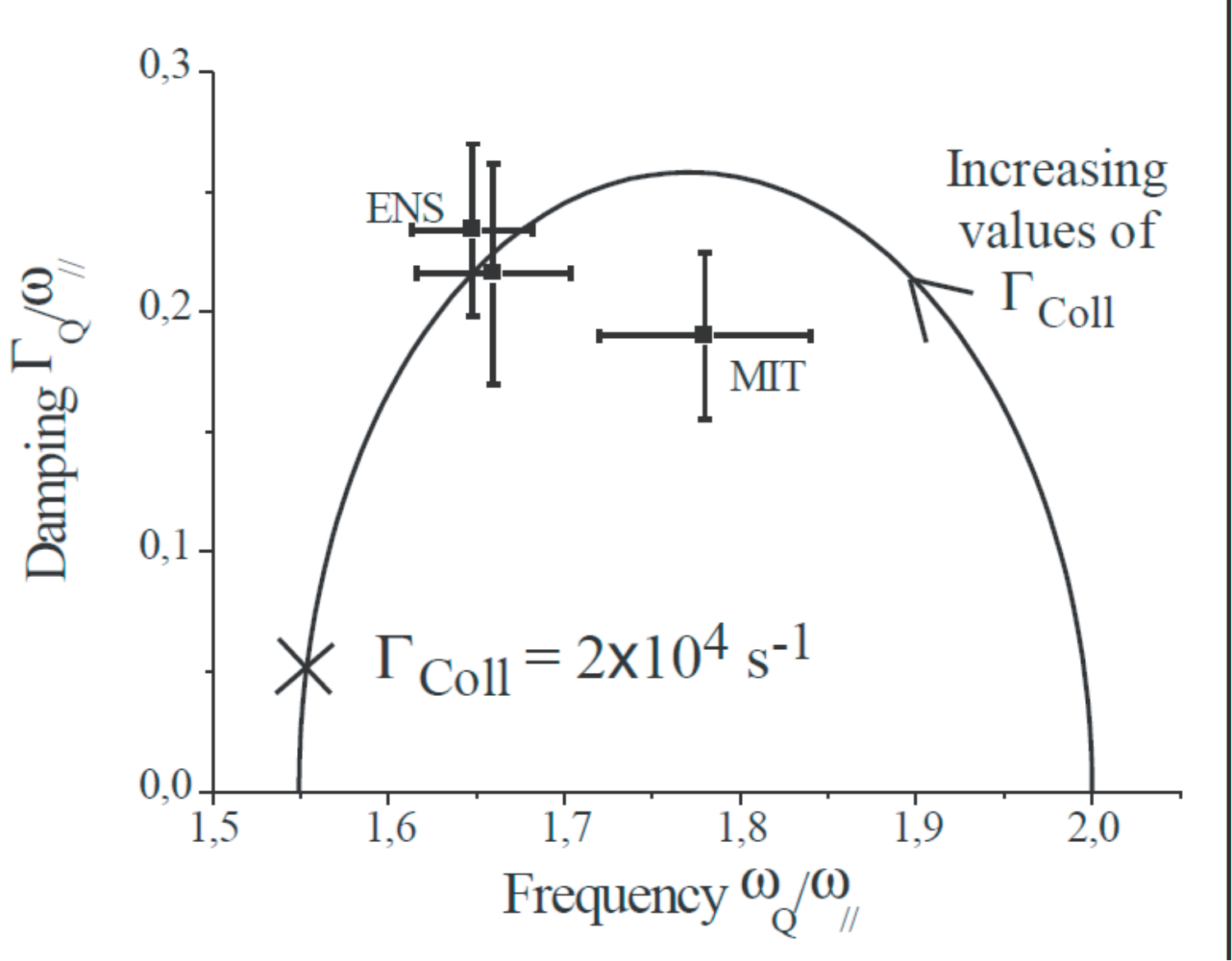}}
\caption{Damping of the quadrupole-monopole mode as a function of its frequency in units of the longitudinal frequency of the trap. In the collisionless regime, there is no damping. As the elastic collision rate increases, the frequency of the mode is shifted and damping is present. (ENS) and (MIT) correspond respectively to the measurements in \citet{Leduc:02} and \citet{Stamper-Kurn:98}. Figure extracted from \citet{Leduc:02}}
\label{fig2_sec6A}
\end{figure}

Usually, ultra-cold clouds are in the collisionless regime in which the atomic
motion is described by a single particle Hamiltonian. On the other hand, if the mean
free path between colliding atoms is small compared to dimensions of the trapped cloud, the atomic cloud is in the hydrodynamic regime. 
In this regime the oscillation frequencies of the excited modes of the gas are modified and
exhibit damping~\cite{Griffin:97,Guery-Odelin:99}. 
Since the scattering length is
notably larger in metastable helium than in alkali atoms, this
regime should be easier to observe in metastable
helium~\cite{Stamper-Kurn:98,Leduc:02}. 
In \citet{Leduc:02}, 
the quadrupole-monopole mode for several elastic collision rates was studied and
indeed a regime close to the hydrodynamic limit was reached in which a shift of $\sim
20\%$ of the mode frequency was observed (see Fig.~\ref{fig2_sec6A}). 
This regime was later also observed in a sodium BEC~\cite{VanderStam:07b}.

%
%
%
%
%
%
%

\section{Outlook}
\label{sec:conclusions}

A significant part of this review was devoted to the substantial body of work on collision processes between metastable atoms. 
This work has led to our ability to routinely cool helium to quantum degeneracy.
It has indicated that for neon, although evaporatively cooled samples of bosonic and fermionic samples have been produced, the conventional path to quantum degeneracy remains difficult though not hopeless. Work is currently concentrating on the detailed investigation and the possible modification of the elastic and inelastic collisional properties.
For the heavier metastable noble gases, inelastic collision rates are so high that we see little hope of achieving degeneracy with them without radical innovations.  
As discussed above, the study of cold collisions in helium has led to the prediction of a magnetic Feshbach resonances. We expect to see attempts to observe and exploit these resonances in the near future. The demonstration of stable mixtures of $^4{\rm He}^*$ and $^{87}{\rm Rb}$~\cite{Byron:10a,Byron:10b} leads us to speculate that useful Feshbach resonances may turn up in this system as well. 

The investigation of scattering resonances in heteronuclear mixtures also opens the possibility of studying Efimov states \cite{Kraemer:06,Knoop:09,Ferlaino:10}
in such mixtures. 
It has been suggested \cite{Incao:06}  that a heteronuclear Efimov state might permit a demonstration of universality in the spacing of Efimov resonances. 
This spacing, which is given by the factor 22.7 in homonuclear systems, is predicted to have a smaller ratio the greater the mass ratio in a heteronuclear system. 
Thus, the observation of a series of states in the same system may be easier. 
A system consisting of one He* and two Rb atoms appears to be a good candidate for such a study \cite{Knoop:11}.

For the purposes of more traditional spectroscopy, we expect that future investigations could further challenge QED by measuring the decay rates of the 3 $^{3}$P states to the ground state using a 389~nm laser to access these states from the 2 $^{3}$S$_{1}$ metastable state for comparison with theoretical calculations \cite{Morton:11}.  Singlet states can be accessed \textit{via} direct laser transition from the 2 $^{3}$S$_{1}$ metastable state to the 2 $^{1}$S$_{0}$ metastable state~\cite{Leeuwen:06, Rooij:11}.  As an alternative, high-lying $l>1$ triplet states near the ionisation limit can be excited; these exhibit a mixed singlet-triplet character and can, as a result of this mixing, decay into the singlet state manifold~\cite{Eyler:08}.  This strategy may allow a more accurate determination of the two-photon decay of the 2 $^{1}$S$_{0}$ state  (19.7(1.0) ms)~\cite{VanDyck:71}, and enable a more stringent test of QED predictions~\cite{Derevianko:97}.

Spectroscopic measurements can also be performed from the
metastable triplet state to singlet states. In particular, two different
proposals have recently been made to measure the 2 $^3$S - 2 $^1$S doubly-forbidden
transition \cite{Leeuwen:06,Eyler:08}. In the first 
\cite{Leeuwen:06}, one photon spectroscopy with a high power CW laser at
1557~nm was proposed to exploit the weak magnetic dipole transition between
these two states. This transition has been observed very recently in Amsterdam in a quantum degenerate gas and its frequency was measured, both for $^3$He and $^4$He, applying an optical frequency comb~\cite{Rooij:11}.
In the second  \cite{Eyler:08}, direct, two photon optical frequency comb
spectroscopy at 886~nm and 2060~nm was proposed, to drive a Raman transition
between the 2 $^3$S $\rightarrow$ 2 $^1$P $\rightarrow$ 2 $^1$S levels. In both cases,
cooled, trapped atoms are needed to ensure narrow linewidths and to
obtain measurable signal to noise ratios. 

From the point of view of atom optics, it seems clear that the three-dimensional
single atom detection capability afforded by metastable noble gas atoms will continue to  inspire 
many investigations. 
We have seen the first successful production of atom pairs and macroscopically occupied
twin beams 
via variants of four-wave mixing of matter waves, but other variants are possible. 
Four-wave mixing enabled by modification of dispersion relations using an 
optical lattice \cite{Hilligsoe:05} has been demonstrated in alkalis \cite{Gemelke:05, Campbell:06}.
In that work twin beams were generated but their degree of correlation, or intensity
squeezing was not measured.
In a more recent experiment using Rb \cite{Bucker:11}, workers used an
intermediate excited state in a nearly one-dimensional gas to produce twin beams
and were able to demonstrate a high degree of relative intensity squeezing between
the beams.
These two techniques can be adapted to metastable atoms and we expect that 
the ability to make observations in three dimensions with good spatial resolution 
will improve our knowledge of the details of four-wave mixing in matter waves.
For example it remains to be seen whether the atoms in the correlated beams are 
sufficiently coherent to permit their use in interferometry experiments.
Recent experiments showing the effects of mean-field interactions \cite{Krachmalnicoff:10} may be showing that uncontrolled phase shifts can be present.

If twin beams can be made sufficiently coherent, the experience of the quantum optics community suggests several possible experiments.
In Sec~\ref{sec:correlations}, we have already alluded to several theoretical proposals to observe the
entanglement which should be present among the atoms
in pair creation experiments.
These proposals seem technically challenging but we believe that some of them
will be realized eventually. 
Another possibilty inspired by quantum optics is to realize
an atomic analog of the famous 
experiment of \citet{Hong:88}.
Using atoms in such experiments will add a new twist on this effect because one can use
either bosons or fermions. 
The results are very different in the two cases.

Another important recent trend in cold atom physics has been the study of strongly correlated many body systems. 
To give some examples, Feshbach resonances in Fermi gases have been used to explore the BEC-BCS crossover regime \cite{Greiner:03, Jochim:03, Bourdel:04}, 
and the implementation of optical lattices has allowed the exploration of other many-body states 
and quantum phase transitions, the first of which was the celebrated superfluid-to-Mott insulator transition \cite{Greiner:02}. 
The physics of many-body systems realized in quantum gases is too rich to summarize here and we refer the reader to a recent review \cite{Bloch:08}.
It seems to us however, that the correlation techniques that have been developed in the context of metastable helium experiments will probably help to shed light on these systems especially in the study of quantum phase transitions where the behavior of correlation functions is often a crucial signature.

\begin{acknowledgments}
This work was initiated by the project CIGMA (Controlled Interactions in quantum Gases of Metastable Atoms), 
partly financed the European Science Foundation EuroQUAM Programme. We thank all students and coworkers
for their contribution to the research presented in this work. W.V. and G.B. acknowledge support from the Dutch Foundation for 
Fundamental Research on Matter (FOM) and the German Research Foundation (DFG) respectively.
\end{acknowledgments}


\bibliographystyle{apsrmp}
\bibliography{RMPMetastable_Atoms}

\end{document}